\documentclass[fleqn,usenatbib]{mnras}

\usepackage{newtxtext,newtxmath}
\usepackage[T1]{fontenc}

\DeclareRobustCommand{\VAN}[3]{#2}
\let\VANthebibliography\thebibliography
\def\thebibliography{\DeclareRobustCommand{\VAN}[3]{##3}\VANthebibliography}

\usepackage{graphicx}	% Including figure files
\usepackage{amsmath}	% Advanced maths commands
\usepackage{todonotes,xcolor}
\usepackage{subcaption}
\usepackage{mhchem}

\definecolor{cobalt}{rgb}{0.0, 0.28, 0.67}

\usepackage[dvipsnames]{xcolor}
\usepackage{amstext} % for \text macro
\usepackage{array}   % for \newcolumntype macro
\newcolumntype{L}{>{$}l<{$}} % math-mode version of "l" column type
\newcolumntype{R}{>{$}r<{$}} % math-mode version of "l" column type
\newcolumntype{C}{>{$}c<{$}} % math-mode version of "l" column type

\usepackage{multirow}

\title[Sense and Sensitivity]{Sense and Sensitivity - I. Uncertainty analysis of the gas-phase chemistry in AGB outflows}

\author[Van de Sande et al.]{
M. Van de Sande,$^{1}$\thanks{E-mail: \href{mailto:mvdsande@strw.leidenuniv.nl}{mvdsande@strw.leidenuniv.nl}}
M. Gueguen,$^{2}$
T. Danilovich,$^{3,4}$
T.\,J. Millar\,$^{5}$
\\
$^{1}$Leiden Observatory, Leiden University, P.O. Box 9513, 2300 RA Leiden, The Netherlands\\
$^{2}$Laboratoire CAPHI, University of Rennes, Rue du Thabor 2, 35000 Rennes, France\\
$^{3}$School of Physics \& Astronomy, Monash University, Wellington Road, Clayton 3800, Victoria, Australia\\
$^{4}$Institute of Astronomy, KU Leuven, Celestĳnenlaan 200D, 3001 Leuven, Belgium\\
$^{5}$Astrophysics Research Centre, School of Mathematics and Physics, Queen's University Belfast, University Road, Belfast BT7 1NN, UK
}

\date{Accepted XXX. Received YYY; in original form ZZZ}

\pubyear{2025}

\begin{document}
\label{firstpage}
\pagerange{\pageref{firstpage}--\pageref{lastpage}}
\maketitle

\begin{abstract}
Chemical reaction networks are central to all chemical models.
Each rate coefficient has an associated uncertainty, which is generally not taken into account when calculating the chemistry.
We performed the first uncertainty analysis of a chemical model of C-rich and O-rich AGB outflows using the \textsc{Rate22} reaction network.
Quantifying the error on the model predictions enables us to determine the need for adding complexity to the model.
Using a Monte Carlo sampling method, we quantified the impact of the uncertainties on the chemical kinetic data on the predicted fractional abundances and column densities.
The errors are caused by a complex interplay of reactions forming and destroying each species.
Parent species show an error on their envelope sizes, which is not caused by the uncertainty on their photodissociation rate, but rather the chemistry reforming the parent after its photodissociation.
Using photodissociation models to estimate the envelope size might be an oversimplification.
The error on the CO envelope impacts retrieved mass-loss rates by up to a factor of two.
For daughter species, the error on the peak fractional abundance ranges from a factor of a few to three orders of magnitude, and is on average about 10\% of its value.
This error is positively correlated with the error on the column density.
The standard model suffices for many species, e.g., the radial distribution of cyanopolyynes and hydrocarbon radicals around IRC\,+10216.
However, including spherical asymmetries, dust-gas chemistry, and photochemistry induced by a close-by stellar companion are still necessary to explain certain observations.
\end{abstract}

%Errors are caused by the interplay of the reactions forming and destroying each species.

% 250 words


%%%%%%%%%%%%%%%%%%%%%%%%%%%%%%%%%%%%%%%%%%%%%%%%%%

%%%%%%%%%%%%%%%%% BODY OF PAPER %%%%%%%%%%%%%%%%%%

%%%%%%%%%%%%%%%%%%%%%%%%%%%%%%%%%%%%%%%%%%%%%%%%%%%%%%%%%%%%%%%%%%%%%%%%%%%%%%%%%%%%%%%%%%%%%%%%%%%%%%%%%%%%%%
\section{Introduction}
%%%%%%%%%%%%%%%%%%%%%%%%%%%%%%%%%%%%%%%%%%%%%%%%%%%%%%%%%%%%%%%%%%%%%%%%%%%%%%%%%%%%%%%%%%%%%%%%%%%%%%%%%%%%%%

The outflows of asymptotic giant branch (AGB) stars are rich astrochemical environments, with around 130 different molecules and some 15 types of newly formed dust species detected so far \citep{Decin2021}.
The type of chemistry in the outflow is set by the star's elemental carbon-to-oxygen ratio (C/O), where stars with C/O < 1 have O-rich outflows and stars with C/O > 1 have C-rich outflows.
Asymmetrical structures are observed to be ubiquitous within their circumstellar envelopes (CSEs), ranging from small-scale structures close to the star, likely caused by its convective motion \citep{Khouri2016,Wittkowski2017,VelillaPrieto2023}, and large-scale structures such as spirals \cite[e.g.,][]{Mauron2006,Maercker2016} and disks \cite[e.g.,][]{Kervella2016,Homan2018,Safonov2025}, which are thought to be caused by binary interaction with a (sub)stellar companion \citep{Decin2020}.
Chemical kinetics models are crucial to understand the close link between chemistry and dynamics throughout the outflow and are essential to interpret the observed abundance distributions.

The most widely used chemical model of AGB outflows describes the gas-phase chemistry from the inner wind outwards, assuming a spherically symmetric outflow with constant expansion velocity \cite[e.g.,][]{Goldreich1976,Scalo1980,Agundez2006,McElroy2013}.
This standard CSE model has successfully explained several observations, especially those of the photochemistry-dominated outer regions of the outflow \cite[e.g.,][]{Huggins1982,Nejad1984,Cherchneff1993,Millar2000,Li2016}.
Nevertheless, important discrepancies with observations remain.

% close the gap with
Additional physics and chemistry has been included in the standard model to better understand observations.
\emph{(i)} Small- and large-scale density structures impact the radiation field throughout the outflow, allowing interstellar UV photons to reach the inner wind. 
These are included in the 1D model by allowing a set fraction of photons to reach the inner wind uninhibited \citep{Agundez2010} or by using the porosity formalism, where the change in opacity of the outflow depends on its specific clumpiness \citep{VandeSande2018}. 
The impact of close-by stellar companions depends on the type of companion and the density structure of the outflow, where the outflow can appear either molecule-poor or show an increase in chemical complexity \citep{VandeSande2022}.
The model can account for the detection of \ce{HC3N} close to IRC\,+10216 \citep{Siebert2022} and that of SiC and SiN in the inner wind of W Aql \citep{Danilovich2024}.
\emph{(iii)} The chemical network can be extended to include simplified dust-gas chemistry \citep{Jura1985,Charnley1993,Dijkstra2003,Dijkstra2006} or comprehensive dust-gas and grain-surface chemistry \citep{VandeSande2019b,VandeSande2020,VandeSande2021}.
This allows for the formation of ices on dust, in particular \ce{H2O} ice as observed around OH/IR stars \citep{Sylvester1999}, and corroborates the observed depletion of SiO and SiS in high-density O-rich outflows \citep{GonzalezDelgado2003,Bujarrabal1989,Decin2010b} and trends in the abundance of refractory species with outflow density \citep{GonzalezDelgado2003,Massalkhi2019,Massalkhi2020}.
The three layers of complexity described above can be combined, which reveals degeneracies in the model predictions \citep{VandeSande2023}.
Including (a combination of) physical and chemical complexities changes the model predictions and improves the agreement with observed abundance profiles.

Chemical models are used in a deterministic manner, giving one solution for a given set of certain input parameters.
This includes the rate coefficients of the reactions in the chemical network, of which each has an associated uncertainty. 
The model predictions hence have an unknown inherent error due to the uncertainty on the kinetic data at the heart of the model.
Assessing the effect of uncertainties on the kinetic data on the model predictions has been done for a variety of astrochemical environments \cite[e.g.,][]{Dobrijevic2003,Vasyunin2004,Wakelam2005,Wakelam2006a,Wakelam2006b,Vasyunin2008,Dobrijevic2010,Penteado2017}, including IRC\,+10216 \citep{Wakelam2010}.
Here, we present the first uncertainty analysis of the standard CSE model, investigating three different outflow densities for both C-rich and O-rich outflows.
Allocating the uncertainty on the model predictions to specific reactions requires a sensitivity analysis, which will be presented in a follow-up paper.

Previously added physics and chemistry are not taken into account, but regarded as perturbations to the standard model.
As we do not vary significantly over the physical description of the outflow, this study complementary to that of \citet{Maes2023}, where the physics was varied but the chemistry was considered to be fixed.
By quantifying the error on the model predictions due to the uncertainties on the input chemical kinetic data, we can establish the need for adding further physical and chemical complexity to the model.
Furthermore, this uncertainty analysis provides insights into both the chemical model and AGB outflows themselves, inasmuch as we can determine what kind of complexities could solve which types of discrepancies with observations.

In Sect. \ref{sect:methods}, the chemical model is described along with the Monte Carlo method used.
We show our results in Sect. \ref{sect:results}, followed by the discussion and conclusions in Sect. \ref{sect:disc} and Sect. \ref{sect:concl}.

\begin{figure}
\centering
\includegraphics[width=\columnwidth]{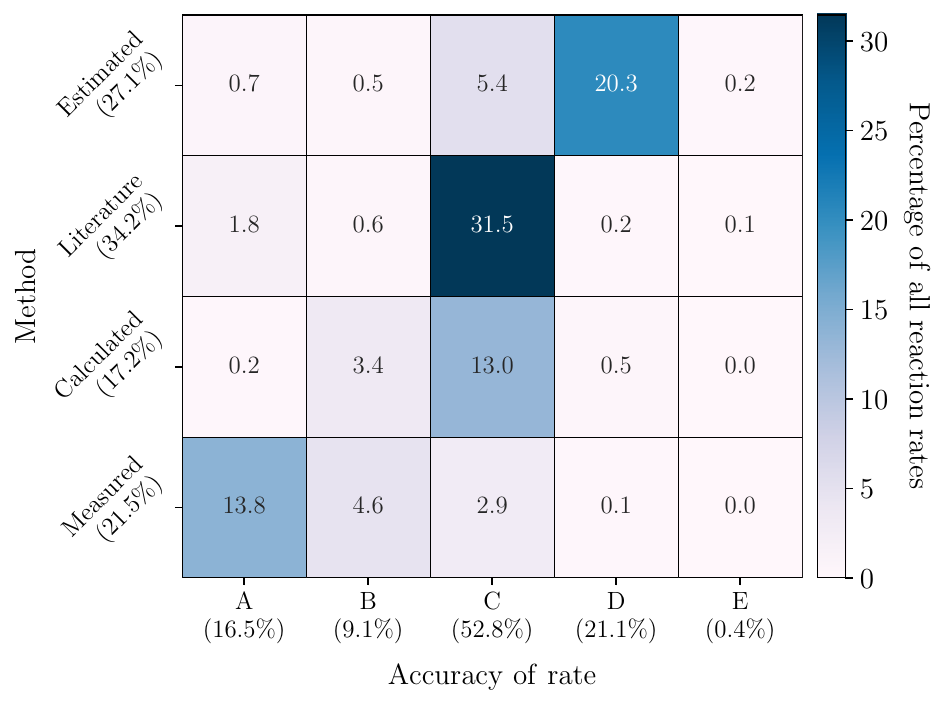}
\caption{Distribution of the \textsc{Rate22} reaction rates per associated accuracy and the method by which the rate has been determined.
Percentages are indicated in colour and as printed in each category.
An accuracy of \emph{A} represents an error of less than 25\%, \emph{B} less than 50\%, \emph{C} within a factor of 2, \emph{D} within an order of magnitude, and \emph{E} a highly uncertain error
The percentage of rates for each method and accuracy are indicated at each row and column, respectively.
}
\label{fig:rate22}
\end{figure}

%%%%%%%%%%%%%%%%%%%%%%%%%%%%%%%%%%%%%%%%%%%%%%%%%%%%%%%%%%%%%%%%%%%%%%%%%%%%%%%%%%%%%%%%%%%%%%%%%%%%%%%%%%%%%%
\section{Chemical model and Monte Carlo method}				\label{sect:methods}
%%%%%%%%%%%%%%%%%%%%%%%%%%%%%%%%%%%%%%%%%%%%%%%%%%%%%%%%%%%%%%%%%%%%%%%%%%%%%%%%%%%%%%%%%%%%%%%%%%%%%%%%%%%%%%

The chemical reaction network and chemical model are described in Sects \ref{subsect:methods:network} and \ref{subsect:methods:model}.
The Monte Carlo sampling method used to take the uncertainties on the rate coefficients into account is described in Sect. \ref{subsect:methods:mc}.
The methods to calculate the model predictions and their errors are described in Sect. \ref{subsect:methods:method}.

%---------------------------------------------------------------------------------------------------------
\subsection{The UDfA \textsc{Rate22} reaction network }		\label{subsect:methods:network}
%---------------------------------------------------------------------------------------------------------

The chemical reaction network used is the publicly available \textsc{Rate22} release of the UMIST Database for Astrochemistry (UDfA)\footnote{\url{https://umistdatabase.net/}}, a gas-phase only network containing 8767 rate coefficients involving 737 species \citep{Millar2024}.
For each rate coefficient, the database lists the method with which it has been determined and an estimate of its accuracy.
The accuracy estimate is divided into five categories, where category \emph{A} represents an error of less than 25\%, \emph{B} an error of less than 50\%, \emph{C} an error within a factor of 2, \emph{D} within an order of magnitude, and \emph{E} a highly uncertain error.

Fig. \ref{fig:rate22} visualises the \textsc{Rate22} reaction network per accuracy and per method.
About one-fifth (21.5\%) of all rate coefficients is measured experimentally, 17.2\% of rate coefficients are calculated. 
The remaining $\sim$60\% of rate coefficients are either estimated or obtained from the literature (referring to coefficients harvested from published networks or when no information on the method could be determined).
Only a quarter of all 8767 rate coefficients (25.6\%) have an error of less than 50\% (accuracy \emph{A} and \emph{B}).
About one-third of all coefficients (31.5\%) are obtained from the literature with an accuracy \emph{C} (factor 2).
On average, the accuracy of the coefficient decreases as the method shifts to estimate or literature.
Measured coefficients are the most accurate, with the majority having an accuracy \emph{A}.
Most calculated coefficients have an accuracy \emph{C}, most estimated coefficients have an accuracy \emph{D}.
Many of the rate coefficients apply only to a restricted range in temperature.
Their uncertainty outside this range is unknown; we assume the same uncertainty as within the range.

The chemical reaction network is closed but is not complete, indeed no network can be complete.
While great effort has gone into including new species detected over the past decade in the \textsc{Rate22} network, we cannot know whether all relevant reactions are covered to sufficiently describe inter- and circumstellar chemistry.
Assessing the effects of the uncertainties on the rate coefficients in the network is one of the first steps to start addressing effects of the (in)completeness of the reaction network on the model predictions.

%---------------------------------------------------------------------------------------------------------
\subsection{Circumstellar envelope chemical model }		\label{subsect:methods:model}
%---------------------------------------------------------------------------------------------------------

We use the publicly available CSE model associated with the UDfA \textsc{Rate22} release\footnote{\url{https://github.com/MarieVdS/rate22_cse_code}}.
The model describes a spherically symmetric outflow with constant mass-loss rate, $\dot{M}$, and expansion velocity, $v_\infty$; density hence falls as $1/r^2$. 
The gas temperature follows a power law.
CO self-shielding is taken into account via a single-band approximation \citep{Morris1983}; \ce{H2} is assumed to be fully self-shielded.

While the \textsc{Rate22} CSE model is able to include porosity and a close-by stellar companion, we only consider a smooth outflow without any internal UV photons.
This standard CSE model is still most commonly used when interpreting observations or predicting observables \cite[e.g.,][]{Huggins1982,Millar1994,Millar2000,Li2016,Agundez2017,Danilovich2017,Saberi2019}.
In our approach, changes to the physics and chemistry of the model are regarded as perturbations to the standard CSE model whose validity is not necessarily justified. 

The physical model parameters are listed in Table \ref{table:model-params}.
We consider a high, intermediate, and low density outflow for both C-rich and O-rich CSEs. 
The temperature profile is the same for all models, as are the start (at $4\times10^{13}$ cm $= 2\ R_*$) and end radii (at $10^{18}$ cm).
The standard interstellar UV field \citep{Draine1978} is assumed to impinge uniformly on the outside of the outflow and is attenuated by dust \citep{McElroy2013}.
The radial range is sampled logarithmically with a step size of 0.03 dex, resulting in 89 radial points.
The parent species, assumed to be present at the start of the model, and their initial abundances are given in Table \ref{table:model-parents}.
As they do not include Ca, Ti or Al, the reactions involving these atoms play no part in the uncertainty analysis.
The computation time of one model is $\sim$25~s.

\begin{table}
	\centering
	\caption{Physical parameters of the chemical model.}
	\begin{tabular}{l l} 
		\hline
    Outflow density, $\dot{M}$ - $v_\infty$        	&    $10^{-5}$  $\mathrm{M}_\odot\ \mathrm{yr}^{-1}$ - 15 km s$^{-1}$  \\
    									&    $10^{-6}$  $\mathrm{M}_\odot\ \mathrm{yr}^{-1}$ - 10 km s$^{-1}$  \\
    									&    $10^{-7}$  $\mathrm{M}_\odot\ \mathrm{yr}^{-1}$ - 5 km s$^{-1}$  \\
    \noalign{\smallskip}
    Stellar temperature, $T_*$        & 2000 K \\
    Exponent $T(r)$, $\epsilon$                    & $-0.7$ \\
    Stellar radius, $R_*$             & 2 $\times 10^{13}$ cm \\
    \noalign{\smallskip}
	Start of the model				& 4 $\times 10^{13}$ cm \\
	End of the model				& 1 $\times 10^{18}$ cm \\
		\hline
	\end{tabular}
    \label{table:model-params}    
\end{table}

\begin{table}
	\caption{Parent species and their abundances relative to \ce{H2} for the C-rich and O-rich outflows, as derived from observations by \citet{Agundez2020}. 
	The CO abundances were retrieved by \citet{Teyssier2006}.
	} 
% 	\centering
    % \resizebox{1.0\columnwidth}{!}{%
    \centering
    \begin{tabular}{l r c  l r }
    \hline  
    \noalign{\smallskip}
    \multicolumn{2}{c}{Carbon-rich} && \multicolumn{2}{c}{Oxygen-rich}  \\  
    \cline{1-2} \cline{4-5} 
    \noalign{\smallskip}
    Species & Abun. & & Species & Abun. \\
    \cline{1-2} \cline{4-5} 
    \noalign{\smallskip}
    He		&  0.17				& & He		& 0.17  \\
    CO		& $8.00\times10^{-4}$	& & CO		& $3.00 \times 10^{-4}$  \\
    N$_2$		& $4.00 \times 10^{-5}$	& & H$_2$O	& $2.15 \times 10^{-4}$  \\
    CH$_4$	& $3.50 \times 10^{-6}$	& & N$_2$ 	& $4.00 \times 10^{-5}$  \\ 
    H$_2$O	& $2.55 \times 10^{-6}$	& & SiO 	& $2.71 \times 10^{-5}$  \\ 
    SiC$_2$	& $1.87 \times 10^{-5}$	& & H$_2$S 	& $1.75 \times 10^{-5}$  \\
    CS		& $1.06 \times 10^{-5}$	& & SO$_2$ 	& $3.72 \times 10^{-6}$  \\
    C$_2$H$_2$& $4.38 \times 10^{-5}$	& & SO 		& $3.06 \times 10^{-6}$  \\
    HCN		& $4.09 \times 10^{-5}$	& & SiS 		& $9.53 \times 10^{-7}$  \\
    SiS   		& $5.98 \times 10^{-6}$	& & NH$_3$ 	& $6.25 \times 10^{-7}$  \\ 
    SiO 		& $5.02 \times 10^{-6}$	& & CO$_2$ 	& $3.00 \times 10^{-7}$  \\   
    HCl		& $3.25 \times 10^{-7}$	& & HCN 	& $2.59 \times 10^{-7}$  \\  
    C$_2$H$_4$& $6.85 \times 10^{-8}$	& & PO 		& $7.75 \times 10^{-8}$  \\ 
    NH$_3$	& $6.00 \times 10^{-8}$	& & CS 		& $5.57 \times 10^{-8}$  \\
    HCP		& $2.50 \times 10^{-8}$	& & PN 		& $1.50 \times 10^{-8}$  \\
    HF    		& $1.70 \times 10^{-8}$	& &  HCl		& $1.00 \times 10^{-8}$  \\  
    H$_2$S	& $4.00 \times 10^{-9}$	& & 	HF	& $1.00 \times 10^{-8}$  \\    
    Na	& $1.00 \times 10^{-8}$	& & 	Na	& $1.00 \times 10^{-8}$	  \\    
    Mg	& $1.00 \times 10^{-8}$	& & 	Mg	& $1.00 \times 10^{-8}$	  \\    
    Fe	& $1.00 \times 10^{-8}$	& & 	Fe	& $1.00 \times 10^{-8}$	  \\    
    \hline 
    \end{tabular}%
    % }
    \label{table:model-parents}    
\end{table}

%---------------------------------------------------------------------------------------------------------
\subsection{Monte Carlo method of including uncertainties of the rates}		\label{subsect:methods:mc}
%---------------------------------------------------------------------------------------------------------

%%%Characterising the model form uncertainty - uncertainty of kinetic data  = validation of the model.

%Local methods to explore the input parameter space, such as one at a time (OAT) methods where one rate coefficient is varied over while the other are kept constant, are invalid for non-linear problems like chemical kinetics.
%These methods cannot account for the interactions between reactions and, moreover, cannot adequately explore the multi-dimensional input parameter space.
%Global methods are necessary to propagate input uncertainties through the model \citep{Saltelli2019}. 

To map the uncertain rate coefficient input space to the uncertain output space, we use a Monte Carlo sampling method following \citet{Dobrijevic1998} and \citet{Dobrijevic2003}, as used by, e.g., \citet{Wakelam2005}, \citet{Wakelam2006a}, and \citet{Vasyunin2008}.
Each rate coefficient $k$ is considered to be a random variable, following a lognormal distribution centred on the listed reaction rate $k_0$.
This assumption is justified in the absence of systematic errors \citep{Thompson1991,Stewart1996}.
The width of the lognormal distribution is given by the reaction's uncertainty factor $F$, determined by the accuracy estimate of the coefficient: $\log(F) = 1.25$ for \emph{A}, 1.5 for \emph{B}, 2 for \emph{C} and 10 for \emph{D}.
For the 32 highly uncertain reactions labeled \emph{E}, we assumed $\log(F) = 10$.

A new set of rate coefficients is generated by randomly sampling each rate coefficient within its associated uncertainty range.
For each rate coefficient $i$, the new rate coefficient is given by
\begin{equation}		\label{eq:k}
    \log k_i = \log k_{i,0} + \epsilon_i \log F_i,
\end{equation}
where $\epsilon_i$ is a random number sampled from a normal distribution with mean of 0 and standard deviation of 1 and $k_{i,0}$ is the rate coefficient as listed in the network.
We therefore adjust the total calculated rate coefficient, calculated at the local temperature of the outflow.
Uncertainties on energy barriers are absorbed into the total rate coefficient. 
However, their values in the network are generally not independent of the choice of the pre-exponential factor, as the adopted form is most often derived from a fit to experimental or theoretical rate coefficients over a specific temperature range.
Photoreaction rates are calculated at each radius, as dust attenuates the unshielded rate (computed using photo-cross-sections where available). 
These attenuated rates are then adjusted using Eq. \ref{eq:k}.

By sampling each rate coefficient $N$ times, $N$ sets of reaction networks are generated.
The value of $\epsilon$ is different for each rate coefficient and each sample.
The chemical model is calculated using each new, randomly sampled reaction network, resulting in $N$ sets of predicted abundance profiles $X_j(r)$ for each species $j$.

Note that local methods, such as one at a time (OAT) methods where one rate coefficient is varied over while the others are kept constant, are invalid for chemical kinetics as it is a highly non-linear problem.
Such methods cannot account for the interactions between reactions and, moreover, cannot adequately explore the multi-dimensional input parameter space.
Global methods, like the Monte Carlo sampling method which we use (where a OAT design is repeated for all rate coefficients), are necessary to propagate input uncertainties through the model \citep{Saltelli2019}. 
Other methods, such the sampling technique of \citet{Morris1991} where the range of possible values for each parameter is sampled by a grid of equally spaced input values, can be used for chemical reaction network studies as well \citep{Bensberg2024}.

We generated $N = $ 10,000 reaction networks via random sampling.
This sample size is large enough to ensure convergence of the model results (see Appendix \ref{app:convergence}).
The chemical model was calculated using each of the newly generated networks, for each of the three outflow densities and for C-rich and O-rich chemistry, resulting in 6 $\times$ 10,000 total models.
The parent species (Table \ref{table:model-parents}) are kept constant when varying the rate coefficients; we do not consider variations in the initial composition in the outflow.

%---------------------------------------------------------------------------------------------------------
\subsection{Methods for model predictions}		\label{subsect:methods:method}
%---------------------------------------------------------------------------------------------------------

The chemical model has two outputs for each species: its fractional abundance profile $X(r)$ w.r.t. \ce{H2} and its column density.
Sect. \ref{subsubsect:uncert:method:mean} describes the calculation of the mean and error of each output.
In Sect. \ref{subsubsect:uncert:method:disp}, we define the dispersion on the column density and peak fractional abundance predictions for the daughter species.
In Sect. \ref{subsubsect:uncert:method:env}, we define the mean envelope size of the parent species and its error.

%--.--.--.--.--.--.--.--.--.--.--.--.--.--.--.--.--.--.--.--.--.--.--.--.--.--.--.--.--.--.--.--.--.--.--.--
\subsubsection{Mean value and error}		\label{subsubsect:uncert:method:mean}
%--.--.--.--.--.--.--.--.--.--.--.--.--.--.--.--.--.--.--.--.--.--.--.--.--.--.--.--.--.--.--.--.--.--.--.--

The mean fractional abundance, $\langle \log X(r)\rangle$, is calculated via
\begin{equation}        \label{eq:mean}
    \langle \log X(r)\rangle = \frac{1}{N} \sum_{i=0}^N \log X_i(r).
\end{equation}
This is the recommended value of the model prediction, which might differ from the fiducial prediction based on the listed rate coefficients.  
The error on the mean abundance, $\Delta \log X(r)$, is defined by 
\begin{equation}		\label{eq:error}
	\Delta \log X(r) = \frac{1}{2} \left(\log X_\mathrm{max} - \log X_\mathrm{min} \right),
\end{equation}
where $\left[\log X_\mathrm{min},\log X_\mathrm{max} \right]$ is the smallest interval that contains $95.4\%$ of all abundance profiles. 
If the distribution of $\log X(r)$ is Gaussian, this interval corresponds to $2 \sigma$ of the distribution.
However, we do not use the standard deviation as a measure for the error, as rapid changes in abundance (e.g., the onset of photodissociation) lead to a non-Gaussian distribution of $\log X$ \citep{Wakelam2005}.

Similarly, the logarithms of the predicted column densities also do not always follow a Gaussian distribution.
We therefore use Eqs (\ref{eq:mean}) and (\ref{eq:error}) to calculate the mean column density and its error.
Such deviations from a Gaussian distribution are illustrated in Appendix \ref{app:gaussian} for both fractional abundance profiles and column densities.

\begin{figure*}
\centering

\begin{subfigure}[h]{1.0\textwidth}
  \centering
	\includegraphics[width=\textwidth]{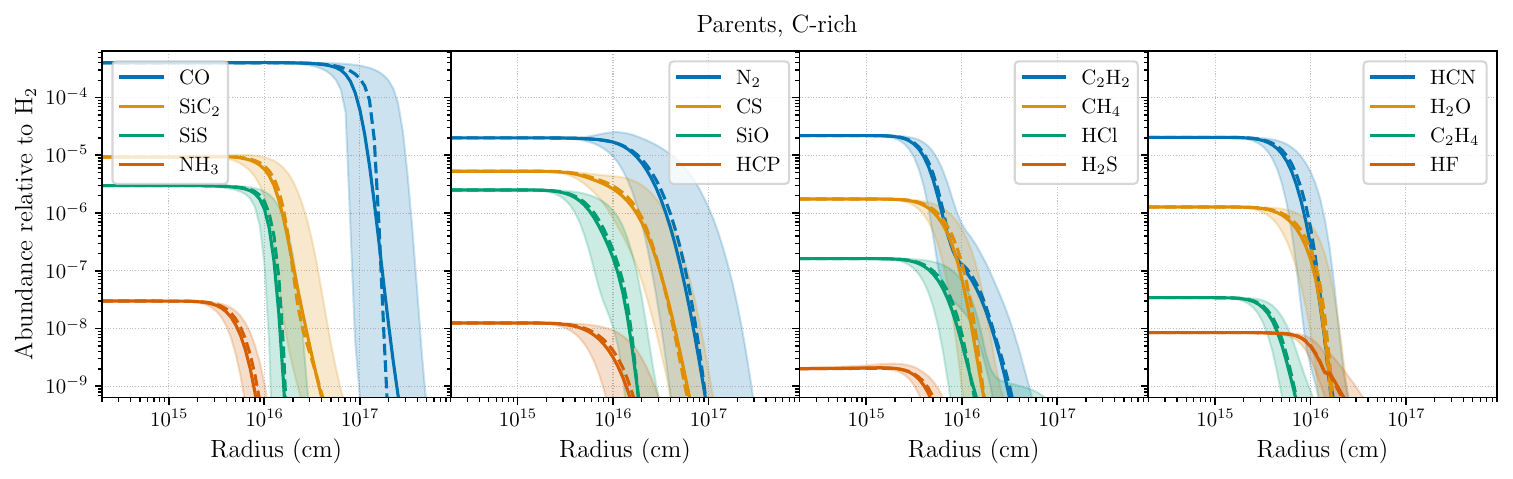}
%  \caption{1a}
  \label{fig:sfig1}
\end{subfigure}%

\vspace{-1em}

\begin{subfigure}[h]{1.0\textwidth}
  \centering
	\includegraphics[width=\textwidth]{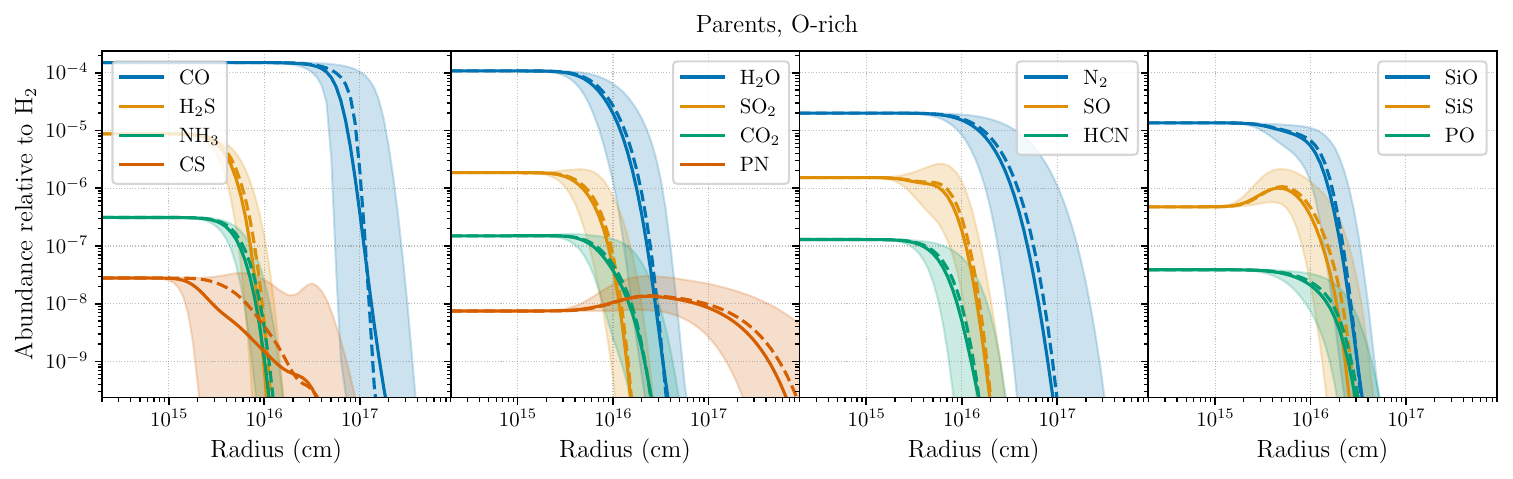}
%  \caption{1a}
  \label{fig:sfig1}
\end{subfigure}%
\caption{
Fractional abundance w.r.t. \ce{H2} of the C-rich \emph{(top)} and O-rich \emph{(bottom)} parent species in an outflow with $\dot{M} = 10^{-6}\ \mathrm{M}_\odot\ \mathrm{yr}^{-1}$. 
The dashed line shows the fiducial model prediction using \textsc{Rate22}. 
The solid line shows the mean abundance $\langle \log X(r)\rangle$ of the Monte Carlo sample of reaction networks.
The shaded region contains 95.4\% of all predicted profiles and corresponds to the error $\Delta \log X(r)$ on the mean abundance.
}
\label{fig:parents-abprof}
\end{figure*}

\begin{figure*}
\centering
\includegraphics[width=0.8\textwidth]{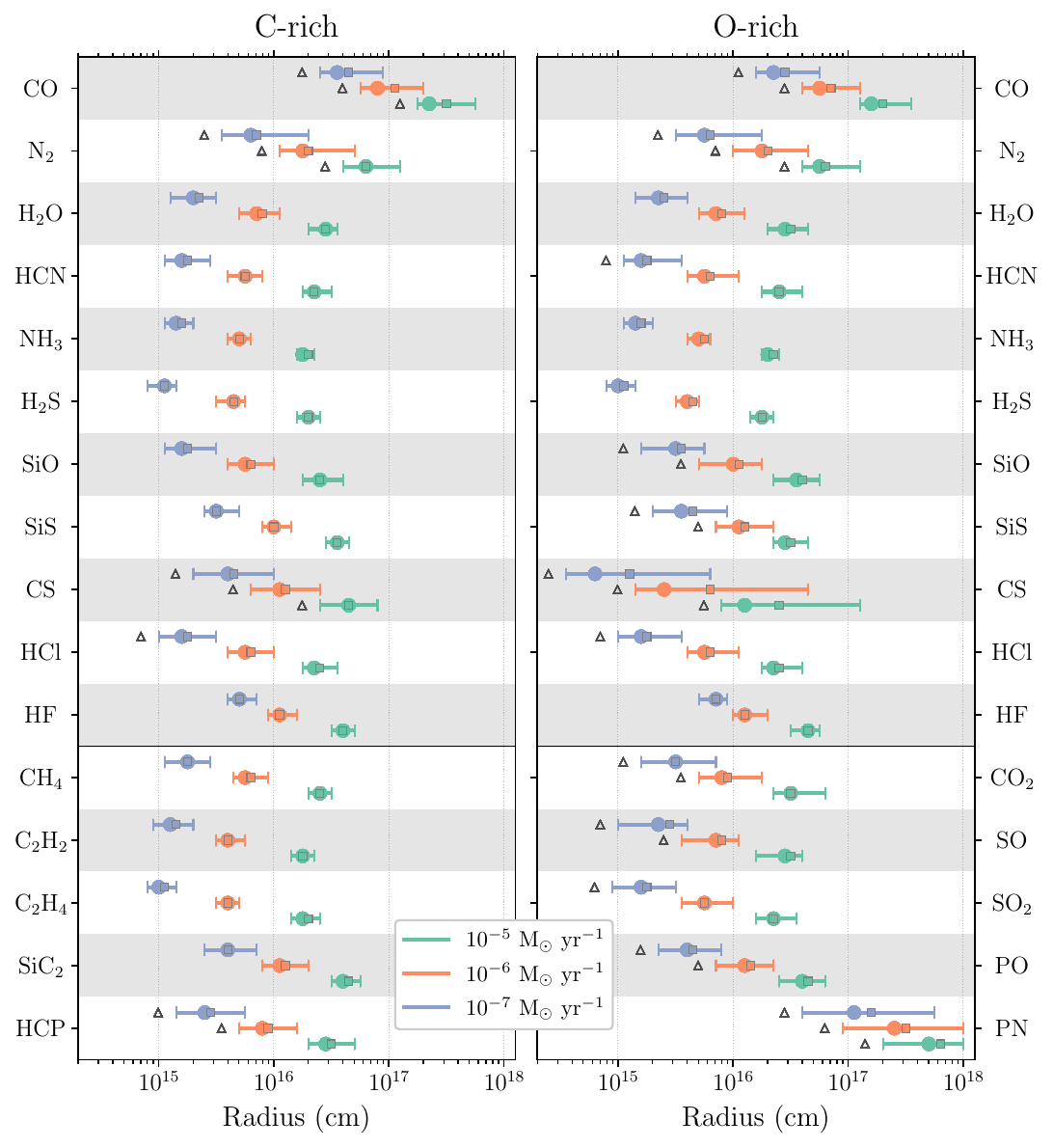}
\caption{
Envelope sizes and their radial extents for the parent species in C-rich (\emph{left}) and O-rich (\emph{right}) outflows.
Different colours indicate different mass-loss rates.
Species in common are shown near the top, differing species near the bottom.
Square: envelope size as predicted using \textsc{Rate22}. 
Circle: mean envelope size obtained from the Monte Carlo sample of reaction networks.
Envelopes that are not well-constrained are marked by a triangle.
These have a range in radial extent (ratio between the largest and smallest radial extent) larger than 3.
}
\label{fig:parents-extent}
\end{figure*}

%--.--.--.--.--.--.--.--.--.--.--.--.--.--.--.--.--.--.--.--.--.--.--.--.--.--.--.--.--.--.--.--.--.--.--.--
\subsubsection{Dispersion of column density and peak abundance}		\label{subsubsect:uncert:method:disp}
%--.--.--.--.--.--.--.--.--.--.--.--.--.--.--.--.--.--.--.--.--.--.--.--.--.--.--.--.--.--.--.--.--.--.--.--

To quantify the precision of the model predictions, we use their dispersion given by
\begin{equation}
	\mathrm{Dispersion} =  \frac{\mathrm{Mean}}{\mathrm{Error}} \times 100 \%,
\end{equation} 
where the mean and error are obtained from Eqs (\ref{eq:mean}) and (\ref{eq:error}).
The dispersion is hence a measure of the spread of the predicted values around their mean value.
A smaller dispersion corresponds to a larger precision of the model predictions. 
Note that this is not  equal to a higher accuracy, as the true value of the model is unknown.

The dispersion can be calculated at each radius of the fractional abundance profiles. 
We focus on the dispersion of the peak abundance both for feasibility of the analysis and relevance to observations.
This is done only for the daughter species.
We divide them into two groups: (i) \emph{all} daughters with a column density of at least $10^{5}$ cm$^{-2}$, removing all species that do not contribute to the overall chemistry, and (ii) the \emph{abundant} daughters, which we define as having a peak fractional abundance of at least $10^{-10}$ w.r.t. \ce{H2} and a column density of at least $10^{12}$ cm$^{-2}$.
Within each group, we define the least precise (most uncertain) daughter species as those with dispersions larger than the mean dispersion plus one standard deviation for both peak abundance and column density. 
Similarly, the most precise (least uncertain) daughters are those with dispersions larger than the mean dispersion minus one standard deviation.

%--.--.--.--.--.--.--.--.--.--.--.--.--.--.--.--.--.--.--.--.--.--.--.--.--.--.--.--.--.--.--.--.--.--.--.--
\subsubsection{Envelope sizes of parent species}		\label{subsubsect:uncert:method:env}
%--.--.--.--.--.--.--.--.--.--.--.--.--.--.--.--.--.--.--.--.--.--.--.--.--.--.--.--.--.--.--.--.--.--.--.--

Similar dispersions are not calculated for the parent species.
Their peak abundances are simply the input abundance at the start of the model, which also dominates their column densities.
For the parent species, the variation on the predicted envelope size is more meaningful.
The mean envelope size is measured by determining the radius where the mean fractional abundance $\langle \log X(r) \rangle$ has decreased by a factor $\log e$. 
Its radial extent corresponds to the radii where $\langle \log X(r) \pm \Delta \log X(r) \rangle$ has decreased by a factor $\log e$.
These values correspond to the recommended value of the predicted envelope size and its error.

As a measure of the precision of the envelope, we use the range in radial extent, defined as the ratio between the largest and smallest predicted radial extents.
We consider an envelope extent to be well-constrained if this range is smaller than a factor three.

%%%%%%%%%%%%%%%%%%%%%%%%%%%%%%%%%%%%%%%%%%%%%%%%%%%%%%%%%%%%%%%%%%%%%%%%%%%%%%%%%%%%%%%%%%%%%%%%%%%%%%%%%%%%%%
\section{Results}		\label{sect:results}
%%%%%%%%%%%%%%%%%%%%%%%%%%%%%%%%%%%%%%%%%%%%%%%%%%%%%%%%%%%%%%%%%%%%%%%%%%%%%%%%%%%%%%%%%%%%%%%%%%%%%%%%%%%%%%

Uncertainties on the kinetic data affect both the column density and abundance predictions, resulting in errors on the model output.
Parent and daughter species are influenced in different ways. 
Their results are shown in Sects \ref{subsect:uncert:parents} and \ref{subsect:uncert:daughter}, respectively.

%The abundance profiles of parent species are unaffected by the uncertainty on the kinetic data before the onset of photodissociation, but show an error on their envelope sizes. 
%We find that the spread in envelope size depends on both the parent species and the chemistry of the outflow. 
%Since all photodissociation reactions having the same uncertainty, the error on the envelope size is due to the reactions reforming the parent after its photodissociation. 
%Most envelopes are well-constrained, with notable exceptions being CO and CS.
%
%
%The abundance profiles of daughter species show significant errors from the start of the model, which leads to errors on their column density predictions. 
%Both the error on the peak fractional abundance and column density generally increase as their value decreases, reaching three and two orders of magnitude, respectively. 
%The dispersion of the peak fractional abundance is on average around 10\%, that of the column density around 3\%.   
%Moreover, these dispersions are correlated, implying that species with a more precise column density estimate tend to have a more precise peak fractional abundance estimate, and vice versa.

\begin{figure*}
\centering

\begin{subfigure}[h]{1.0\textwidth}
  \centering
	\includegraphics[width=\textwidth]{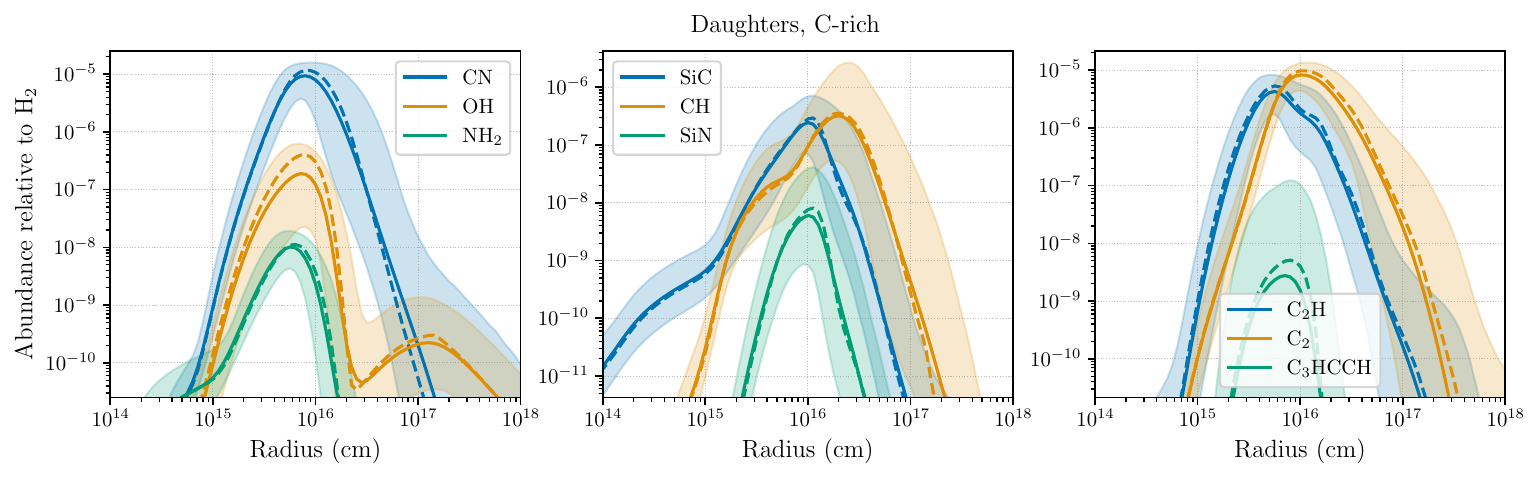}
%  \caption{1a}
  \label{fig:sfig1}
\end{subfigure}%

\vspace{-1em}

\begin{subfigure}[h]{1.0\textwidth}
  \centering
	\includegraphics[width=\textwidth]{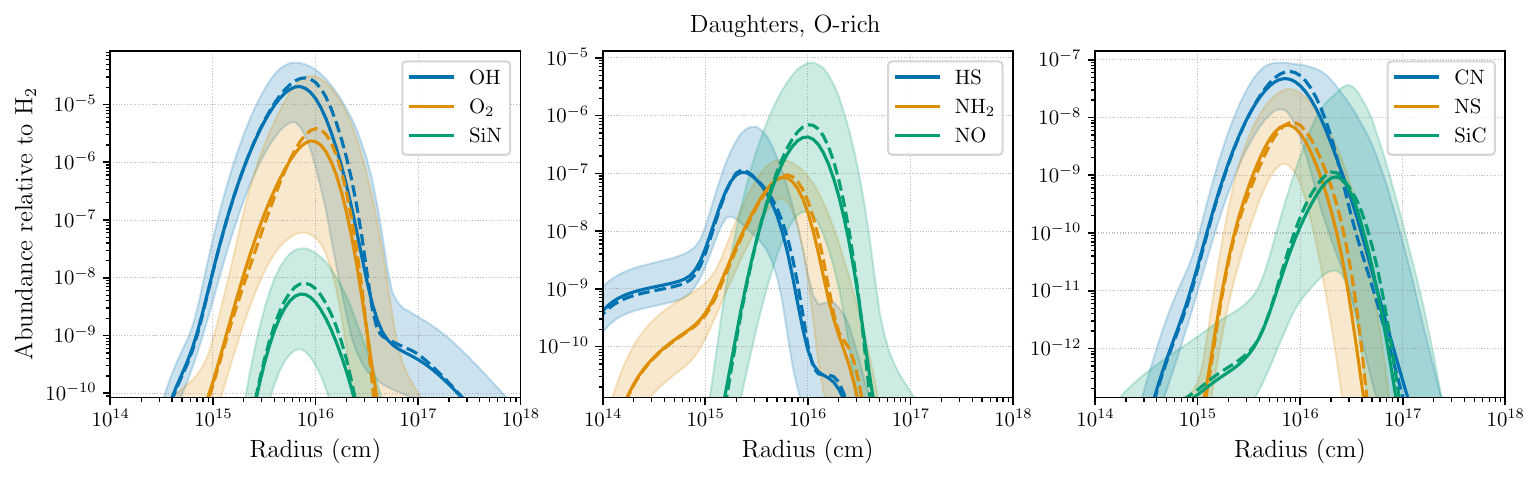}
%  \caption{1a}
  \label{fig:sfig1}
\end{subfigure}%
\caption{
Fractional abundance w.r.t. \ce{H2} of a selection of daughter species in a C-rich \emph{(top)} and O-rich \emph{(bottom)} outflow with $\dot{M} = 10^{-6}\ \mathrm{M}_\odot\ \mathrm{yr}^{-1}$. 
The dashed line shows the fiducial model prediction using \textsc{Rate22}. 
The solid line shows the mean abundance $\langle \log X(r)\rangle$ of the Monte Carlo sample of reaction networks.
The shaded region contains 95.4\% of all predicted profiles and corresponds to the error $\Delta \log X(r)$ on the mean abundance.
}
\label{fig:daughters-abprof}
\end{figure*}

%---------------------------------------------------------------------------------------------------------
\subsection{Uncertainty on parents' molecular envelope sizes}		\label{subsect:uncert:parents}
%---------------------------------------------------------------------------------------------------------

Fig. \ref{fig:parents-abprof} shows the mean fractional abundance profiles and their errors for the C-rich and O-rich parents in an outflow with $\dot{M} = 10^{-6}\ \mathrm{M}_\odot\ \mathrm{yr}^{-1}$.
In the inner wind, their predicted abundances are unaffected by the uncertainties on the reaction rates.
Significant errors appear only after the onset photodissociation, leading to a spread in envelope size.

In O-rich outflows (bottom panel, Fig.~\ref{fig:parents-abprof}), the abundance profiles of SO, SiS, and to a lesser extent \ce{SO2} show an increase in abundance before photodissociation destroys them. 
This bump is facilitated by OH, the photodissociation product of the parent \ce{H2O} which peaks at this location.
The increase in SO is caused by OH + S, where additional S is released from the parent \ce{H2S}.
\ce{SO2} is formed via OH + SO, but the bump is not as pronounced as it photodissociates into SO.  
The increase in SiS is mostly due to SO + Si, with some contribution of \ce{SO2} + Si, where Si is mostly formed by the photodissociation of the parent SiO.

Fig.~\ref{fig:parents-extent} shows the mean envelope sizes and radial extents together with the fiducial sizes for C-rich and O-rich outflows for all outflow densities. 
The values are listed in Table~\ref{table:uncert-parents}.
The fiducial envelope sizes do not differ significantly from the mean predicted size; the discrepancy is larger than a factor 1.5 only for CS in O-rich outflows.
As expected, the envelopes shift outward with increasing mass-loss rate. 
The mean envelope size is set by the onset of photodissociation, shifting the envelope outward with increasing outflow density.
For each outflow density, the relative locations of the envelopes are determined by their specific photodissociation rates: the larger the rate, the smaller the envelope.

The radial extent of the envelope depends both on the species and the chemistry of the outflow.
Envelope sizes that are not well-constrained, with a range in extent larger than a factor three, are indicated by a triangle.
These are CO, \ce{N2}, and CS for both C-rich and O-rich outflows, SiO, SiS, PO, and PN for O-rich outflows, and HCP for C-rich outflows.
The error on the envelope size is not caused by uncertainty the parents' photodissociation, as their rate coefficients all have accuracy \emph{C}.
Rather, the error on the envelope size is determined by reactions reforming the parent after the onset of its photodissociation. 
This depends on the specific molecule and chemical composition of the outflow.
Fig. \ref{fig:parents-extent} shows that HCN, SiO, SiS, and CS have a larger range in O-rich outflows compared than in C-rich outflows. 
\ce{H2O}, \ce{NH3}, \ce{H2S} and HF are well constrained for both outflows chemistries.

\begin{table}
	\centering
	\caption{Number of daughter species in the two groups for each mass-loss rate and outflow chemistry. 
	\emph{All} daughter species have column density $\geq 10^{5}$ cm$^{-2}$.
	\emph{Abundant} daughter species have peak fractional abundance $\geq 10^{-10}$ w.r.t. \ce{H2} and column density $ \geq 10^{12}$ cm$^{-2}$.
	}
	\begin{tabular}{c c cc c c} 
	\hline
     & \multicolumn{2}{c}{Carbon-rich} && \multicolumn{2}{c}{Oxygen-rich}  \\  
    \cline{2-3} \cline{5-6} 
    \noalign{\smallskip}
    $\dot{M}$ & All & Abun. && All & Abun. \\  
    \cline{2-3} \cline{5-6} 
    \noalign{\smallskip}
$10^{-5}\ \mathrm{M}_\odot\ \mathrm{yr}^{-1}$	&	529	&	78	&&	201	&	26	\\
$10^{-6}\ \mathrm{M}_\odot\ \mathrm{yr}^{-1}$	&	520	&	71	&&	193	&	25	\\
$10^{-7}\ \mathrm{M}_\odot\ \mathrm{yr}^{-1}$	&	506	&	63	&&	181	&	23	\\
		\hline
	\end{tabular}
    \label{table:numberofdaughters}    
\end{table}

%---------------------------------------------------------------------------------------------------------
\subsection{Uncertainty on daughter species }		\label{subsect:uncert:daughter}
%---------------------------------------------------------------------------------------------------------

Table \ref{table:numberofdaughters} lists the number of daughter species in the \emph{all} and \emph{abundant} daughter groups for each outflow density and composition. 
These are the species with a column density of at least $10^{5}$ cm$^{-2}$, removing all species that do not contribute to the overall chemistry, and those with a peak fractional abundance of at least $10^{-10}$ w.r.t. \ce{H2} and a column density of at least $10^{12}$ cm$^{-2}$.
The number of molecules within each group increases with outflow density for both O-rich and C-rich outflows.
Thanks to the reactivity of carbon, C-rich outflows have a larger number of daughters than the O-rich outflows in both categories (a factor of $\sim 3$ more).

In Sect. \ref{subsubsect:uncert:daughter:profiles}, we present the abundance profiles of a selection of daughter species.
To condense the results, we use mean peak fractional abundances and mean column densities, their errors, and their dispersions.
Sect. \ref{subsubsect:uncert:daughter:correlations} presents the correlations between the peak fractional abundances and mean column densities and their errors, Sect. \ref{subsubsect:uncert:daughter:disp} presents the correlation between their dispersions and determines the most and least certain daughter species.

%--.--.--.--.--.--.--.--.--.--.--.--.--.--.--.--.--.--.--.--.--.--.--.--.--.--.--.--.--.--.--.--.--.--.--.--
\subsubsection{Mean abundance profiles and errors}		\label{subsubsect:uncert:daughter:profiles}
%--.--.--.--.--.--.--.--.--.--.--.--.--.--.--.--.--.--.--.--.--.--.--.--.--.--.--.--.--.--.--.--.--.--.--.--

Fig. \ref{fig:daughters-abprof} shows the mean fractional abundance profiles and their errors for a selection of daughter species in a C-rich and O-rich outflow with $\dot{M} = 10^{-6}\ \mathrm{M}_\odot\ \mathrm{yr}^{-1}$. 
Unlike for parent species, the predicted abundances of daughter species have a non-negligible error from the start of the model.
Furthermore, the error does not necessarily grow throughout the outflow.
While the error increases throughout the outflow for some species, e.g., CN and \ce{C2H} in the C-rich outflow and OH and CN in the O-rich outflow, the rate with which it increases varies throughout. 
Other daughters show a different behaviour.
The error on, e.g., HS in the O-rich outflow and \ce{NH2} in the C-rich outflow decreases around $10^{15}$ cm, after which it increases again.

%--.--.--.--.--.--.--.--.--.--.--.--.--.--.--.--.--.--.--.--.--.--.--.--.--.--.--.--.--.--.--.--.--.--.--.--
\subsubsection{Correlations in peak fractional abundance, column density, and their dispersions}		\label{subsubsect:uncert:daughter:correlations}
%--.--.--.--.--.--.--.--.--.--.--.--.--.--.--.--.--.--.--.--.--.--.--.--.--.--.--.--.--.--.--.--.--.--.--.--

\begin{figure*}
\centering
\includegraphics[width=\textwidth]{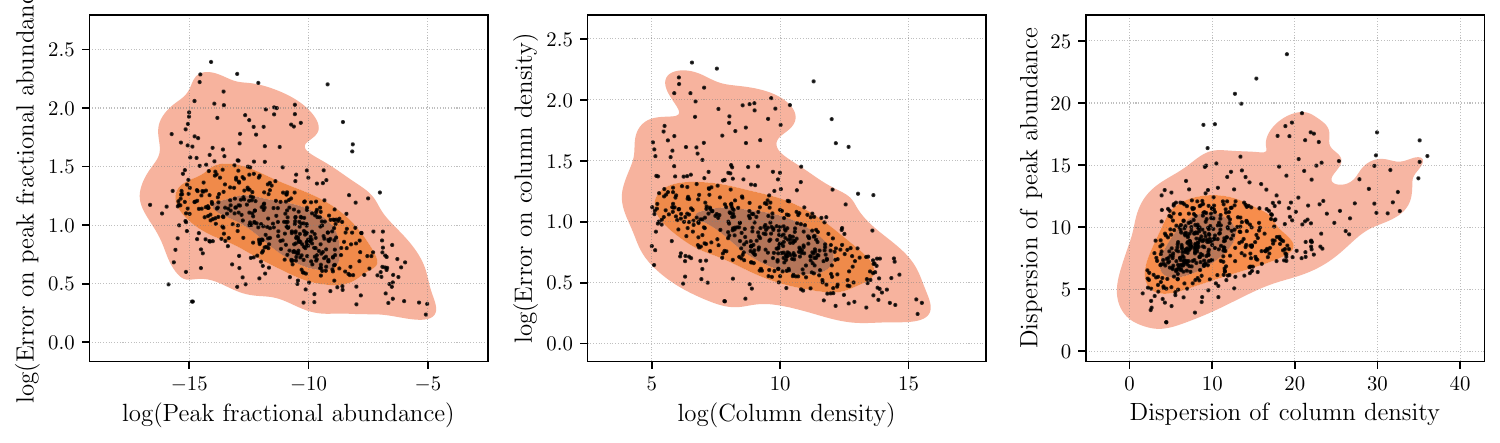}
\caption{
Scatter plots for all daughter species in a C-rich outflow with $\dot{M} = 10^{-6}\ \mathrm{M}_\odot\ \mathrm{yr}^{-1}$.
\emph{Left:} mean peak fractional abundance versus the error on the peak abundance. \emph{Middle:} mean column density versus its error. 
\emph{Right:} dispersion of the column density versus the dispersion of the peak fractional abundance.
The species have a column density of at least $10^{5}$ cm$^{-2}$, removing those that do not contribute to the overall chemistry. 
The shaded contours show the kernel density estimation of the point distribution, with three intensity levels indicating regions of increasing point concentration.
}
\label{fig:crich-allstats-all}
\end{figure*}

Fig.~\ref{fig:crich-allstats-all} shows scatter plots of peak fractional abundance versus its error, mean column density versus its error, and  dispersion of the column density versus dispersion of the peak abundance for all daughter species in a C-rich outflow with $\dot{M} = 10^{-6}\ \mathrm{M}_\odot\ \mathrm{yr}^{-1}$.
The peak abundances and column densities and their respective errors show a moderate negative monotonic correlation (left and middle panels). 
The corresponding Spearman rank correlation coefficients (RCCs) are $-0.53$ and $-0.55$, respectively. 
Larger values hence tend to have smaller errors.
The dispersion of the column density and the dispersion of the peak fractional abundance has a moderate positive correlation (right panel, RCC $= 0.44$).
Species with a smaller dispersion of their peak abundance also have smaller dispersions of their column density, and vice versa. 

Figures \ref{fig:app-crich-relevant}, \ref{fig:app-orich-relevant}, \ref{fig:app-crich-all}, and \ref{fig:app-orich-all} show similar scatter plots for all mass-loss rates and groups of daughter species.
The corresponding RCCs and their p-values (the probability of observing a correlation as strong as or stronger than the calculated one if no correlation exists) are listed in Table \ref{table:RCCs}. 
There is a negative correlation between peak abundance and its error and between column density and its error for both groups of daughter species, except for the abundant O-rich daughters.
For this subgroup, there is no statistically significant correlation, which could be due to the smaller number of abundant daughters ($\sim$25).
A positive correlation between the dispersion of the peak abundance and the dispersion of the column density exists for both groups of daughter species.

%--.--.--.--.--.--.--.--.--.--.--.--.--.--.--.--.--.--.--.--.--.--.--.--.--.--.--.--.--.--.--.--.--.--.--.--
\subsubsection{Mean dispersions of peak abundance and column density}		\label{subsubsect:uncert:daughter:disp}
%--.--.--.--.--.--.--.--.--.--.--.--.--.--.--.--.--.--.--.--.--.--.--.--.--.--.--.--.--.--.--.--.--.--.--.--

Table \ref{table:meandispersions} lists the mean dispersions of the peak abundance and column density and their respective standard deviations, for both groups of daughter species, where we averaged over different mass-loss rates within each group. 
The mean dispersion of the peak abundance is $\sim$10\% for both groups of daughter species, with a standard deviation of $\sim$3.5\% for C-rich daughters and $\sim$5\% for O-rich daughters. 
Hence, the error on the mean peak abundance is generally 10\% of its value.
The mean dispersion of the column density depends on the group of daughter species, where the abundant daughters have smaller values: from $\sim$11$\pm$7\% for all C-rich daughters to $\sim$5$\pm$2\% for the abundant, and from $\sim$14$\pm$10\% for all O-rich daughters to $\sim$5$\pm$2\% for the abundant.
The outflow density causes only small variations in mean dispersion.
The difference in dispersion on peak abundance is at most $0.5\%$ (for abundant O-rich daughters) and on column density is at most $0.4\%$ (for all O-rich daughters).

Fig. \ref{fig:disp-relevant} shows scatter plots of dispersion of the peak fractional abundance versus the dispersion of the column density for the abundant daughter species for all mass-loss rates in C-rich and O-rich outflows, with the most and least uncertain daughters labelled.
The most and least uncertain daughters species are summarised in Table \ref{table:mostleastcertaindaughters}.
There is a clear positive correlation between the two dispersions.
Species with a more precise prediction of their peak abundance hence also tend to have a more precise column density prediction.
The corresponding RCCs and p-values are listed in Table \ref{table:RCCs}.

% Comment Tom
%Has this correlation got something to do with the fact that the column density for a daughter with a 'precise' peak radius is given by the product of that distance, r_p, with f(X,r_p) * n(H2,r_p) ?  At least to a factor of 2-5.  Once a sharp peak is passed, there is very little column density added due to the 1/r2 fall in H2 and the decreasing f(X,r).  And by a similar argement, a sharp fall in f(X) at smaller radii may offset the rise in n(H2).  It would be interesting to look at the cumultive growth in N(X,r) at some point.

%%%%%%%%%%%%%%%%%%%%%%%%%%%%%%%%%%%%%%%%%%%%%%%%%%%%%%%%%%%%%%%%%%%%%%%%%%%%%%%%%%%%%%%%%%%%%%%%%%%%%%%%%%%%%%
\section{Discussion}							\label{sect:disc}

In Sect. \ref{subsect:disc:env}, we discuss the errors on the parents' envelope sizes and the impact of the error on the CO envelope size on retrieved mass-loss rates, and compare our model results with observations.
In Sect. \ref{subsect:disc:error}, we elaborate on the uncertainties on the daughter species and discuss their impact on the interpretation of observations.
Finally, in Sect. \ref{subsect:disc:compl}, we assess the need for adding complexity to the standard chemical kinetics model, for the particular case of cyanopolyynes and hydrocarbon radicals in the outflow of IRC\,+10216 and in general.

%---------------------------------------------------------------------------------------------------------
\subsection{Error on parents' envelope sizes}		\label{subsect:disc:env}
%---------------------------------------------------------------------------------------------------------

The error on the predicted envelope sizes of parent species depends on the specific molecule, the outflow's chemical composition, and, to a lesser extent, the outflow density (Fig. \ref{fig:parents-extent}). 
This error is caused by chemistry, as discussed in Sect. \ref{subsubsect:disc:env:error}.
The impact of the error on the CO envelope size on retrieved mass-loss rates is discussed in Sect. \ref{subsubsect:disc:env:CO}.
We compare our predicted envelope sizes to previous models and observations in Sect. \ref{subsubsect:disc:env:obs}.

%--.--.--.--.--.--.--.--.--.--.--.--.--.--.--.--.--.--.--.--.--.--.--.--.--.--.--.--.--.--.--.--.--.--.--.--
\subsubsection{Chemistry causing the errors on molecular envelopes}		\label{subsubsect:disc:env:error}
%--.--.--.--.--.--.--.--.--.--.--.--.--.--.--.--.--.--.--.--.--.--.--.--.--.--.--.--.--.--.--.--.--.--.--.--

The variation in radial range of the parents' envelope sizes is not due to different errors on their photodissociation rates, as these all have accuracy \emph{C}.
Rather, it is caused by the chemistry reforming the parent after the onset of its photodissociation.
Using photodissociation models to estimate the envelope size hence might be an oversimplification. 

To first order, the error on the envelope increases with the number of reactions involved in the parent's reformation.
For example, the parent \ce{H2O} has a well-constrained range in radial extent and a straightforward reformation pathway: \ce{H2O} photodissociates into OH and predominantly reformed through the hydrogenation of OH. 

Other parents are reformed through a more complex sequence of reactions, involving not only parents and direct photodissociation products, but also $n$th-generation daughter species. 
This is most noticeable for CS, with a radial extent of more than an order of magnitude in O-rich outflows and a factor $\sim$5 in C-rich outflows.
In C-rich outflows, CS reformation occurs via one main channel: \ce{C2} + S, where \ce{C2} is generated by the photodissociation of the parent \ce{SiC2} and \ce{C2H}, a photodissociation product of the parent \ce{C2H2}.
In O-rich outflows, the main reformation pathways varies with radius.
First, CS is reformed via C + HCS, the latter originating from C and the parent \ce{H2S}.
Further out, the reaction between C and the parent SO takes over. 
Finally, the bump in abundance around $~5 \times 10^{16}$ cm (bottom panel, Fig. \ref{fig:parents-abprof}) is caused by the dissociative recombination of \ce{HCS+} with electrons.
\ce{HCS+} arises from a sequence of reactions, ending with \ce{H2} + \ce{CS+}, which in turn is produced via \ce{S+} + CH.
The CH radical is formed by the dissociative recombination of \ce{CH3+} with electrons, itself produced by successive reactions of \ce{H2} with \ce{C+}.

Analysing the reaction pathways only gives us a rough sense of the general mechanism behind the origin of the error on the envelope sizes \citep{Saltelli2019}.
A sensitivity analysis is needed to determine which specific reactions are the main contributors to each error. 
This will be done in a follow-up paper.

\begin{table}
	\centering
	\caption{Mean values and standard deviations of the dispersions on peak fractional abundance and column density for the different groups of daughter species.
	The average is taken over the three different outflow densities for each outflow chemistry for both mean and standard deviations.
	}
	\resizebox{1.0\columnwidth}{!}{%
	\begin{tabular}{c c cc c c} 
	\hline
     & \multicolumn{2}{c}{Carbon-rich} && \multicolumn{2}{c}{Oxygen-rich}  \\  
    \cline{2-3} \cline{5-6} 
    \noalign{\smallskip}
      & All & Abun. && All & Abun. \\  
    \cline{2-3} \cline{5-6} 
    \noalign{\smallskip}
Mean dispersion peak abundance		& 9.47\% & 9.94\% && 10.26\% & 9.84\% \\
Standard deviation					& 3.02\% & 3.62\% && 4.56\% & 4.99\% \\
    \noalign{\smallskip}
Mean dispersion column density		& 11.53\% & 5.18\% && 14.06\% & 5.17\% \\
Standard deviation    				& 6.65\% & 2.43\% && 9.67\% & 2.37\% \\
	\hline
	\end{tabular}
	}%
    \label{table:meandispersions}    
\end{table}

\begin{table}
	\centering
	\caption{Most and least precise abundant daughters species for each outflow chemistry. 
	The most/least certain species have a dispersion of both peak abundance and column density larger/smaller than their mean plus/minus standard deviation in any outflow density.
	}
	\begin{tabular}{l l} 
	\hline
     \multicolumn{2}{c}{Carbon-rich} \\  
%    \noalign{\smallskip}
	\hline
	Most uncertain daughters & \ce{CH2C3N}, \ce{CH}, \ce{NO}, \ce{SiH}, \ce{HC4S}, \\
				& \ce{C3HCCH}, \ce{C5N}, \ce{C4H2}, \ce{C9H}, \ce{C5N-}, \\
				& \ce{OCS}, \ce{NCCN}, \ce{NC4N}, \ce{C3CCH} \\
	Least uncertain daughters & \ce{C6H2}, \ce{C4H}, \ce{HCNH+}, \ce{NH2}, \ce{C2}, \\
					& \ce{C7H2}, \ce{NaCl} \\
    \noalign{\smallskip}
	\hline
    \multicolumn{2}{c}{Oxygen-rich} \\  
	\hline
	Most uncertain daughters & \ce{HSiS}, \ce{CH}, \ce{O2}, \ce{NO}, \ce{OCS}, \ce{HNSi} \\
	Least uncertain daughters & \ce{NH2}, \ce{NH4+}, \ce{NH} \\
	\hline
	\end{tabular}
    \label{table:mostleastcertaindaughters}    
\end{table}

\begin{figure*}
\centering
\includegraphics[width=\textwidth]{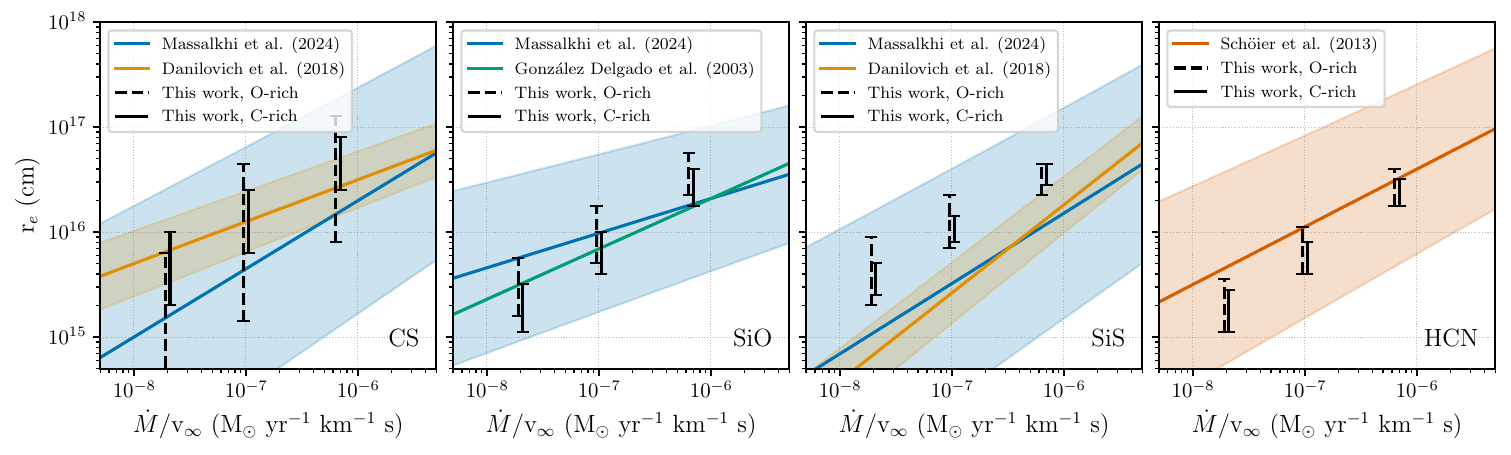}
\caption{
Variation of envelope size, $r_e$, with outflow density as retrieved from observations for CS, SiO, SiS, and HCN.
Blue and orange: relationships and errors retrieved by \citet{Massalkhi2024} and \citet{Danilovich2018}, respectively.
Green: relationship retrieved by \citet{GonzalezDelgado2003}, without errors.
Red: relationship and error retrieved by \citet{Schoier2013}.
Our predicted ranges in envelope size are shown in black, for both O-rich (dashed) and C-rich (solid) outflows.
}
\label{fig:parents-obs}
\end{figure*}

%--.--.--.--.--.--.--.--.--.--.--.--.--.--.--.--.--.--.--.--.--.--.--.--.--.--.--.--.--.--.--.--.--.--.--.--
\subsubsection{Impact on retrieved mass-loss rates}		\label{subsubsect:disc:env:CO}
%--.--.--.--.--.--.--.--.--.--.--.--.--.--.--.--.--.--.--.--.--.--.--.--.--.--.--.--.--.--.--.--.--.--.--.--

CO is an important density tracer for AGB outflows. 
Mass-loss rates are commonly determined using radiative transfer models to reproduce observed CO lines \cite[e.g.,][]{Schoier2001,Danilovich2015}. 
Especially when using unresolved (single-dish telescope) line observations, assumptions are made about the CO envelope size. 
These are based on CO photodissociation and self-shielding, with \citet{Mamon1988} being the most used prescription.
\citet{Saberi2019} developed a more complex model including more and updated spectroscopic data, resulting in envelope sizes up to 40\% smaller than those of \citet{Mamon1988}.
We find that uncertainties on the chemical kinetic data cause an uncertainty on the CO envelope, which has a radial extent larger than a factor of three in all but the lowest density O-rich outflow (Fig. \ref{fig:parents-extent}).
Different CO envelope extents can impact the calculated mass-loss rates, especially when only low-energy CO lines are used in the mass-loss calculations.

To determine the impact on the retrieved mass-loss rates, we calculated radiative transfer models for the C-rich and O-rich outflows (Table \ref{table:model-params}), using the smallest and largest predicted CO envelope sizes along with the standard \citet{Mamon1988} sizes. 
Dust optical depths were estimated from the correlations between mass-loss rate and optical depth for the O-rich and C-rich outflows separately \citep{Danilovich2015}.
Mass-loss rates were then estimated from the CO $J=2 \to 1$ and $J=1 \to 0$ line fluxes using the formula from \citet{Ramstedt2008}.

For each outflow density and chemistry, we calculate a percentage difference using $(\dot{M}_\mathrm{max} - \dot{M}_\mathrm{min})/\dot{M}_\mathrm{Mamon}$,
where $\dot{M}_\mathrm{max, min}$ is the mass-loss rates retrieved from our largest and smallest envelope sizes and $\dot{M}_\mathrm{Mamon}$ the mass-loss rate retrieved using \citet{Mamon1988}. 
A similar percentage difference is calculated for the predicted line fluxes.
The predicted line fluxes, mass-loss rates, and their respective percentage differences are elaborated on in Appendix \ref{app:RT}.

The CO $J=1 \to 0$ line flux uncertainty ranges from 2-155\%, with a corresponding mass-loss rate uncertainty between 1.5 and 92\%.
We find that larger envelope sizes predict brighter low-$J$ CO lines, as expected. 
The \citet{Mamon1988} envelope sizes all lie between our smallest and largest extents, but not always at the mean of our models.
%The retrieved uncertainty on the CO $J=2 \to 1$ lines fluxes lies in the range 3 to 58\%, with corresponding mass-loss rate uncertainty between 3 and 46\%.
The CO $J=2 \to 1$ line flux uncertainty ranges from 3-58\%, with corresponding mass-loss rate uncertainty between 3 and 46\%.
The largest differences are for the low mass-loss rate O-rich models and the smallest for the high mass-loss rate models, because the latter tend towards higher optical depths..
For all models, the relationship between change in line flux and change in apparent mass-loss rate is not a 1:1 correspondence, since excitation and optical depth effects are partially taken into account in the data the \citet{Ramstedt2008} formula was fit to.

The radial extent of the CO envelope due to chemistry is larger than the difference in envelope size between \citet{Mamon1988} and a more complex treatment of CO photodissociation (\citealt{Saberi2019}, factor three versus 40\%), but leads to at most a factor two difference on the  mass-loss rate.
Therefore, the simpler \citet{Mamon1988} prescription can still be used when retrieving mass-loss rates from unresolved, low-energy CO lines.

\begin{figure*}
\centering

\begin{subfigure}[h]{1.0\textwidth}
  \centering
	\includegraphics[width=\textwidth]{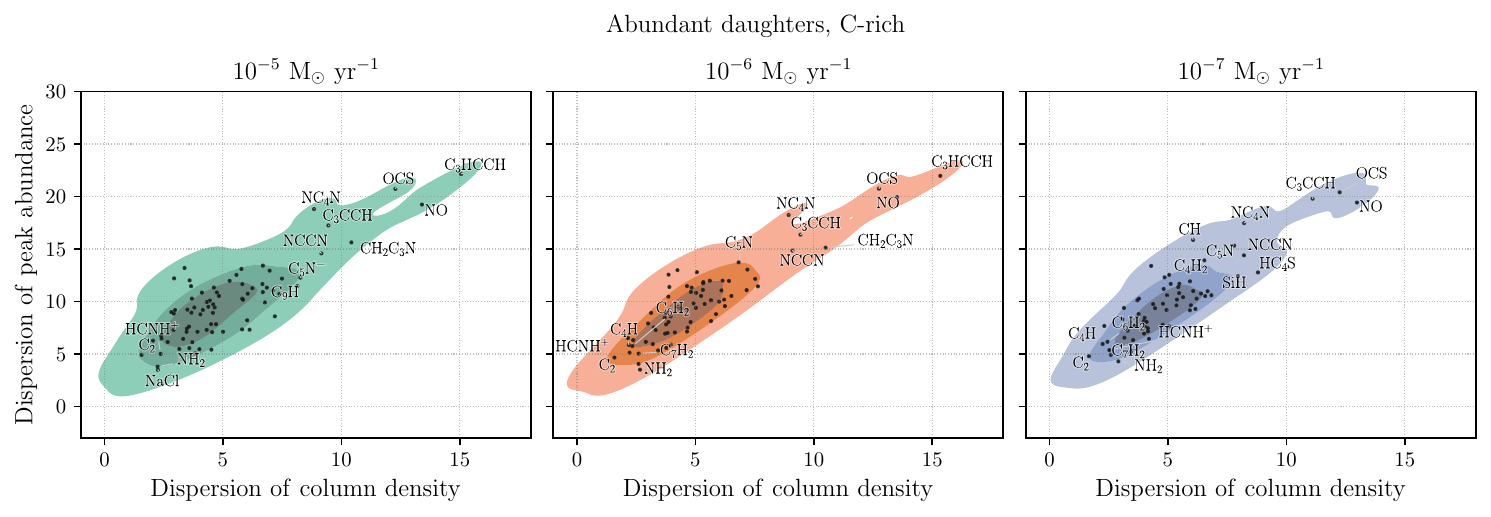}
%  \caption{1a}
  \label{fig:sfig1}
\end{subfigure}%

\vspace{0em}

\begin{subfigure}[h]{1.0\textwidth}
  \centering
	\includegraphics[width=\textwidth]{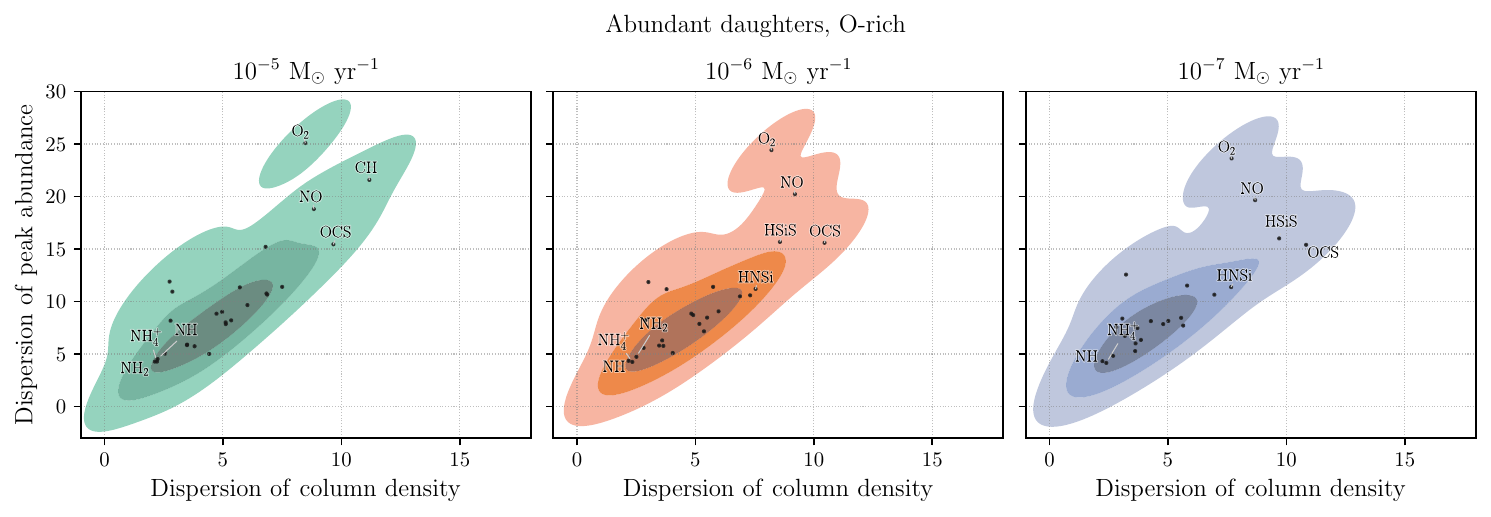}
%  \caption{1a}
  \label{fig:sfig1}
\end{subfigure}%
\caption{
Scatter plots showing the dispersion of the column density versus the dispersion of the peak fractional abundance of the abundant daughter species in all C-rich \emph{(top)} and O-rich \emph{(bottom)} outflows. 
\emph{Left}: outflow with $\dot{M} = 10^{-5}\ \mathrm{M}_\odot\ \mathrm{yr}^{-1}$, \emph{middle}: $\dot{M} = 10^{-6}\ \mathrm{M}_\odot\ \mathrm{yr}^{-1}$, \emph{right:} $\dot{M} = 10^{-7}\ \mathrm{M}_\odot\ \mathrm{yr}^{-1}$
The shaded contours show the kernel density estimation of the point distribution, with three intensity levels indicating regions of increasing point concentration.
In each panel, the top right shows the least precise daughters with a dispersion of both peak abundance and column density larger than their mean plus standard deviation (per mass-loss rate).
The bottom left shows most precise daughters with dispersions smaller than their mean minus standard deviation (per mass-loss rate).
}
\label{fig:disp-relevant}
\end{figure*}

%--.--.--.--.--.--.--.--.--.--.--.--.--.--.--.--.--.--.--.--.--.--.--.--.--.--.--.--.--.--.--.--.--.--.--.--
\subsubsection{Comparison to models and observations}		\label{subsubsect:disc:env:obs}
%--.--.--.--.--.--.--.--.--.--.--.--.--.--.--.--.--.--.--.--.--.--.--.--.--.--.--.--.--.--.--.--.--.--.--.--

\citet{Maes2023} determined the influence of errors on the mass-loss rate, outflow velocity, and temperature profile on the model output using the same chemical model and  a reaction network based on the \textsc{Rate12} network \citep{McElroy2013}. 
They found that the error on the predicted envelope size caused by uncertainty on outflow temperature increases with decreasing outflow density.
Generally, this error is comparable to the chemical error for lower density outflows. 
For our high-density outflow, the chemical error is several factors larger.

The dependence of the CS, SiO, SiS, and HCN envelope sizes on outflow density has been determined observationally.
Figure \ref{fig:parents-obs} shows the retrieved envelope size-outflow density relationships together with their error, when available, and our model predictions.
The relationships retrieved by \citet{Massalkhi2024} have an error of at least one order of magnitude for all densities; our retrieved envelope sizes lie within their errors. 
For CS, our models agree with the relationship retrieved by \citet{Danilovich2018} and the retrieved C-rich envelope sizes of \citet{Unnikrishnan2025}.
For HCN, our models agree with the relationship retrieved by \citet{Schoier2013}.
For SiO, our models reproduce the relationship for SiO of \citet{GonzalezDelgado2003} (without errors) relatively well, with only the high-density O-rich outflow having clearly a larger envelope size.
For SiS, our model predicts larger envelope sizes than those retrieved by \citet{Danilovich2018}.
This was also found by \citet{Maes2023}, indicating that its photodissociation rate is likely too small and lies outside its associated uncertainty range of a factor two (category \emph{C}).
The photodissociation cross section of SiS has never been measured or calculated. 
Its photodissociation rate might be similar to those of SiO and CS \citep{vanDishoeck1988}, which have unshielded rates of $\sim 10^{-9}$ s$^{-1}$ in the network, compared to $10^{-10}$ s$^{-1}$ for SiS.
For HCN, our models agree with the relationship retrieved by \citet{Schoier2013}.

Photodissociation models have been used to determine the envelope sizes of species other than CO.
Given that chemical reactions cause the error on the envelope size, this is an oversimplification.
\citet{Maercker2008} used the prescription of \citet{Netzer1987} for the location of the peak in OH abundance to estimate the \ce{H2O} envelope size.
We find that the \citet{Netzer1987} envelope sizes are slightly smaller than our predicted ranges for all outflows and chemistries.
%\citet{Maercker2016} retrieved the \ce{H2O} envelope size from \emph{Herschel} observations.  
%Their retrieved size is larger than the \citet{Netzer1987} size for the lower-density outflows.
\citet{Maercker2016} retrieved the \ce{H2O} envelope size from \emph{Herschel} observations. 
Their retrieved size is not well constrained, but tends to be larger than the \citet{Netzer1987} size for lower-density outflows
The retrieved size matches the \citet{Netzer1987} size for the high-density outflow of IK~Tau.
The most \ce{H2O} transitions were detected in IK~Tau's outflow, indicating that more observational data covering a variety of energies is necessary to observationally constrain envelope sizes.

%A few points here. 1) they've also used some interferometric data to measure the envelope sizes (IK Tau and TX Cam - I think someone also did this for CW Leo using old interferometer data).
%2) they also fit an envelope size similar to the other results in your fig 7. Might be worth also including? see their eq 3. (they even included uncertainties) - oh, unless that's your fig. appendix?

%---------------------------------------------------------------------------------------------------------
\subsection{Average uncertainties and model precision}		\label{subsect:disc:error}
%---------------------------------------------------------------------------------------------------------

The positive correlation between the dispersion of the peak fractional abundance and the dispersion of the column density of the daughter species implies that daughters with a more precise peak fractional abundance generally have a more precise column density.
The average dispersion of the column density is smaller when considering only the abundant daughters (Table \ref{table:meandispersions}): the largest errors on the column density are associated with daughters with a column density below our threshold of $10^{12}$ cm$^{-2}$.
Because the higher-density inner regions contribute more to this integrated quantity, species with large uncertainties on their abundances in this region will have larger errors on their column densities. 

The average uncertainties on the model predictions are discussed in Sect. \ref{subsubsect:disc:error:average}.
The radial behaviour of error on the fractional abundance profiles is addressed in Sect. \ref{subsubsect:disc:error:nonprop}.
The impact of errors on the interpretation of observations is examined in Sect. \ref{subsubsect:disc:error:interp}.

%--.--.--.--.--.--.--.--.--.--.--.--.--.--.--.--.--.--.--.--.--.--.--.--.--.--.--.--.--.--.--.--.--.--.--.--
\subsubsection{Average uncertainties}		\label{subsubsect:disc:error:average}
%--.--.--.--.--.--.--.--.--.--.--.--.--.--.--.--.--.--.--.--.--.--.--.--.--.--.--.--.--.--.--.--.--.--.--.--

The average dispersion of the peak fractional abundance is $\sim$10\% for all outflows and sets of daughter species.
Therefore, the error on the predicted peak abundance is on average about 10\% of its value.
In absolute terms, this varies from a factor of a few for the most precise daughters to more than two orders of magnitude for the least precise daughters (Table \ref{table:mostleastcertaindaughters}).
Interestingly, some of the most uncertain daughters are simple species, while some of the least uncertain are longer carbon-chains.
Similar to the error on the parents' envelope sizes (Sect. \ref{subsubsect:disc:env:error}), we find that the size of the error on a predicted peak abundance is related to the number of reactions involved in the formation of the daughter species.
Carbon-chains are only a few generations removed from the parent \ce{C2H2} in the C-rich outflow, leading to a comparatively small error on \ce{C2, C4H, C6H2}, and \ce{C7H2}.
On the other hand, several pathways contribute to the formation of NO, all of which require the liberation of N or O from via the photodissociation of parent species and their photoproducts.
However, an uncertainty analysis does not allow us to determine the mechanisms by which the error accumulates \citep{Saltelli2019}.
To determine which specific reactions are the main contributors to the error, we will perform a sensitivity analysis in a follow-up paper.

The error on the peak abundance affects the classification of HCCO as an \emph{abundant} daughter in the highest density C-rich outflow and of \ce{CH4} and \ce{CH3CN} in the highest density O-rich outflow.
The mean peak abundances of HCCO and \ce{CH4} drop below $10^{-12}$ w.r.t. \ce{H2}, but their maximum abundance is still above our threshold.
The maximum peak abundance of \ce{CH3CN} increases to $15 \times 10^{-12}$ w.r.t. \ce{H2}, but its mean abundance is below this threshold.
None of the most precise species are first-generation daughter species (\ce{C2} is mostly produced by the photodissociation of \ce{C2H} and not by photodissociation of the C-rich parent \ce{SiC2}).
This indicates that the error on the model predictions is highly dependent on the reactions forming and destroying each species.

%--.--.--.--.--.--.--.--.--.--.--.--.--.--.--.--.--.--.--.--.--.--.--.--.--.--.--.--.--.--.--.--.--.--.--.--
\subsubsection{Non-propagation of errors}		\label{subsubsect:disc:error:nonprop}
%--.--.--.--.--.--.--.--.--.--.--.--.--.--.--.--.--.--.--.--.--.--.--.--.--.--.--.--.--.--.--.--.--.--.--.--

We find that errors on predicted abundances do not propagate throughout the outflow, but fluctuate locally without systematic accumulation.
The error on the daughter abundance profiles and the radial spread on the parents' abundance profiles does not necessarily grow with increasing radius.

This is due to the nature of our chemical system. 
The dominant chemical processes change throughout the outflow in response to its gradient in density and temperature, effectively generating a new set of ordinary differential equations at each radius.
Consequently, the error on the predicted abundance of a species reflects how well the model captures the dominant chemical processes of that species at a particular location within the outflow, rather than being accumulated from previous computational steps.

%--.--.--.--.--.--.--.--.--.--.--.--.--.--.--.--.--.--.--.--.--.--.--.--.--.--.--.--.--.--.--.--.--.--.--.--
\subsubsection{Impact on the interpretation of observations}		\label{subsubsect:disc:error:interp}
%--.--.--.--.--.--.--.--.--.--.--.--.--.--.--.--.--.--.--.--.--.--.--.--.--.--.--.--.--.--.--.--.--.--.--.--

Given the average dispersion of $\sim 10\%$ on the predicted peak abundance, caution is needed when fitting predicted (peak) abundances to abundances retrieved from observations. 
Ad hoc modifications to the physics or chemistry of a model should generally be avoided, especially for the least certain species (Table \ref{table:mostleastcertaindaughters}).
While these could explain away disagreement with observations, it might not to so in a meaningful way as the ``improved'' prediction could still lie within the (unquantified) uncertainty of the fiducial model.

Caution is especially needed when retrieving or refining reaction rates based on observations, as the error on the retrieved abundances is generally not taken into account.
%: radiative transfer models are generally used in a deterministic way.
The improved agreement could therefore lie within the (unquantified) error of the chemical model and/or the  (unquantified) error of the retrieved abundance.

%---------------------------------------------------------------------------------------------------------
\subsection{Assessing the need for added complexity}		\label{subsect:disc:compl}
%---------------------------------------------------------------------------------------------------------

Observations have driven the development of the standard CSE model, as gas-phase only chemistry in a smooth outflow does not always reproduce retrieved abundance profiles.
This uncertainty analysis allows us to address which discrepancies with observations can be solved by an increase in complexity of the model.
Our criteria are based on differences between predicted and retrieved abundances and the general behaviour of the abundance profile.

In Sect. \ref{subsubsect:disc:compl:cwleo}, we discuss the need for tailoring a chemical model to the outflow of the C-rich AGB star IRC\,+10216 to describe the observed radial distribution of cyanopolyynes and hydrocarbons. 
Sect. \ref{subsubsect:disc:compl:dust} examines the general need for dust-gas chemistry, porosity, and the UV field of a close-by stellar companion.
%We find that these added complexities are necessary to explain certain anomalous observations, especially  when an outflow shows clear deviations from spherical symmetry.

%\green{Not an academic exercise, real insight!}
%\green{We do not need to eliminate uncertainties to already get a better understandig of the model.
%By quantifying uncertainty, we can better understand of what the model is sensitive to... SEE RECORDING
%What can be solved by a gain in accuracy.}
%\green{Shape, not just a bit different!}
%\green{Our modelling strategy rests on the following criteria: }

\begin{figure*}
\centering
\includegraphics[width=\textwidth]{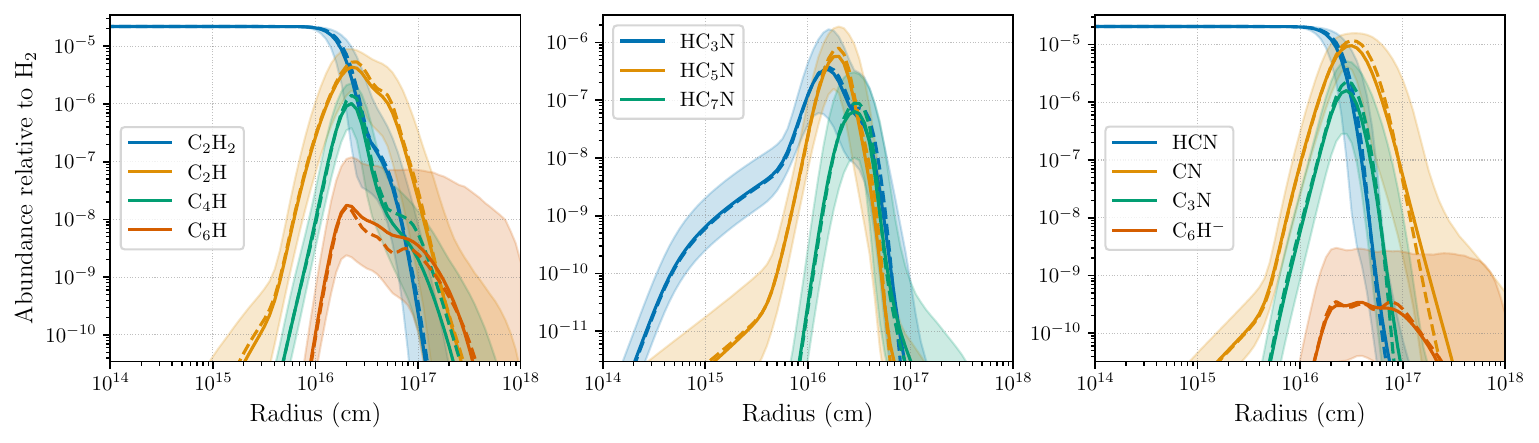}
\caption{
Fractional abundance w.r.t. \ce{H2} of selected carbon chains in a C-rich outflow with $\dot{M} = 10^{-5}\ \mathrm{M}_\odot\ \mathrm{yr}^{-1}$. 
The dashed line shows the fiducial model prediction using \textsc{Rate22}. 
The solid line shows the mean abundance $\langle \log X(r)\rangle$ of the Monte Carlo sample of reaction networks.
The shaded region contains 95.4\% of all predicted profiles and corresponds to the error $\Delta \log X(r)$ on the mean abundance.
}
\label{fig:cwleo-frac}
\end{figure*}

\begin{figure*}
\centering
\includegraphics[width=\textwidth]{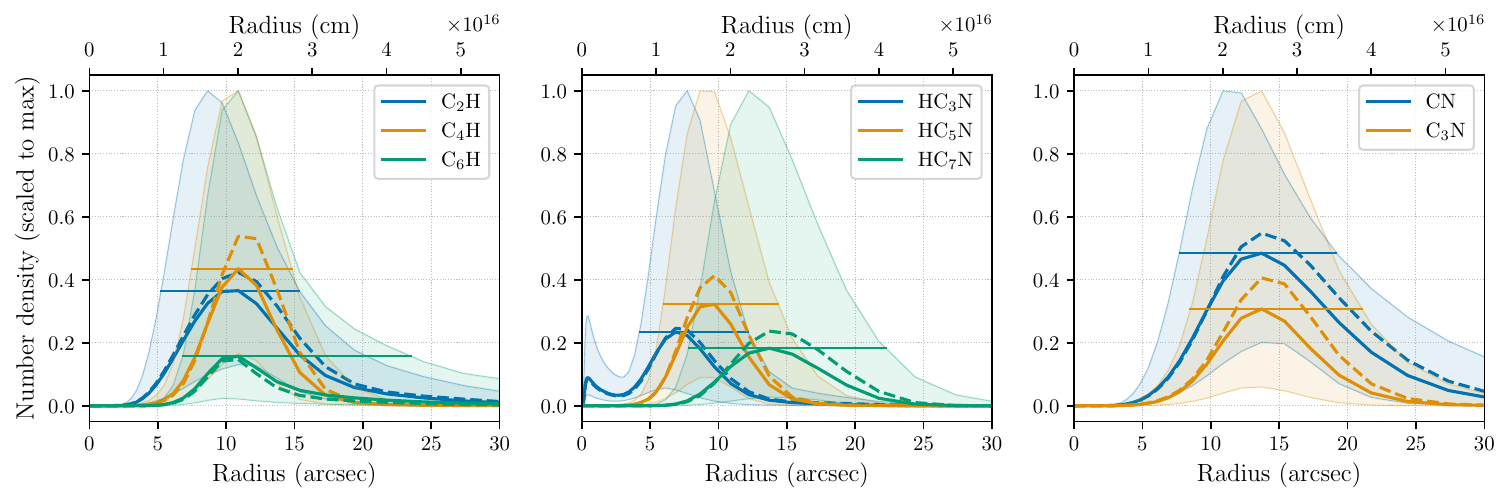}
\caption{
Number density scaled to the mean peak value of selected carbon chains in a C-rich outflow with $\dot{M} = 10^{-5}\ \mathrm{M}_\odot\ \mathrm{yr}^{-1}$. 
Radius in arcsec assumes a distance of 123 pc (the distance to IRC\,+10216, \citealt{Groenewegen2012}).
The dashed line shows the fiducial model prediction using \textsc{Rate22}. 
The solid line shows the mean abundance $\langle \log X(r)\rangle$ of the Monte Carlo sample of reaction networks.
The shaded region contains 95.4\% of all predicted profiles and corresponds to the error $\Delta \log X(r)$ on the mean abundance.
The horizontal line shows the range where the abundance is at least the mean peak abundance.
}
\label{fig:cwleo-numdens}
\end{figure*}

%--.--.--.--.--.--.--.--.--.--.--.--.--.--.--.--.--.--.--.--.--.--.--.--.--.--.--.--.--.--.--.--.--.--.--.--
\subsubsection{Cyanopolyynes and hydrocarbons around IRC\,+10216}		\label{subsubsect:disc:compl:cwleo}
%--.--.--.--.--.--.--.--.--.--.--.--.--.--.--.--.--.--.--.--.--.--.--.--.--.--.--.--.--.--.--.--.--.--.--.--

Chemical models have been specifically tailored to the outflow of the C-rich AGB star IRC\,+10216, which contains multiple broken shell-like structures \cite[e.g.,][]{Mauron2006,DeBeck2012,Cernicharo2015}.
Although IRC\,+10216 is the most studied AGB outflow, the distribution of cyanopolyynes (HC$_\mathrm{2n+1}$N, $n=3,5,7,...$) and hydrocarbon radicals (C$_\mathrm{2n}$H, $n=1,2,3,...$) in its outer regions is a long-standing puzzle. 
Standard chemical models predict a radial sequence with molecule length. 
While the cyanopolyynes show such a radial sequence, the hydrocarbon radicals are cospatial \citep{Guelin1999,Agundez2017,Keller2017}.
To solve these discrepancies, density-enhanced dust shells have been included \citep{Brown2003,Cordiner2009} and the extinction throughout the inhomogeneous outflow has been lowered \citep{Agundez2017}.
Despite these efforts, the different spatial behaviour of these molecules remains unexplained.

Figure \ref{fig:cwleo-frac} shows the fractional abundances of cyanopolyynes and hydrocarbons in the C-rich outflow with $\dot{M} = 10^{-5}\ \mathrm{M}_\odot\ \mathrm{yr}^{-1}$, which is closest in outflow density to IRC\,+10216.
From the mean abundance profiles, we find a radial sequence of the cyanopolyynes and cospatial hydrocarbons, in agreement with observations.
However, uncertainties on the chemical kinetic data also lead to a radial spread of the peak abundances; \ce{C6H} and \ce{C6H-} are particularly ill-constrained.
To better compare to Figure 6 of \citet{Agundez2017}, Figure \ref{fig:cwleo-numdens} shows the predicted number densities of these molecules, with radius converted to arcsec using a distance to IRC\,+10216 of 123 pc \citep{Groenewegen2012}.
We find that CN and \ce{C3N} peak at the same radius, as observed and unlike the model predictions of \citet{Agundez2017}.

Within the uncertainties on the chemical kinetic data, the classic spherically symmetric outflow model is hence able to reproduce the observed cospatial distribution of the cyanopolyynes and hydrocarbons.
This implies that the effects of the observed density distributions are smaller than those caused by the uncertainties on the kinetic data.
Therefore, adding physical or chemical complexity to the standard model does not aid in the interpretation of the relative locations of these molecules.
%Accounting for the radial extent of the peak abundances caused by uncertainties on the chemical kinetic data, the classic spherically symmetric outflow reproduces the cospatial distribution of the cyanopolyynes and hydrocarbons.
%Therefore, adding physical or chemical complexity to the standard model does not aid in the interpretation of the relative locations of these molecules.
%As these are obtained from azimuthally averaged observations, a 3D approach for both radiative transfer and chemical models might be necessary to determine the impact of the asymmetries that have been observed in IRC\,+10216's outflow on its composition. 

%--.--.--.--.--.--.--.--.--.--.--.--.--.--.--.--.--.--.--.--.--.--.--.--.--.--.--.--.--.--.--.--.--.--.--.--
\subsubsection{Dust-gas chemistry, porosity, and companion UV radiation}		\label{subsubsect:disc:compl:dust}
%--.--.--.--.--.--.--.--.--.--.--.--.--.--.--.--.--.--.--.--.--.--.--.--.--.--.--.--.--.--.--.--.--.--.--.--

Dust-gas chemistry was included \citep{VandeSande2019b,VandeSande2020,VandeSande2021} to assess whether it could indeed lead to a decrease in abundance of SiO and SiS before their photodissociation, as observed in high density O-rich outflows \cite[e.g.][]{Bujarrabal1989,Decin2010b,Verbena2019}.
Further clues for gas-phase depletion onto dust grains are found in trends in abundance with outflow density, where the abundance of SiO, SiS, and CS with is found to decrease with increasing outflow density \cite[e.g.,][]{GonzalezDelgado2003,Massalkhi2019,Danilovich2019,Massalkhi2024}.
%Trends in abundance with outflow density, where the abundance of SiO, SiS, and CS with is found to decrease with increasing outflow density \cite[e.g.,][]{GonzalezDelgado2003,Massalkhi2019,Danilovich2019,Massalkhi2024}, provide further clues of gas-phase depletion onto dust grains.

Uncertainties on the chemical kinetic data cannot reproduce the observed abundance profiles of SiO and SiS; the uncertainty on their envelope sizes does not extend into the region in the intermediate wind where SiO and SiS are observed to decrease \citep{Decin2010b}.
Including dust-gas chemistry is hence necessary to close the gap with observations, especially for higher-density outflows: \cite{VandeSande2019b} included a comprehensive dust-gas chemistry and was able to reproduce the observed depletion of SiO and SiS.
Additionally, \ce{H2O} ice is observed around OH/IR stars \cite{Sylvester1999}. 
Dust-gas chemistry, either simple \cite[e.g.][]{Jura1985,Dijkstra2003} or complex \citep{VandeSande2019b,VandeSande2020}, is necessary include the formation of ices and refractory organics \citep{VandeSande2021} on grain surfaces.

The porosity formalism can be used to approximate non-spherically symmetric outflows \citep{VandeSande2018}.
A porous outflow does not significantly impact the composition of the outflow and mostly affects the radial extent of species.
It can enhance the influence of the UV field of a close-by stellar companion.
This can increase the chemical complexity of the inner wind \citep{VandeSande2022}. 
This includes the species SiN, SiC, and NS, which were observed to have large abundances in the inner region of W~Aql's outflow  \citep{Danilovich2024}, which is shaped by binary interaction \cite[e.g.,][]{Ramstedt2017}. 
Uncertainties on the kinetic data cannot account for their large observed abundances and asymmetric distributions in the intermediate outflow.
Including the effects of its known F9 companion is essential to do so.

\ce{HC3N} is observed to have an unexpectedly large abundance in IRC\,+10216's inner wind \cite{Siebert2022}. 
Including a solar-like companion and porosity increased its predicted abundance by an order of magnitude.
Additionally, different parent abundances affects the \ce{HC3N} abundance in the inner wind. 
The initial \ce{C2H2} and HCN abundances used in the chemical model presented in \citet{Siebert2022} are a factor two lower than those used here to ensure agreement with observational results of \cite{Fonfria2008}. 
This leads to an \ce{HC3N} abundance about one order of magnitude smaller than the abundance shown in Fig. \ref{fig:cwleo-frac}. 

A porous outflow, especially combined with a stellar companion, can increase the SO and \ce{SO2} abundance before their photodissociation.
This is observed for higher density O-rich outflows \citep{Danilovich2016,Danilovich2020}. 
Uncertainty on the kinetic data allows for rates that would increase their abundance before photodissociation, but only by a factor of a few (Fig. \ref{fig:parents-abprof}).
This is linked to the uncertainty on the \ce{H2O} envelope.
The increases predicted by our model are smaller than those observed, added chemical complexity is hence necessary to interpret the SO and \ce{SO2} observations. 
We note that SiS also allows for an increase in abundance before photodissociation in O-rich outflows (bottom panel, Fig. \ref{fig:parents-abprof}); this has not been observed so far.

%%%%%%%%%%%%%%%%%%%%%%%%%%%%%%%%%%%%%%%%%%%%%%%%%%%%%%%%%%%%%%%%%%%%%%%%%%%%%%%%%%%%%%%%%%%%%%%%%%%%%%%%%%%%%%
\section{Conclusions}						\label{sect:concl}
%%%%%%%%%%%%%%%%%%%%%%%%%%%%%%%%%%%%%%%%%%%%%%%%%%%%%%%%%%%%%%%%%%%%%%%%%%%%%%%%%%%%%%%%%%%%%%%%%%%%%%%%%%%%%%

We presented the first uncertainty analysis of the standard CSE model, where we quantify the effect of uncertainties in the chemical kinetic data on the model output, for both C-rich and O-rich outflows at different outflow densities.
This is also the first such analysis of the \textsc{Rate22} reaction network. 
We used a Monte Carlo sampling method to propagate the uncertainties, generating 10,000 reaction networks, and assessed their effects on the fractional abundance and column density predictions, quantifying their mean values and errors.

Uncertainties on the chemical kinetic data lead to errors on the size of the parents' molecular envelope.
The size of the error is not due to the uncertainty on their photodissociation rates, as these are all a factor of two for the relevant reactions, but rather due to the chemistry reforming the parent after its photodissociation.
Therefore, using photodissociation models to constrain envelope sizes might be an oversimplification.
In particular, the error on the CO envelope is larger than the improvement on its size achieved when using more complex photodissociation models. 
Although the CO envelope is not well constrained, its error only affects the retrieved mass-loss rates by at most a factor of two.

The error on the fractional abundance does not propagate throughout the outflow.
The non-accumulation of errors is due to the nature of the outflow, as the gradients in temperature and density set the main chemical processes throughout the outflow.
Hence, it is a reflection of how well the model captures the chemistry of each species at each radius.
For the daughter species, we find that the dispersion of the peak fractional abundance is correlated with that of the column density.
Species with a more precise peak fractional abundance estimate hence tend to have a more precise column density estimate.
The average dispersion is about 10\% for the peak fractional abundance and 5\% for the column density.
This leads us to caution against refining or deriving rate coefficients from observations of CSEs.

The standard CSE model of gas-phase chemistry in a smooth outflow is still widely applicable to observations. 
For example, the radial distribution of the cyanopolyynes and hydrocarbon radicals around IRC\,+10216 can be reproduced without adding complexity to the standard model of gas-phase chemistry in a smooth outflow, despite observed density structure, as uncertainties in the kinetic data dominate over these structures. 
Including complexities as spherical asymmetries, dust-gas chemistry, and/or the UV field of a close-by stellar companion, is only necessary to explain certain observations.
These include gas-phase depletion onto dust grains and photo-induced chemical complexity taking place in the inner regions.

Our uncertainty analysis only characterises the uncertainty in the model predictions due to the uncertainty on the input chemical kinetic data.
To determine which reactions contribute the most to the uncertainty on the model predictions, we will perform a sensitivity analysis in a follow-up paper.

%%%%%%%%%%%%%%%%%%%%%%%%%%%%%%%%%%%%%%%%%%%%%%%%%%
\section*{Acknowledgements}

MVdS acknowledges support from the European Union’s Horizon 2020 research and innovation programme under the Marie Skłodowska-Curie grant agreement No 882991 and the Oort Fellowship at Leiden Observatory, and thanks the Munich Institute for Astro-, Particle and BioPhysics (MIAPbP), which is funded by the Deutsche Forschungsgemeinschaft (DFG, German Research Foundation) under Germany's Excellence Strategy – EXC-2094 – 390783311, for fostering insightful discussions.
MG acknowledges support from the European Union’s Horizon 2020 research and innovation programme under the Marie Skłodowska-Curie grant agreement No 101026214. 
TJM gratefully acknowledges the STFC for support under grant reference ST/T000198/1.
TD is supported in part by the Australian Research Council through a Discovery Early Career Researcher Award (DE230100183).

%%%%%%%%%%%%%%%%%%%%%%%%%%%%%%%%%%%%%%%%%%%%%%%%%%
\section*{Data Availability}

%The inclusion of a Data Availability Statement is a requirement for articles published in MNRAS. Data Availability Statements provide a standardised format for readers to understand the availability of data underlying the research results described in the article. The statement may refer to original data generated in the course of the study or to third-party data analysed in the article. The statement should describe and provide means of access, where possible, by linking to the data or providing the required accession numbers for the relevant databases or DOIs.

The data underlying this article will be shared on reasonable request to the corresponding author.



\bibliographystyle{mnras}
\bibliography{chemistry} % if your bibtex file is called example.bib

@ARTICLE{Agundez2006,
   author = {{Ag{\'u}ndez}, M. and {Cernicharo}, J.},
    title = "{Oxygen Chemistry in the Circumstellar Envelope of the Carbon-Rich Star IRC +10216}",
  journal = {\apj},
   eprint = {astro-ph/0605645},
     year = 2006,
    month = oct,
   volume = 650,
    pages = {374-393},
      doi = {10.1086/506313},
   adsurl = {http://adsabs.harvard.edu/abs/2006ApJ...650..374A}
}

@ARTICLE{Agundez2010,
   author = {{Ag{\'u}ndez}, M. and {Cernicharo}, J. and {Gu{\'e}lin}, M.},
    title = "{Photochemistry in the Inner Layers of Clumpy Circumstellar Envelopes: Formation of Water in C-rich Objects and of C-bearing Molecules in O-rich Objects}",
  journal = {\apjl},
archivePrefix = "arXiv",
   eprint = {1010.2093},
     year = 2010,
    month = dec,
   volume = 724,
    pages = {L133-L136},
      doi = {10.1088/2041-8205/724/2/L133},
   adsurl = {http://adsabs.harvard.edu/abs/2010ApJ...724L.133A}
}

@ARTICLE{Agundez2017,
   author = {{Ag{\'u}ndez}, M. and {Cernicharo}, J. and {Quintana-Lacaci}, G. and 
	{Castro-Carrizo}, A. and {Velilla Prieto}, L. and {Marcelino}, N. and 
	{Gu{\'e}lin}, M. and {Joblin}, C. and {Mart{\'{\i}}n-Gago}, J.~A. and 
	{Gottlieb}, C.~A. and {Patel}, N.~A. and {McCarthy}, M.~C.},
    title = "{Growth of carbon chains in IRC +10216 mapped with ALMA}",
  journal = {\aap},
archivePrefix = "arXiv",
   eprint = {1702.04429},
 primaryClass = "astro-ph.SR",
     year = 2017,
    month = apr,
   volume = 601,
      eid = {A4},
    pages = {A4},
      doi = {10.1051/0004-6361/201630274},
   adsurl = {http://adsabs.harvard.edu/abs/2017A%26A...601A...4A}
}

@ARTICLE{Agundez2020,
       author = {{Ag{\'u}ndez}, M. and {Mart{\'\i}nez}, J.~I. and {de Andres}, P.~L. and {Cernicharo}, J. and {Mart{\'\i}n-Gago}, J.~A.},
        title = "{Chemical equilibrium in AGB atmospheres: successes, failures, and prospects for small molecules, clusters, and condensates}",
      journal = {\aap},
         year = 2020,
        month = may,
       volume = {637},
          eid = {A59},
        pages = {A59},
          doi = {10.1051/0004-6361/202037496},
archivePrefix = {arXiv},
       eprint = {2004.00519},
 primaryClass = {astro-ph.SR},
       adsurl = {https://ui.adsabs.harvard.edu/abs/2020A&A...637A..59A}
}

@article{Bensberg2024,
author = {Bensberg, Moritz and Reiher, Markus},
title = {Uncertainty-Aware First-Principles Exploration of Chemical Reaction Networks},
journal = {The Journal of Physical Chemistry A},
volume = {128},
number = {22},
pages = {4532-4547},
year = {2024},
doi = {10.1021/acs.jpca.3c08386},
    note ={PMID: 38787736},
URL = {https://doi.org/10.1021/acs.jpca.3c08386},
eprint = {https://doi.org/10.1021/acs.jpca.3c08386}
}

@ARTICLE{Brown2003,
   author = {{Brown}, J.~M. and {Millar}, T.~J.},
    title = "{Modelling enhanced density shells in the circumstellar envelope of IRC +10216}",
  journal = {\mnras},
     year = 2003,
    month = mar,
   volume = 339,
    pages = {1041-1047},
      doi = {10.1046/j.1365-8711.2003.06252.x},
   adsurl = {http://adsabs.harvard.edu/abs/2003MNRAS.339.1041B}
}

@ARTICLE{Bujarrabal1989,
   author = {{Bujarrabal}, V. and {Gomez-Gonzalez}, J. and {Planesas}, P.
	},
    title = "{CO and SiO thermal emission in evolved stars}",
  journal = {\aap},
     year = 1989,
    month = jul,
   volume = 219,
    pages = {256-264},
   adsurl = {https://ui.adsabs.harvard.edu/abs/1989A%26A...219..256B}
}

@ARTICLE{Cernicharo2015,
   author = {{Cernicharo}, J. and {Marcelino}, N. and {Ag{\'u}ndez}, M. and 
	{Gu{\'e}lin}, M.},
    title = "{Molecular shells in IRC+10216: tracing the mass loss history}",
  journal = {\aap},
archivePrefix = "arXiv",
   eprint = {1412.1948},
 primaryClass = "astro-ph.SR",
     year = 2015,
    month = mar,
   volume = 575,
      eid = {A91},
    pages = {A91},
      doi = {10.1051/0004-6361/201424565},
   adsurl = {http://adsabs.harvard.edu/abs/2015A%26A...575A..91C}
}

@INPROCEEDINGS{Charnley1993,
       author = {{Charnley}, S.~B. and {Smith}, R.~G.},
        title = "{Ice Mantle Formation in the Envelopes of OH/IR Stars}",
    booktitle = {Planetary Nebulae},
         year = "1993",
       editor = {{Weinberger}, R. and {Acker}, Agn{\`e}s},
       series = {IAU Symposium},
       volume = {155},
        month = "Jan",
        pages = {329},
       adsurl = {https://ui.adsabs.harvard.edu/abs/1993IAUS..155..329C}
}

@article{Cherchneff1993,
       author = {{Cherchneff}, Isabelle and {Glassgold}, Alfred E.},
        title = "{The Formation of Carbon Chain Molecules in IRC +10216}",
      journal = {\apjl},
         year = 1993,
        month = dec,
       volume = {419},
        pages = {L41},
          doi = {10.1086/187132},
       adsurl = {https://ui.adsabs.harvard.edu/abs/1993ApJ...419L..41C}
}

@ARTICLE{Cordiner2009,
   author = {{Cordiner}, M.~A. and {Millar}, T.~J.},
    title = "{Density-Enhanced Gas and Dust Shells in a New Chemical Model for IRC+10216}",
  journal = {\apj},
archivePrefix = "arXiv",
   eprint = {0903.0890},
 primaryClass = "astro-ph.GA",
     year = 2009,
    month = may,
   volume = 697,
    pages = {68-78},
      doi = {10.1088/0004-637X/697/1/68},
   adsurl = {http://adsabs.harvard.edu/abs/2009ApJ...697...68C}
}

@ARTICLE{Danilovich2015,
       author = {{Danilovich}, T. and {Teyssier}, D. and {Justtanont}, K. and
         {Olofsson}, H. and {Cerrigone}, L. and {Bujarrabal}, V. and
         {Alcolea}, J. and {Cernicharo}, J. and {Castro-Carrizo}, A. and
         {Garc{\'\i}a-Lario}, P. and {Marston}, A.},
        title = "{New observations and models of circumstellar CO line emission of AGB stars in the Herschel SUCCESS programme}",
      journal = {\aap},
         year = "2015",
        month = "Sep",
       volume = {581},
          eid = {A60},
        pages = {A60},
          doi = {10.1051/0004-6361/201526705},
archivePrefix = {arXiv},
       eprint = {1506.09065},
 primaryClass = {astro-ph.SR},
       adsurl = {https://ui.adsabs.harvard.edu/abs/2015A&A...581A..60D}
}

@ARTICLE{Danilovich2016,
   author = {{Danilovich}, T. and {De Beck}, E. and {Black}, J.~H. and {Olofsson}, H. and 
	{Justtanont}, K.},
    title = "{Sulphur molecules in the circumstellar envelopes of M-type AGB stars}",
  journal = {\aap},
archivePrefix = "arXiv",
   eprint = {1602.00517},
 primaryClass = "astro-ph.SR",
     year = 2016,
    month = apr,
   volume = 588,
      eid = {A119},
    pages = {A119},
      doi = {10.1051/0004-6361/201527943},
   adsurl = {http://adsabs.harvard.edu/abs/2016A%26A...588A.119D}
}

@ARTICLE{Danilovich2017,
       author = {{Danilovich}, T. and {Van de Sande}, M. and {De Beck}, E. and {Decin}, L. and {Olofsson}, H. and {Ramstedt}, S. and {Millar}, T.~J.},
        title = "{Sulphur-bearing molecules in AGB stars. I. The occurrence of hydrogen sulphide}",
      journal = {\aap},
         year = 2017,
        month = oct,
       volume = {606},
          eid = {A124},
        pages = {A124},
          doi = {10.1051/0004-6361/201731203},
archivePrefix = {arXiv},
       eprint = {1707.06003},
 primaryClass = {astro-ph.SR},
       adsurl = {https://ui.adsabs.harvard.edu/abs/2017A&A...606A.124D}
}

@ARTICLE{Danilovich2018,
       author = {{Danilovich}, T. and {Ramstedt}, S. and {Gobrecht}, D. and {Decin}, L. and {De Beck}, E. and {Olofsson}, H.},
        title = "{Sulphur-bearing molecules in AGB stars. II. Abundances and distributions of CS and SiS}",
      journal = {\aap},
         year = 2018,
        month = oct,
       volume = {617},
          eid = {A132},
        pages = {A132},
          doi = {10.1051/0004-6361/201833317},
archivePrefix = {arXiv},
       eprint = {1807.05144},
 primaryClass = {astro-ph.SR},
       adsurl = {https://ui.adsabs.harvard.edu/abs/2018A&A...617A.132D}
}

@article{Danilovich2019,
    author = {Danilovich, T. and Richards, A.~M.~S. and Karakas, A.~I. and Van de Sande, M. and Decin, L. and De Ceuster, F.},
    title = "{An ALMA view of CS and SiS around oxygen-rich AGB stars}",
    journal = {\mnras},
    volume = {484},
    number = {1},
    pages = {494-509},
    year = {2019},
    month = {01},
    issn = {0035-8711},
    doi = {10.1093/mnras/stz002},
    url = {https://doi.org/10.1093/mnras/stz002},
    eprint = {http://oup.prod.sis.lan/mnras/article-pdf/484/1/494/27512363/stz002.pdf},
}

@ARTICLE{Danilovich2020,
       author = {{Danilovich}, T. and {Richards}, A.~M.~S. and {Decin}, L. and {Van de Sande}, M. and {Gottlieb}, C.~A.},
        title = "{An ALMA view of SO and SO$_{2}$ around oxygen-rich AGB stars}",
      journal = {\mnras},
         year = 2020,
        month = may,
       volume = {494},
       number = {1},
        pages = {1323-1347},
          doi = {10.1093/mnras/staa693},
archivePrefix = {arXiv},
       eprint = {2003.04334},
 primaryClass = {astro-ph.SR},
       adsurl = {https://ui.adsabs.harvard.edu/abs/2020MNRAS.494.1323D}
}

@ARTICLE{Danilovich2024,
       author = {{Danilovich}, T. and {Malfait}, J. and {Van de Sande}, M. and {Montarg{\`e}s}, M. and {Kervella}, P. and {De Ceuster}, F. and {Coenegrachts}, A. and {Millar}, T.~J. and {Richards}, A.~M.~S. and {Decin}, L. and {Gottlieb}, C.~A. and {Pinte}, C. and {De Beck}, E. and {Price}, D.~J. and {Wong}, K.~T. and {Bolte}, J. and {Menten}, K.~M. and {Baudry}, A. and {de Koter}, A. and {Etoka}, S. and {Gobrecht}, D. and {Gray}, M. and {Herpin}, F. and {Jeste}, M. and {Lagadec}, E. and {Maes}, S. and {McDonald}, I. and {Marinho}, L. and {M{\"u}ller}, H.~S.~P. and {Pimpanuwat}, B. and {Plane}, J.~M.~C. and {Sahai}, R. and {Wallstr{\"o}m}, S.~H.~J. and {Yates}, J. and {Zijlstra}, A.},
        title = "{Chemical tracers of a highly eccentric AGB-main-sequence star binary}",
      journal = {Nature Astronomy},
         year = 2024,
        month = mar,
       volume = {8},
        pages = {308-327},
          doi = {10.1038/s41550-023-02154-y},
archivePrefix = {arXiv},
       eprint = {2407.16979},
 primaryClass = {astro-ph.SR},
       adsurl = {https://ui.adsabs.harvard.edu/abs/2024NatAs...8..308D}
}

@ARTICLE{DeBeck2012,
   author = {{De Beck}, E. and {Lombaert}, R. and {Ag{\'u}ndez}, M. and {Daniel}, F. and 
	{Decin}, L. and {Cernicharo}, J. and {M{\"u}ller}, H.~S.~P. and 
	{Min}, M. and {Royer}, P. and {Vandenbussche}, B. and {de Koter}, A. and 
	{Waters}, L.~B.~F.~M. and {Groenewegen}, M.~A.~T. and {Barlow}, M.~J. and 
	{Gu{\'e}lin}, M. and {Kahane}, C. and {Pearson}, J.~C. and {Encrenaz}, P. and 
	{Szczerba}, R. and {Schmidt}, M.~R.},
    title = "{On the physical structure of IRC +10216. Ground-based and Herschel observations of CO and C$_{2}$H}",
  journal = {\aap},
archivePrefix = "arXiv",
   eprint = {1201.1850},
 primaryClass = "astro-ph.SR",
     year = 2012,
    month = mar,
   volume = 539,
      eid = {A108},
    pages = {A108},
      doi = {10.1051/0004-6361/201117635},
   adsurl = {http://adsabs.harvard.edu/abs/2012A%26A...539A.108D}
}

@ARTICLE{Decin2010b,
   author = {{Decin}, L. and {Ag{\'u}ndez}, M. and {Barlow}, M.~J. and {Daniel}, F. and 
	{Cernicharo}, J. and {Lombaert}, R. and {De Beck}, E. and {Royer}, P. and 
	{Vandenbussche}, B. and {Wesson}, R. and {Polehampton}, E.~T. and 
	{Blommaert}, J.~A.~D.~L. and {De Meester}, W. and {Exter}, K. and 
	{Feuchtgruber}, H. and {Gear}, W.~K. and {Gomez}, H.~L. and 
	{Groenewegen}, M.~A.~T. and {Gu{\'e}lin}, M. and {Hargrave}, P.~C. and 
	{Huygen}, R. and {Imhof}, P. and {Ivison}, R.~J. and {Jean}, C. and 
	{Kahane}, C. and {Kerschbaum}, F. and {Leeks}, S.~J. and {Lim}, T. and 
	{Matsuura}, M. and {Olofsson}, G. and {Posch}, T. and {Regibo}, S. and 
	{Savini}, G. and {Sibthorpe}, B. and {Swinyard}, B.~M. and {Yates}, J.~A. and 
	{Waelkens}, C.},
    title = "{Warm water vapour in the sooty outflow from a luminous carbon star}",
  journal = {\nat},
archivePrefix = "arXiv",
   eprint = {1104.2316},
 primaryClass = "astro-ph.SR",
     year = 2010,
    month = sep,
   volume = 467,
    pages = {64-67},
      doi = {10.1038/nature09344},
   adsurl = {http://adsabs.harvard.edu/abs/2010Natur.467...64D}
}

@ARTICLE{Decin2020,
       author = {{Decin}, L. and {Montarg{\`e}s}, M. and {Richards}, A.~M.~S. and {Gottlieb}, C.~A. and {Homan}, W. and {McDonald}, I. and {El Mellah}, I. and {Danilovich}, T. and {Wallstr{\"o}m}, S.~H.~J. and {Zijlstra}, A. and {Baudry}, A. and {Bolte}, J. and {Cannon}, E. and {De Beck}, E. and {De Ceuster}, F. and {de Koter}, A. and {De Ridder}, J. and {Etoka}, S. and {Gobrecht}, D. and {Gray}, M. and {Herpin}, F. and {Jeste}, M. and {Lagadec}, E. and {Kervella}, P. and {Khouri}, T. and {Menten}, K. and {Millar}, T.~J. and {M{\"u}ller}, H.~S.~P. and {Plane}, J.~M.~C. and {Sahai}, R. and {Sana}, H. and {Van de Sande}, M. and {Waters}, L.~B.~F.~M. and {Wong}, K.~T. and {Yates}, J.},
        title = "{(Sub)stellar companions shape the winds of evolved stars}",
      journal = {Science},
         year = 2020,
        month = sep,
       volume = {369},
       number = {6510},
        pages = {1497-1500},
          doi = {10.1126/science.abb1229},
archivePrefix = {arXiv},
       eprint = {2009.11694},
 primaryClass = {astro-ph.SR},
       adsurl = {https://ui.adsabs.harvard.edu/abs/2020Sci...369.1497D}
}

@ARTICLE{Decin2021,
       author = {{Decin}, Leen},
title = {Evolution and Mass Loss of Cool Aging Stars: A Daedalean Story},
journal = {\araa},
volume = {59},
number = {1},
pages = {337-389},
year = {2021},
doi = {10.1146/annurev-astro-090120-033712},

URL = { 
        https://doi.org/10.1146/annurev-astro-090120-033712
    
},
eprint = { 
        https://doi.org/10.1146/annurev-astro-090120-033712
    
}}

@ARTICLE{Dijkstra2003,
   author = {{Dijkstra}, C. and {Dominik}, C. and {Hoogzaad}, S.~N. and {de Koter}, A. and 
	{Min}, M.},
    title = "{Water ice growth around evolved stars}",
  journal = {\aap},
   eprint = {astro-ph/0301569},
     year = 2003,
    month = apr,
   volume = 401,
    pages = {599-611},
      doi = {10.1051/0004-6361:20030102},
   adsurl = {http://adsabs.harvard.edu/abs/2003A%26A...401..599D}
}

@ARTICLE{Dijkstra2006,
   author = {{Dijkstra}, C. and {Dominik}, C. and {Bouwman}, J. and {de Koter}, A.
	},
    title = "{Water ice growth around evolved stars. II. Modeling infrared spectra}",
  journal = {\aap},
     year = 2006,
    month = apr,
   volume = 449,
    pages = {1101-1116},
      doi = {10.1051/0004-6361:20054053},
   adsurl = {https://ui.adsabs.harvard.edu/abs/2006A%26A...449.1101D}
}

@ARTICLE{Dobrijevic1998,
       author = {{Dobrijevic}, M. and {Parisot}, J.~P.},
        title = "{Effect of chemical kinetics uncertainties on hydrocarbon production in the stratosphere of Neptune}",
      journal = {\planss},
         year = 1998,
        month = may,
       volume = {46},
       number = {5},
        pages = {491-505},
          doi = {10.1016/S0032-0633(97)00176-1},
       adsurl = {https://ui.adsabs.harvard.edu/abs/1998P&SS...46..491D}
}

@ARTICLE{Dobrijevic2003,
       author = {{Dobrijevic}, M. and {Ollivier}, J.~L. and {Billebaud}, F. and {Brillet}, J. and {Parisot}, J.~P.},
        title = "{Effect of chemical kinetic uncertainties on photochemical modeling results: Application to Saturn's atmosphere}",
      journal = {\aap},
         year = 2003,
        month = jan,
       volume = {398},
        pages = {335-344},
          doi = {10.1051/0004-6361:20021659},
       adsurl = {https://ui.adsabs.harvard.edu/abs/2003A&A...398..335D}
}

@article{Dobrijevic2010,
title = {Comparison of methods for the determination of key reactions in chemical systems: Application to Titan’s atmosphere},
journal = {Advances in Space Research},
volume = {45},
number = {1},
pages = {77-91},
year = {2010},
issn = {0273-1177},
doi = {https://doi.org/10.1016/j.asr.2009.06.005},
url = {https://www.sciencedirect.com/science/article/pii/S0273117709003834},
author = {M. Dobrijevic and E. Hébrard and S. Plessis and N. Carrasco and P. Pernot and M. Bruno-Claeys}
}

@ARTICLE{Draine1978,
   author = {{Draine}, B.~T.},
    title = "{Photoelectric heating of interstellar gas}",
  journal = {\apjs},
     year = 1978,
    month = apr,
   volume = 36,
    pages = {595-619},
      doi = {10.1086/190513},
   adsurl = {http://adsabs.harvard.edu/abs/1978ApJS...36..595D}
}

@ARTICLE{Fonfria2008,
       author = {{Fonfr{\'\i}a}, J.~P. and {Cernicharo}, J. and {Richter}, M.~J. and
         {Lacy}, J.~H.},
        title = "{A Detailed Analysis of the Dust Formation Zone of IRC +10216 Derived from Mid-Infrared Bands of C$_{2}$H$_{2}$ and HCN}",
      journal = {\apj},
         year = 2008,
        month = jan,
       volume = {673},
       number = {1},
        pages = {445-469},
          doi = {10.1086/523882},
archivePrefix = {arXiv},
       eprint = {0709.4390},
 primaryClass = {astro-ph},
       adsurl = {https://ui.adsabs.harvard.edu/abs/2008ApJ...673..445F}
}

@ARTICLE{GonzalezDelgado2003,
   author = {{Gonz{\'a}lez Delgado}, D. and {Olofsson}, H. and {Kerschbaum}, F. and 
	{Sch{\"o}ier}, F.~L. and {Lindqvist}, M. and {Groenewegen}, M.~A.~T.
	},
    title = "{``Thermal'' SiO radio line emission towards M-type AGB stars: A probe of circumstellar dust formation and dynamics}",
  journal = {\aap},
   eprint = {astro-ph/0302179},
     year = 2003,
    month = nov,
   volume = 411,
    pages = {123-147},
      doi = {10.1051/0004-6361:20031068},
   adsurl = {http://adsabs.harvard.edu/abs/2003A%26A...411..123G}
}

@ARTICLE{Goldreich1976,
   author = {{Goldreich}, P. and {Scoville}, N.},
    title = "{OH-IR stars. I - Physical properties of circumstellar envelopes}",
  journal = {\apj},
     year = 1976,
    month = apr,
   volume = 205,
    pages = {144-154},
      doi = {10.1086/154257},
   adsurl = {http://adsabs.harvard.edu/abs/1976ApJ...205..144G}
}

@ARTICLE{Groenewegen2012,
       author = {{Groenewegen}, M.~A.~T. and {Barlow}, M.~J. and {Blommaert}, J.~A.~D.~L. and {Cernicharo}, J. and {Decin}, L. and {Gomez}, H.~L. and {Hargrave}, P.~C. and {Kerschbaum}, F. and {Ladjal}, D. and {Lim}, T.~L. and {Matsuura}, M. and {Olofsson}, G. and {Sibthorpe}, B. and {Swinyard}, B.~M. and {Ueta}, T. and {Yates}, J.},
        title = "{An independent distance estimate to CW Leonis}",
      journal = {\aap},
         year = 2012,
        month = jul,
       volume = {543},
          eid = {L8},
        pages = {L8},
          doi = {10.1051/0004-6361/201219604},
archivePrefix = {arXiv},
       eprint = {1206.5982},
 primaryClass = {astro-ph.SR},
       adsurl = {https://ui.adsabs.harvard.edu/abs/2012A&A...543L...8G}
}

@INPROCEEDINGS{Guelin1999,
   author = {{Gu{\'e}lin}, M. and {Neininger}, N. and {Lucas}, R. and {Cernicharo}, J.
	},
    title = "{Carbon-chain molecules as tracers of time-dependent chemistry}",
booktitle = {The Physics and Chemistry of the Interstellar Medium},
     year = 1999,
   editor = {{Ossenkopf}, V. and {Stutzki}, J. and {Winnewisser}, G.},
    month = aug,
   adsurl = {http://adsabs.harvard.edu/abs/1999pcim.conf..326G}
}

@ARTICLE{Homan2018,
       author = {{Homan}, Ward and {Danilovich}, Taissa and {Decin}, Leen and {de Koter}, Alex and {Nuth}, Joseph and {Van de Sande}, Marie},
        title = "{ALMA detection of a tentative nearly edge-on rotating disk around the nearby AGB star R Doradus}",
      journal = {\aap},
         year = 2018,
        month = jun,
       volume = {614},
          eid = {A113},
        pages = {A113},
          doi = {10.1051/0004-6361/201732246},
archivePrefix = {arXiv},
       eprint = {1803.06207},
 primaryClass = {astro-ph.SR},
       adsurl = {https://ui.adsabs.harvard.edu/abs/2018A&A...614A.113H}
}

@ARTICLE{Huggins1982,
   author = {{Huggins}, P.~J. and {Glassgold}, A.~E.},
    title = "{The photochemistry of carbon-rich circumstellar shells}",
  journal = {\apj},
     year = 1982,
    month = jan,
   volume = 252,
    pages = {201-207},
      doi = {10.1086/159547},
   adsurl = {http://adsabs.harvard.edu/abs/1982ApJ...252..201H}
}

@ARTICLE{Jura1985,
   author = {{Jura}, M. and {Morris}, M.},
    title = "{Condensation onto grains in the outflows from mass-losing red giants}",
  journal = {\apj},
     year = 1985,
    month = may,
   volume = 292,
    pages = {487-493},
      doi = {10.1086/163180},
   adsurl = {https://ui.adsabs.harvard.edu/abs/1985ApJ...292..487J}
}

@PHDTHESIS{Keller2017,
       author = {{Keller}, Denise},
        title = "{Molecules in the circumstellar envelope of the evolved carbon-rich star IRC+10216}",
       school = {Rheinische Friedrich Wilhelms University of Bonn, Germany},
         year = 2017,
        month = jan,
       adsurl = {https://ui.adsabs.harvard.edu/abs/2017PhDT.......447K}
}

@ARTICLE{Kervella2016,
       author = {{Kervella}, P. and {Homan}, W. and {Richards}, A.~M.~S. and {Decin}, L. and {McDonald}, I. and {Montarg{\`e}s}, M. and {Ohnaka}, K.},
        title = "{ALMA observations of the nearby AGB star L$_{2}$ Puppis. I. Mass of the central star and detection of a candidate planet}",
      journal = {\aap},
         year = 2016,
        month = dec,
       volume = {596},
          eid = {A92},
        pages = {A92},
          doi = {10.1051/0004-6361/201629877},
archivePrefix = {arXiv},
       eprint = {1611.06231},
 primaryClass = {astro-ph.SR},
       adsurl = {https://ui.adsabs.harvard.edu/abs/2016A&A...596A..92K}
}

@ARTICLE{Khouri2016,
   author = {{Khouri}, T. and {Maercker}, M. and {Waters}, L.~B.~F.~M. and 
	{Vlemmings}, W.~H.~T. and {Kervella}, P. and {de Koter}, A. and 
	{Ginski}, C. and {De Beck}, E. and {Decin}, L. and {Min}, M. and 
	{Dominik}, C. and {O'Gorman}, E. and {Schmid}, H.-M. and {Lombaert}, R. and 
	{Lagadec}, E.},
    title = "{Study of the inner dust envelope and stellar photosphere of the AGB star R Doradus using SPHERE/ZIMPOL}",
  journal = {\aap},
archivePrefix = "arXiv",
   eprint = {1605.05504},
 primaryClass = "astro-ph.SR",
     year = 2016,
    month = jun,
   volume = 591,
      eid = {A70},
    pages = {A70},
      doi = {10.1051/0004-6361/201628435},
   adsurl = {http://adsabs.harvard.edu/abs/2016A%26A...591A..70K}
}

@ARTICLE{Li2016,
   author = {{Li}, X. and {Millar}, T.~J. and {Heays}, A.~N. and {Walsh}, C. and 
	{van Dishoeck}, E.~F. and {Cherchneff}, I.},
    title = "{Chemistry and distribution of daughter species in the circumstellar envelopes of O-rich AGB stars}",
  journal = {\aap},
     year = 2016,
    month = apr,
   volume = 588,
      eid = {A4},
    pages = {A4},
      doi = {10.1051/0004-6361/201525739},
   adsurl = {http://adsabs.harvard.edu/abs/2016A%26A...588A...4L}
}

@ARTICLE{Maercker2008,
   author = {{Maercker}, M. and {Sch{\"o}ier}, F.~L. and {Olofsson}, H. and 
	{Bergman}, P. and {Ramstedt}, S.},
    title = "{Circumstellar water vapour in M-type AGB stars: radiative transfer models, abundances, and predictions for HIFI}",
  journal = {\aap},
archivePrefix = "arXiv",
   eprint = {0801.0971},
     year = 2008,
    month = mar,
   volume = 479,
    pages = {779-791},
      doi = {10.1051/0004-6361:20078680},
   adsurl = {http://adsabs.harvard.edu/abs/2008A%26A...479..779M}
}

@ARTICLE{Maercker2016,
   author = {{Maercker}, M. and {Danilovich}, T. and {Olofsson}, H. and {De Beck}, E. and 
	{Justtanont}, K. and {Lombaert}, R. and {Royer}, P.},
    title = "{A HIFI view on circumstellar H$_{2}$O in M-type AGB stars: radiative transfer, velocity profiles, and H$_{2}$O line cooling}",
  journal = {\aap},
archivePrefix = "arXiv",
   eprint = {1605.00504},
 primaryClass = "astro-ph.SR",
     year = 2016,
    month = jun,
   volume = 591,
      eid = {A44},
    pages = {A44},
      doi = {10.1051/0004-6361/201628310},
   adsurl = {http://adsabs.harvard.edu/abs/2016A%26A...591A..44M}
}

@ARTICLE{Maes2023,
       author = {{Maes}, S. and {Van de Sande}, M. and {Danilovich}, T. and {De Ceuster}, F. and {Decin}, L.},
        title = "{Sensitivity study of chemistry in AGB outflows using chemical kinetics}",
      journal = {\mnras},
         year = 2023,
        month = jul,
       volume = {522},
       number = {3},
        pages = {4654-4673},
          doi = {10.1093/mnras/stad1152},
archivePrefix = {arXiv},
       eprint = {2304.05924},
 primaryClass = {astro-ph.GA},
       adsurl = {https://ui.adsabs.harvard.edu/abs/2023MNRAS.522.4654M}
}

@ARTICLE{Mamon1988,
   author = {{Mamon}, G.~A. and {Glassgold}, A.~E. and {Huggins}, P.~J.},
    title = "{The photodissociation of CO in circumstellar envelopes}",
  journal = {\apj},
     year = 1988,
    month = may,
   volume = 328,
    pages = {797-808},
      doi = {10.1086/166338},
   adsurl = {http://adsabs.harvard.edu/abs/1988ApJ...328..797M}
}

@ARTICLE{Massalkhi2019,
       author = {{Massalkhi}, S. and {Ag{\'u}ndez}, M. and {Cernicharo}, J.},
        title = "{Study of CS, SiO, and SiS abundances in carbon star envelopes: assessing their role as gas-phase precursors of dust}",
      journal = {\aap},
         year = 2019,
        month = aug,
       volume = {628},
          eid = {A62},
        pages = {A62},
          doi = {10.1051/0004-6361/201935069},
archivePrefix = {arXiv},
       eprint = {1906.09461},
 primaryClass = {astro-ph.SR},
       adsurl = {https://ui.adsabs.harvard.edu/abs/2019A&A...628A..62M}
}

@ARTICLE{Massalkhi2020,
       author = {{Massalkhi}, S. and {Ag{\'u}ndez}, M. and {Cernicharo}, J. and {Velilla-Prieto}, L.},
        title = "{The abundance of S- and Si-bearing molecules in O-rich circumstellar envelopes of AGB stars}",
      journal = {\aap},
         year = 2020,
        month = sep,
       volume = {641},
          eid = {A57},
        pages = {A57},
          doi = {10.1051/0004-6361/202037900},
archivePrefix = {arXiv},
       eprint = {2007.00572},
 primaryClass = {astro-ph.SR},
       adsurl = {https://ui.adsabs.harvard.edu/abs/2020A&A...641A..57M}
}

@ARTICLE{Massalkhi2024,
       author = {{Massalkhi}, S. and {Ag{\'u}ndez}, M. and {Fonfr{\'\i}a}, J.~P. and {Pardo}, J.~R. and {Velilla-Prieto}, L. and {Cernicharo}, J.},
        title = "{Multiline study of the radial extent of SiO, CS, and SiS in asymptotic giant branch envelopes}",
      journal = {\aap},
         year = 2024,
        month = aug,
       volume = {688},
          eid = {A16},
        pages = {A16},
          doi = {10.1051/0004-6361/202450188},
archivePrefix = {arXiv},
       eprint = {2405.19922},
 primaryClass = {astro-ph.SR},
       adsurl = {https://ui.adsabs.harvard.edu/abs/2024A&A...688A..16M}
}

@ARTICLE{Mauron2006,
   author = {{Mauron}, N. and {Huggins}, P.~J.},
    title = "{Imaging the circumstellar envelopes of AGB stars}",
  journal = {\aap},
   eprint = {astro-ph/0602623},
     year = 2006,
    month = jun,
   volume = 452,
    pages = {257-268},
      doi = {10.1051/0004-6361:20054739},
   adsurl = {http://adsabs.harvard.edu/abs/2006A%26A...452..257M}
}

@ARTICLE{McElroy2013,
   author = {{McElroy}, D. and {Walsh}, C. and {Markwick}, A.~J. and {Cordiner}, M.~A. and 
	{Smith}, K. and {Millar}, T.~J.},
    title = "{The UMIST database for astrochemistry 2012}",
  journal = {\aap},
archivePrefix = "arXiv",
   eprint = {1212.6362},
 primaryClass = "astro-ph.SR",
     year = 2013,
    month = feb,
   volume = 550,
      eid = {A36},
    pages = {A36},
      doi = {10.1051/0004-6361/201220465},
   adsurl = {http://adsabs.harvard.edu/abs/2013A%26A...550A..36M}
}

@ARTICLE{Millar1994,
   author = {{Millar}, T.~J. and {Herbst}, E.},
    title = "{A new chemical model of the circumstellar envelope surrounding IRC+10216}",
  journal = {\aap},
     year = 1994,
    month = aug,
   volume = 288,
    pages = {561-571},
   adsurl = {http://adsabs.harvard.edu/abs/1994A%26A...288..561M}
}

@ARTICLE{Millar2000,
   author = {{Millar}, T.~J. and {Herbst}, E. and {Bettens}, R.~P.~A.},
    title = "{Large molecules in the envelope surrounding IRC+10216}",
  journal = {\mnras},
     year = 2000,
    month = jul,
   volume = 316,
    pages = {195-203},
      doi = {10.1046/j.1365-8711.2000.03560.x},
   adsurl = {http://adsabs.harvard.edu/abs/2000MNRAS.316..195M}
}

@ARTICLE{Millar2024,
       author = {{Millar}, T.~J. and {Walsh}, C. and {Van de Sande}, M. and {Markwick}, A.~J.},
        title = "{The UMIST Database for Astrochemistry 2022}",
      journal = {\aap},
         year = 2024,
        month = feb,
       volume = {682},
          eid = {A109},
        pages = {A109},
          doi = {10.1051/0004-6361/202346908},
archivePrefix = {arXiv},
       eprint = {2311.03936},
 primaryClass = {astro-ph.GA},
       adsurl = {https://ui.adsabs.harvard.edu/abs/2024A&A...682A.109M}
}

@ARTICLE{Morris1983,
   author = {{Morris}, M. and {Jura}, M.},
    title = "{Molecular self-shielding in the outflows from late-type stars}",
  journal = {\apj},
     year = 1983,
    month = jan,
   volume = 264,
    pages = {546-553},
      doi = {10.1086/160622},
   adsurl = {http://adsabs.harvard.edu/abs/1983ApJ...264..546M}
}

@article{Morris1991,
 ISSN = {00401706},
 URL = {http://www.jstor.org/stable/1269043},
 author = {Max D. Morris},
 journal = {Technometrics},
 number = {2},
 pages = {161--174},
 publisher = {[Taylor & Francis, Ltd., American Statistical Association, American Society for Quality]},
 title = {Factorial Sampling Plans for Preliminary Computational Experiments},
 urldate = {2025-03-10},
 volume = {33},
 year = {1991}
}

@ARTICLE{Nejad1984,
   author = {{Nejad}, L.~A.~M. and {Millar}, T.~J. and {Freeman}, A.},
    title = "{Chemical modelling of molecular sources. III - C3H in IRC + 10216}",
  journal = {\aap},
     year = 1984,
    month = may,
   volume = 134,
    pages = {129-133},
   adsurl = {http://adsabs.harvard.edu/abs/1984A%26A...134..129N}
}

@ARTICLE{Netzer1987,
       author = {{Netzer}, Nathan and {Knapp}, G.~R.},
        title = "{Mass Loss from Evolved Stars. VII. OH Maser Shell Radii and Mass-Loss Rates for OH/IR Stars}",
      journal = {\apj},
         year = 1987,
        month = dec,
       volume = {323},
        pages = {734},
          doi = {10.1086/165867},
       adsurl = {https://ui.adsabs.harvard.edu/abs/1987ApJ...323..734N}
}

@ARTICLE{Penteado2017,
   author = {{Penteado}, E.~M. and {Walsh}, C. and {Cuppen}, H.~M.},
    title = "{Sensitivity Analysis of Grain Surface Chemistry to Binding Energies of Ice Species}",
  journal = {\apj},
archivePrefix = "arXiv",
   eprint = {1708.01450},
     year = 2017,
    month = jul,
   volume = 844,
      eid = {71},
    pages = {71},
      doi = {10.3847/1538-4357/aa78f9},
   adsurl = {http://adsabs.harvard.edu/abs/2017ApJ...844...71P}
}

@ARTICLE{Ramstedt2008,
       author = {{Ramstedt}, S. and {Sch{\"o}ier}, F.~L. and {Olofsson}, H. and {Lundgren}, A.~A.},
        title = "{On the reliability of mass-loss-rate estimates for AGB stars}",
      journal = {\aap},
         year = 2008,
        month = aug,
       volume = {487},
       number = {2},
        pages = {645-657},
          doi = {10.1051/0004-6361:20078876},
archivePrefix = {arXiv},
       eprint = {0806.0517},
 primaryClass = {astro-ph},
       adsurl = {https://ui.adsabs.harvard.edu/abs/2008A&A...487..645R}
}

@ARTICLE{Ramstedt2017,
       author = {{Ramstedt}, S. and {Mohamed}, S. and {Vlemmings}, W.~H.~T. and {Danilovich}, T. and {Brunner}, M. and {De Beck}, E. and {Humphreys}, E.~M.~L. and {Lindqvist}, M. and {Maercker}, M. and {Olofsson}, H. and {Kerschbaum}, F. and {Quintana-Lacaci}, G.},
        title = "{The circumstellar envelope around the S-type AGB star W Aql. Effects of an eccentric binary orbit}",
      journal = {\aap},
         year = 2017,
        month = sep,
       volume = {605},
          eid = {A126},
        pages = {A126},
          doi = {10.1051/0004-6361/201730934},
archivePrefix = {arXiv},
       eprint = {1709.07327},
 primaryClass = {astro-ph.SR},
       adsurl = {https://ui.adsabs.harvard.edu/abs/2017A&A...605A.126R}
}

@ARTICLE{Saberi2019,
       author = {{Saberi}, M. and {Vlemmings}, W.~H.~T. and {De Beck}, E.},
        title = "{Photodissociation of CO in the outflow of evolved stars}",
      journal = {\aap},
         year = 2019,
        month = may,
       volume = {625},
          eid = {A81},
        pages = {A81},
          doi = {10.1051/0004-6361/201935309},
archivePrefix = {arXiv},
       eprint = {1904.05425},
 primaryClass = {astro-ph.SR},
       adsurl = {https://ui.adsabs.harvard.edu/abs/2019A&A...625A..81S}
}

@ARTICLE{Safonov2025,
       author = {{Safonov}, Boris S. and {Zheltoukhov}, Sergey G. and {Tatarnikov}, Andrey M. and {Strakhov}, Ivan A. and {Shenavrin}, Victor I.},
        title = "{Disk in the Circumstellar Envelope of Carbon Mira V Cygni}",
      journal = {\aj},
         year = 2025,
        month = mar,
       volume = {169},
       number = {3},
          eid = {140},
        pages = {140},
          doi = {10.3847/1538-3881/adaaf3},
archivePrefix = {arXiv},
       eprint = {2501.10092},
 primaryClass = {astro-ph.SR},
       adsurl = {https://ui.adsabs.harvard.edu/abs/2025AJ....169..140S}
}

@article{Saltelli2019,
title = {Why so many published sensitivity analyses are false: A systematic review of sensitivity analysis practices},
journal = {Environmental Modelling & Software},
volume = {114},
pages = {29-39},
year = {2019},
issn = {1364-8152},
doi = {https://doi.org/10.1016/j.envsoft.2019.01.012},
url = {https://www.sciencedirect.com/science/article/pii/S1364815218302822},
author = {Andrea Saltelli and Ksenia Aleksankina and William Becker and Pamela Fennell and Federico Ferretti and Niels Holst and Sushan Li and Qiongli Wu}
}

@ARTICLE{Scalo1980,
   author = {{Scalo}, J.~M. and {Slavsky}, D.~B.},
    title = "{Chemical structure of circumstellar shells}",
  journal = {\apjl},
     year = 1980,
    month = jul,
   volume = 239,
    pages = {L73-L77},
      doi = {10.1086/183295},
   adsurl = {http://adsabs.harvard.edu/abs/1980ApJ...239L..73S}
}

@ARTICLE{Schoier2001,
       author = {{Sch{\"o}ier}, F.~L. and {Olofsson}, H.},
        title = "{Models of circumstellar molecular radio line emission. Mass loss rates for a sample of bright carbon stars}",
      journal = {\aap},
         year = 2001,
        month = mar,
       volume = {368},
        pages = {969-993},
          doi = {10.1051/0004-6361:20010072},
archivePrefix = {arXiv},
       eprint = {astro-ph/0101477},
 primaryClass = {astro-ph},
       adsurl = {https://ui.adsabs.harvard.edu/abs/2001A&A...368..969S}
}

@ARTICLE{Schoier2013,
   author = {{Sch{\"o}ier}, F.~L. and {Ramstedt}, S. and {Olofsson}, H. and 
	{Lindqvist}, M. and {Bieging}, J.~H. and {Marvel}, K.~B.},
    title = "{The abundance of HCN in circumstellar envelopes of AGB stars of different chemical type}",
  journal = {\aap},
archivePrefix = "arXiv",
   eprint = {1301.2129},
 primaryClass = "astro-ph.SR",
     year = 2013,
    month = feb,
   volume = 550,
      eid = {A78},
    pages = {A78},
      doi = {10.1051/0004-6361/201220400},
   adsurl = {http://cdsads.u-strasbg.fr/abs/2013A%26A...550A..78S}
}

@ARTICLE{Siebert2022,
       author = {{Siebert}, Mark A. and {Van de Sande}, Marie and {Millar}, Thomas J. and {Remijan}, Anthony J.},
        title = "{Investigating Anomalous Photochemistry in the Inner Wind of IRC+10216 through Interferometric Observations of HC$_{3}$N}",
      journal = {Astrophys. J.},
         year = 2022,
        month = dec,
       volume = {941},
       number = {1},
          eid = {90},
        pages = {90},
          doi = {10.3847/1538-4357/ac9e52},
archivePrefix = {arXiv},
       eprint = {2210.14941},
 primaryClass = {astro-ph.SR},
       adsurl = {https://ui.adsabs.harvard.edu/abs/2022ApJ...941...90S}
}

@ARTICLE{Stewart1996,
       author = {{Stewart}, Richard W. and {Thompson}, Anne M.},
        title = "{Kinetic data imprecisions in photochemical rate calculations: Means, medians, and temperature dependence}",
      journal = {\jgr},
         year = 1996,
        month = sep,
       volume = {101},
       number = {D15},
        pages = {20,953-20,964},
          doi = {10.1029/96JD01708},
       adsurl = {https://ui.adsabs.harvard.edu/abs/1996JGR...10120953S}
}

@ARTICLE{Sylvester1999,
       author = {{Sylvester}, R.~J. and {Kemper}, F. and {Barlow}, M.~J. and
         {de Jong}, T. and {Waters}, L.~B.~F.~M. and {Tielens}, A.~G.~G.~M. and
         {Omont}, A.},
        title = "{2.4-197 mu m spectroscopy of OH/IR stars: the IR characteristics of circumstellar dust in O-rich environments}",
      journal = {\aap},
         year = "1999",
        month = "Dec",
       volume = {352},
        pages = {587-599},
archivePrefix = {arXiv},
       eprint = {astro-ph/9910368},
 primaryClass = {astro-ph},
       adsurl = {https://ui.adsabs.harvard.edu/abs/1999A&A...352..587S}
}

@ARTICLE{Teyssier2006,
   author = {{Teyssier}, D. and {Hernandez}, R. and {Bujarrabal}, V. and 
	{Yoshida}, H. and {Phillips}, T.~G.},
    title = "{CO line emission from circumstellar envelopes}",
  journal = {\aap},
     year = 2006,
    month = apr,
   volume = 450,
    pages = {167-179},
      doi = {10.1051/0004-6361:20053759},
   adsurl = {http://adsabs.harvard.edu/abs/2006A%26A...450..167T}
}

@ARTICLE{Thompson1991,
       author = {{Thompson}, Anne M. and {Stewart}, Richard W.},
        title = "{Effect of chemical kinetics uncertainties on calculated constituents in a tropospheric photochemical model}",
      journal = {\jgr},
         year = 1991,
        month = jul,
       volume = {96},
       number = {D7},
        pages = {13,089-13,108},
          doi = {10.1029/91JD01056},
       adsurl = {https://ui.adsabs.harvard.edu/abs/1991JGR....9613089T}
}

@ARTICLE{Unnikrishnan2025,
       author = {{Unnikrishnan}, R. and {Andriantsaralaza}, M. and {De Beck}, E. and {Nyman}, L.-{\r{A}}. and {Olofsson}, H. and {Vlemmings}, W.~H.~T. and {Maercker}, M. and {Van de Sande}, M. and {Danilovich}, T. and {Millar}, T.~J. and {Charnley}, S.~B. and {Rawlings}, M.~G.},
        title = "{Charting circumstellar chemistry of carbon-rich asymptotic giant branch stars: II. Abundances and spatial distributions of CS}",
      journal = {\aap},
         year = 2025,
        month = jul,
       volume = {699},
          eid = {A48},
        pages = {A48},
          doi = {10.1051/0004-6361/202554996},
archivePrefix = {arXiv},
       eprint = {2505.22173},
 primaryClass = {astro-ph.SR},
       adsurl = {https://ui.adsabs.harvard.edu/abs/2025A&A...699A..48U}
}

@ARTICLE{VandeSande2018,
       author = {{Van de Sande}, M. and {Sundqvist}, J.~O. and {Millar}, T.~J. and
         {Keller}, D. and {Homan}, W. and {de Koter}, A. and {Decin}, L. and
         {De Ceuster}, F.},
        title = "{Determining the effects of clumping and porosity on the chemistry in a non-uniform AGB outflow}",
      journal = {\aap},
         year = "2018",
        month = "Aug",
       volume = {616},
          eid = {A106},
        pages = {A106},
          doi = {10.1051/0004-6361/201732276},
archivePrefix = {arXiv},
       eprint = {1803.01796},
 primaryClass = {astro-ph.SR},
       adsurl = {https://ui.adsabs.harvard.edu/\#abs/2018A&A...616A.106V}
}

@ARTICLE{VandeSande2019b,
       author = {{Van de Sande}, M. and {Walsh}, C. and {Mangan}, T.~P. and {Decin}, L.},
        title = "{Chemical modelling of dust-gas chemistry within AGB outflows I. Effect on the gas-phase chemistry}",
      journal = {Mon. Not. R. Astron. Soc.},
         year = "2019",
        month = "Sep",
        pages = {2325},
          doi = {10.1093/mnras/stz2702},
archivePrefix = {arXiv},
       eprint = {1909.10410},
 primaryClass = {astro-ph.SR},
       adsurl = {https://ui.adsabs.harvard.edu/abs/2019MNRAS.tmp.2325V}
}

@ARTICLE{VandeSande2020,
       author = {{Van de Sande}, M. and {Walsh}, C. and {Danilovich}, T.},
        title = "{Chemical modelling of dust-gas chemistry within AGB outflows - II. Effect of the dust-grain size distribution}",
      journal = {Mon. Not. R. Astron. Soc.},
         year = 2020,
        month = jun,
       volume = {495},
       number = {2},
        pages = {1650-1665},
          doi = {10.1093/mnras/staa1270},
archivePrefix = {arXiv},
       eprint = {2005.01553},
 primaryClass = {astro-ph.SR},
       adsurl = {https://ui.adsabs.harvard.edu/abs/2020MNRAS.495.1650V}
}

@ARTICLE{VandeSande2021,
       author = {{Van de Sande}, M. and {Walsh}, C. and {Millar}, T.~J.},
        title = "{Chemical modelling of dust-gas chemistry within AGB outflows - III. Photoprocessing of the ice and return to the ISM}",
      journal = {Mon. Not. R. Astron. Soc.},
         year = 2021,
        month = feb,
       volume = {501},
       number = {1},
        pages = {491-506},
          doi = {10.1093/mnras/staa3689},
archivePrefix = {arXiv},
       eprint = {2011.11563},
 primaryClass = {astro-ph.SR},
       adsurl = {https://ui.adsabs.harvard.edu/abs/2021MNRAS.501..491V}
}

@ARTICLE{VandeSande2022,
       author = {{Van de Sande}, M. and {Millar}, T.~J.},
        title = "{The impact of stellar companion UV photons on the chemistry of the circumstellar environments of AGB stars}",
      journal = {Mon. Not. R. Astron. Soc.},
         year = 2022,
        month = feb,
       volume = {510},
       number = {1},
        pages = {1204-1222},
          doi = {10.1093/mnras/stab3282},
archivePrefix = {arXiv},
       eprint = {2111.05053},
 primaryClass = {astro-ph.SR},
       adsurl = {https://ui.adsabs.harvard.edu/abs/2022MNRAS.510.1204V}
}

@ARTICLE{VandeSande2023,
       author = {{Van de Sande}, Marie and {Walsh}, Catherine and {Millar}, Tom J.},
        title = "{Disentangling physics and chemistry in AGB outflows: revealing degeneracies when adding complexity}",
      journal = {Faraday Discussions},
         year = 2023,
        month = sep,
       volume = {245},
        pages = {586-608},
          doi = {10.1039/D3FD00039G},
archivePrefix = {arXiv},
       eprint = {2306.05252},
 primaryClass = {astro-ph.SR},
       adsurl = {https://ui.adsabs.harvard.edu/abs/2023FaDi..245..586V}
}

@INPROCEEDINGS{vanDishoeck1988,
       author = {{van Dishoeck}, Ewine F. and {Black}, John H.},
        title = "{Diffuse Cloud Chemistry}",
    booktitle = {Rate Coefficients in Astrochemistry},
         year = 1988,
       editor = {{Millar}, T.~J. and {Williams}, D.~A.},
       series = {Astrophysics and Space Science Library},
       volume = {146},
        month = jan,
        pages = {209},
          doi = {10.1007/978-94-009-3007-0_15},
       adsurl = {https://ui.adsabs.harvard.edu/abs/1988ASSL..146..209V}
}

@ARTICLE{Vasyunin2004,
       author = {{Vasyunin}, A.~I. and {Sobolev}, A.~M. and {Wiebe}, D.~S. and
         {Semenov}, D.~A.},
        title = "{Influence of Uncertainties in the Rate Constants of Chemical Reactions on Astrochemical Modeling Results}",
      journal = {Astronomy Letters},
         year = 2004,
        month = aug,
       volume = {30},
        pages = {566-576},
          doi = {10.1134/1.1784498},
archivePrefix = {arXiv},
       eprint = {astro-ph/0311450},
 primaryClass = {astro-ph},
       adsurl = {https://ui.adsabs.harvard.edu/abs/2004AstL...30..566V}
}

@ARTICLE{Vasyunin2008,
       author = {{Vasyunin}, A.~I. and {Semenov}, D. and {Henning}, Th. and
         {Wakelam}, V. and {Herbst}, Eric and {Sobolev}, A.~M.},
        title = "{Chemistry in Protoplanetary Disks: A Sensitivity Analysis}",
      journal = {\apj},
         year = 2008,
        month = jan,
       volume = {672},
       number = {1},
        pages = {629-641},
          doi = {10.1086/523887},
archivePrefix = {arXiv},
       eprint = {0709.3323},
 primaryClass = {astro-ph},
       adsurl = {https://ui.adsabs.harvard.edu/abs/2008ApJ...672..629V}
}

@ARTICLE{VelillaPrieto2023,
       author = {{Velilla-Prieto}, L. and {Fonfr{\'\i}a}, J.~P. and {Ag{\'u}ndez}, M. and {Castro-Carrizo}, A. and {Gu{\'e}lin}, M. and {Quintana-Lacaci}, G. and {Cherchneff}, I. and {Joblin}, C. and {McCarthy}, M.~C. and {Mart{\'\i}n-Gago}, J.~A. and {Cernicharo}, J.},
        title = "{Atmospheric molecular blobs shape up circumstellar envelopes of AGB stars}",
      journal = {\nat},
         year = 2023,
        month = may,
       volume = {617},
       number = {7962},
        pages = {696-700},
          doi = {10.1038/s41586-023-05917-9},
archivePrefix = {arXiv},
       eprint = {2412.04027},
 primaryClass = {astro-ph.SR},
       adsurl = {https://ui.adsabs.harvard.edu/abs/2023Natur.617..696V}
}

@ARTICLE{Verbena2019,
       author = {{Verbena}, J.~L. and {Bujarrabal}, V. and {Alcolea}, J. and
         {G{\'o}mez-Garrido}, M. and {Castro-Carrizo}, A.},
        title = "{Interferometric observations of SiO thermal emission in the inner wind of M-type AGB stars IK Tauri and IRC+10011}",
      journal = {\aap},
         year = "2019",
        month = "Apr",
       volume = {624},
          eid = {A107},
        pages = {A107},
          doi = {10.1051/0004-6361/201834864},
archivePrefix = {arXiv},
       eprint = {1902.10933},
 primaryClass = {astro-ph.SR},
       adsurl = {https://ui.adsabs.harvard.edu/abs/2019A&A...624A.107V}
}

@ARTICLE{Wakelam2005,
       author = {{Wakelam}, V. and {Selsis}, F. and {Herbst}, E. and {Caselli}, P.},
        title = "{Estimation and reduction of the uncertainties in chemical models: application to hot core chemistry}",
      journal = {\aap},
         year = 2005,
        month = dec,
       volume = {444},
       number = {3},
        pages = {883-891},
          doi = {10.1051/0004-6361:20053673},
archivePrefix = {arXiv},
       eprint = {astro-ph/0509194},
 primaryClass = {astro-ph},
       adsurl = {https://ui.adsabs.harvard.edu/abs/2005A&A...444..883W}
}

@ARTICLE{Wakelam2006a,
       author = {{Wakelam}, V. and {Herbst}, E. and {Selsis}, F.},
        title = "{The effect of uncertainties on chemical models of dark clouds}",
      journal = {\aap},
         year = 2006,
        month = may,
       volume = {451},
       number = {2},
        pages = {551-562},
          doi = {10.1051/0004-6361:20054682},
archivePrefix = {arXiv},
       eprint = {astro-ph/0601611},
 primaryClass = {astro-ph},
       adsurl = {https://ui.adsabs.harvard.edu/abs/2006A&A...451..551W}
}

@ARTICLE{Wakelam2006b,
       author = {{Wakelam}, V. and {Herbst}, E. and {Selsis}, F. and {Massacrier}, G.},
        title = "{Chemical sensitivity to the ratio of the cosmic-ray ionization rates of He and H$_{2}$ in dense clouds}",
      journal = {\aap},
         year = 2006,
        month = dec,
       volume = {459},
       number = {3},
        pages = {813-820},
          doi = {10.1051/0004-6361:20065472},
archivePrefix = {arXiv},
       eprint = {astro-ph/0608551},
 primaryClass = {astro-ph},
       adsurl = {https://ui.adsabs.harvard.edu/abs/2006A&A...459..813W}
}

@ARTICLE{Wakelam2010,
       author = {{Wakelam}, V. and {Herbst}, E. and {Le Bourlot}, J. and {Hersant}, F. and
         {Selsis}, F. and {Guilloteau}, S.},
        title = "{Sensitivity analyses of dense cloud chemical models}",
      journal = {\aap},
         year = 2010,
        month = jul,
       volume = {517},
          eid = {A21},
        pages = {A21},
          doi = {10.1051/0004-6361/200913856},
archivePrefix = {arXiv},
       eprint = {1004.1902},
 primaryClass = {astro-ph.GA},
       adsurl = {https://ui.adsabs.harvard.edu/abs/2010A&A...517A..21W}
}

@article{Wittkowski2017,
	author = {{Wittkowski}, M. and {Hofmann, K.-H.} and {Höfner, S.} and {Le Bouquin, J. B.} and {Nowotny, W.} and {Paladini, C.} and {Young, J.} and {Berger, J.-P.} and {Brunner, M.} and {de Gregorio-Monsalvo, I.} and {Eriksson, K.} and {Hron, J.} and {Humphreys, E. M. L.} and {Lindqvist, M.} and {Maercker, M.} and {Mohamed, S.} and {Olofsson, H.} and {Ramstedt, S.} and {Weigelt, G.}},
	title = {Aperture synthesis imaging of the carbon AGB star R Sculptoris⋆ - Detection of a complex structure and a dominating spot on the stellar disk},
	DOI= "10.1051/0004-6361/201630214",
	url= "https://doi.org/10.1051/0004-6361/201630214",
	journal = {A&A},
	year = 2017,
	volume = 601,
	pages = "A3",
}

%%%%%%%%%%%%%%%%%%%%%%%%%%%%%%%%%%%%%%%%%%%%%%%%%%

%%%%%%%%%%%%%%%%% APPENDICES %%%%%%%%%%%%%%%%%%%%%
\newpage
\appendix

%%%%%%%%%%%%%%%%%%%%%%%%%%%%%%%%%%%%%%%%%%%%%%%%%%%%%%%%%%%%%%%%%%%%%%%%%%%%%%%%%%%%%%%%%%%%%%%%%%%%%%%%%%%%%%
\section{Deviations from a lognormal distribution}			\label{app:gaussian}
%%%%%%%%%%%%%%%%%%%%%%%%%%%%%%%%%%%%%%%%%%%%%%%%%%%%%%%%%%%%%%%%%%%%%%%%%%%%%%%%%%%%%%%%%%%%%%%%%%%%%%%%%%%%%%

Fig. \ref{fig:app-gaussian-FA} illustrates the deviation from a Gaussian distribution of the logarithm of the predicted fractional abundances.
Rapid changes in abundance, caused by e.g., the onset of photodissociation, lead to a non-Gaussian distribution of  $\log X$ \citep{Wakelam2005}.
Using the standard deviation as a measure of the error is in these cases not appropriate.
We therefore use the smallest interval that contains $95.4\%$ of all abundance profiles (Eq. \ref{eq:error}) to determine the error. 

Similarly, Fig. \ref{fig:app-gaussian-CD} illustrates how the logarithm of the predicted column densities can also deviate from a Gaussian distribution.
For most species, the smallest interval that contains $95.4\%$ of all column densities corresponds to 2 times the standard deviation.
This is not the case for species with a non-lognormal distribution of their predicted column densities (bottom panel).

\begin{figure*}
\centering
\includegraphics[width=\textwidth]{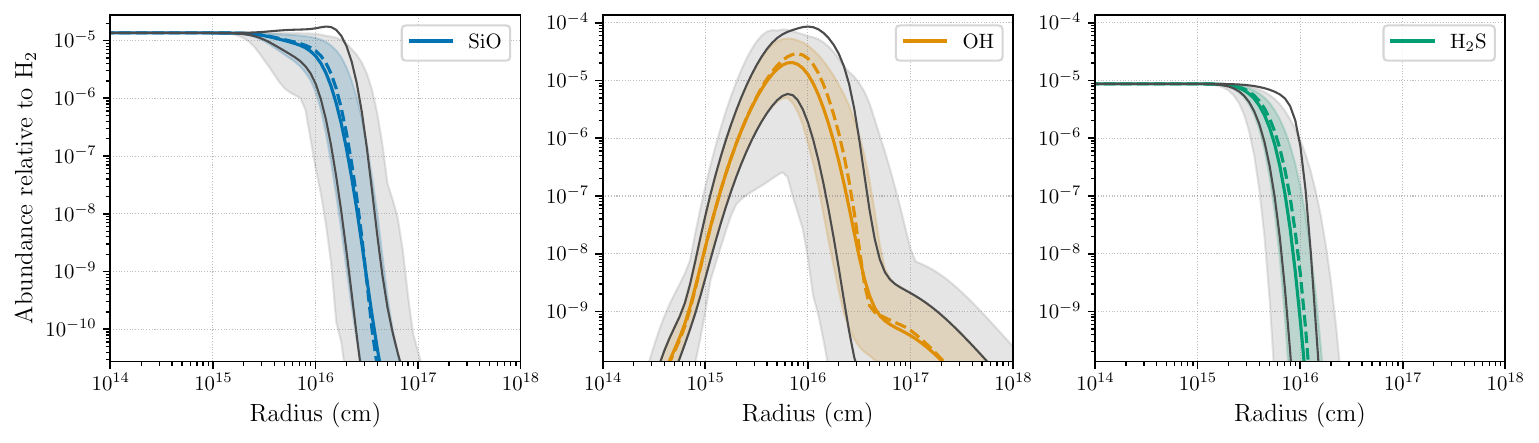}
\caption{Fractional abundance w.r.t. \ce{H2} of SiO, OH, and \ce{H2S} in the O-rich outflow with $\dot{M} = 10^{-6}\ \mathrm{M}_\odot\ \mathrm{yr}^{-1}$. 
The dashed coloured line shows the fiducial model prediction using \textsc{Rate22}. 
The solid coloured line shows the mean abundance $\langle \log X(r)\rangle$ of the Monte Carlo sample of reaction networks.
The coloured shaded region contains 95.4\% of all predicted profiles and corresponds to the error $\Delta \log X(r)$ on the mean abundance.
The grey shaded region contains all 10,000 predicted profiles.
The black lines show the mean abundance $\pm$ 2 $\times$ the standard deviation of the distribution of $\log X$. 
}
\label{fig:app-gaussian-FA}
\end{figure*}

\begin{figure}
\centering
\includegraphics[width=0.8\columnwidth]{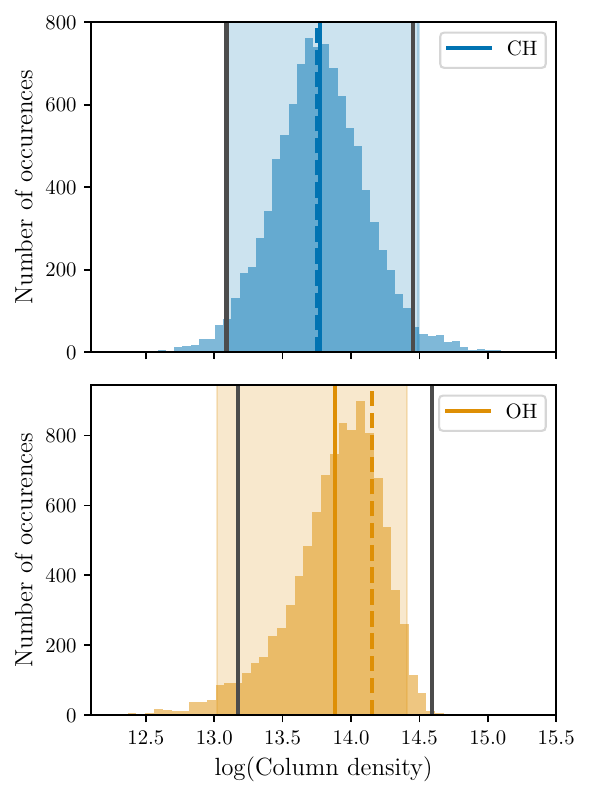}
\caption{Histogram of the predicted column densities of CH and OH in the C-rich outflow with $\dot{M} = 10^{-6}\ \mathrm{M}_\odot\ \mathrm{yr}^{-1}$. 
The dashed coloured line shows the fiducial model prediction using \textsc{Rate22}. 
The solid coloured line shows the mean column density of the Monte Carlo sample of reaction networks.
The shaded region contains 95.4\% of all predicted profiles and corresponds to the error on the column density.
The black lines show the mean column density $\pm$ 2 $\times$ the standard deviation of the distribution of predicted column densities. 
}
\label{fig:app-gaussian-CD}
\end{figure}

%%%%%%%%%%%%%%%%%%%%%%%%%%%%%%%%%%%%%%%%%%%%%%%%%%%%%%%%%%%%%%%%%%%%%%%%%%%%%%%%%%%%%%%%%%%%%%%%%%%%%%%%%%%%%%
\section{Convergence of the Monte Carlo sampling method}				\label{app:convergence}
%%%%%%%%%%%%%%%%%%%%%%%%%%%%%%%%%%%%%%%%%%%%%%%%%%%%%%%%%%%%%%%%%%%%%%%%%%%%%%%%%%%%%%%%%%%%%%%%%%%%%%%%%%%%%%

To determine the convergence of the Monte Carlo sampling method results, we assessed the stability of our predictions with increasing sample size.
Convergence was deemed adequate if increasing the sample size does not produce significant changes in the final predictions.
Specifically, we evaluated the relative change in the mean and error of the column densities and peak fractional abundances as the sample size increases.
The predictions were calculated for increasing sample sizes, using increments of 200.

For all outflow chemistries and densities, we find that the relative change in all estimates with increasing the sample size lies well below 1\%.
%Hence, all estimates are clearly converged after 10,000 samples.
The convergence of the Monte Carlo modelling is illustrated in Fig. \ref{fig:app-conv-cd}, which shows the convergence trajectories of the mean column densities and their dispersions normalised to their final values for all daughter species in the C-rich outflow with $\mathrm{10^{-6}\ M_\odot\ yr^{-1}}$.
These convergence trajectories for the column density and its dispersion and the peak fractional abundance and its error for all outflow chemistries and densities.
Our sample size of 10,000 models is therefore sufficient.

%
% for the C-rich outflow with $\mathrm{10^{-6}\ M_\odot\ yr^{-1}}$.
%Fig. XX and YY show the mean column density and its dispersion and the mean peak fractional abundance and its error, respectively, evaluated at increasing sample sizes normalised to their final values for all daughter species.

\begin{figure}
\centering
\includegraphics[width=\columnwidth]{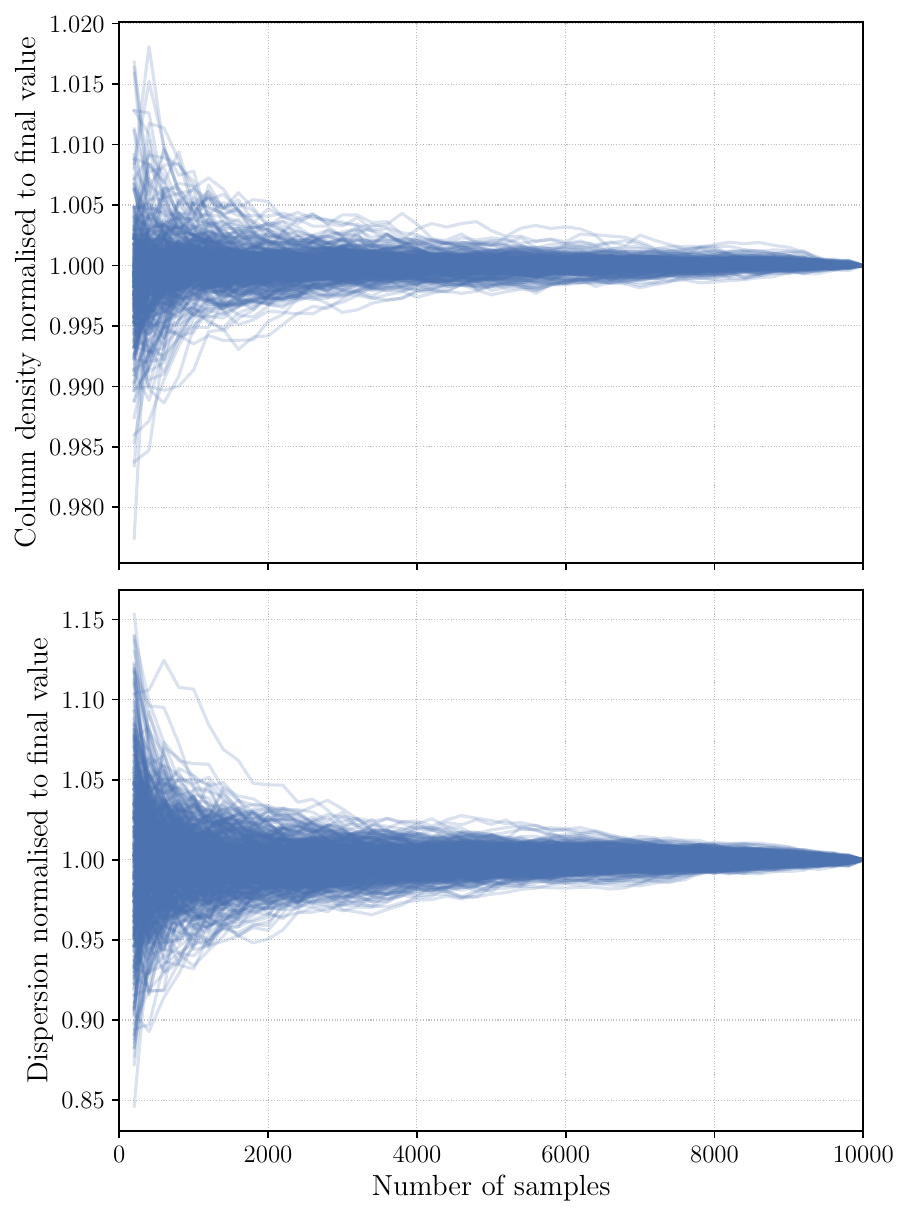}
\caption{Convergence trajectories for \emph{(top)} column density and \emph{(bottom)} its dispersion for all daughter species in the C-rich outflow with $\mathrm{10^{-6}\ M_\odot\ yr^{-1}}$.
The predictions are calculated for increasing sample sizes, using increments of 200, and normalised to their final values. 
}
\label{fig:app-conv-cd}
\end{figure}

\newpage

%%%%%%%%%%%%%%%%%%%%%%%%%%%%%%%%%%%%%%%%%%%%%%%%%%%%%%%%%%%%%%%%%%%%%%%%%%%%%%%%%%%%%%%%%%%%%%%%%%%%%%%%%%%%%%
\section{Radial extent parents}				\label{app:extent}
%%%%%%%%%%%%%%%%%%%%%%%%%%%%%%%%%%%%%%%%%%%%%%%%%%%%%%%%%%%%%%%%%%%%%%%%%%%%%%%%%%%%%%%%%%%%%%%%%%%%%%%%%%%%%%

Table \ref{table:uncert-parents} lists the radial extent of the molecular envelopes of the parent species in C-rich and O-rich outflows. 
These are shown in Fig. \ref{fig:parents-extent}. 
The mean radial extent is given along with the smallest and largest predicted values.

\begin{table*}
	\caption{Radial extent of the molecular envelopes of the parent species in C-rich (\emph{top}) and O-rich (\emph{bottom}) outflows for different mass-loss rates.
	For each mass-loss rate, the mean radial extent of the molecular envelope is listed, along with the smallest and largest predicted values (min and max, respectively). 
	All values are in cm.
	} 
% 	\centering
    % \resizebox{1.0\columnwidth}{!}{%
    \centering
    \begin{tabular}{l c c c c c c c c c c c }
    \hline  
    \noalign{\smallskip}
     \multicolumn{12}{c}{Carbon-rich}  \\
    \noalign{\smallskip}
    \hline
    \noalign{\smallskip}
	 &     \multicolumn{3}{c}{$\mathrm{10^{-5}\ M_\odot\ yr^{-1}}$} & & \multicolumn{3}{c}{$\mathrm{10^{-6}\ M_\odot\ yr^{-1}}$}  & & \multicolumn{3}{c}{$\mathrm{10^{-7}\ M_\odot\ yr^{-1}}$}  \\  
    \cline{2-4} \cline{6-8} \cline{10-12}
    \noalign{\smallskip}
    Species & Min & Mean & Max & & Min & Mean & Max & & Min & Mean & Max  \\
    \cline{2-4} \cline{6-8} \cline{10-12}
    \noalign{\smallskip}
\ce{CO} & $1.79 \times 10^{17}$ & $2.25 \times 10^{17}$ & $5.65 \times 10^{17}$ &  & $5.65 \times 10^{16}$ & $7.98 \times 10^{16}$ & $2.00 \times 10^{17}$ &  & $2.52 \times 10^{16}$ & $3.56 \times 10^{16}$ & $8.96 \times 10^{16}$ \\
\ce{N2} & $4.00 \times 10^{16}$ & $6.34 \times 10^{16}$ & $1.27 \times 10^{17}$ &  & $1.13 \times 10^{16}$ & $1.79 \times 10^{16}$ & $5.04 \times 10^{16}$ &  & $3.57 \times 10^{15}$ & $6.34 \times 10^{15}$ & $2.00 \times 10^{16}$ \\
\ce{H2O} & $2.00 \times 10^{16}$ & $2.83 \times 10^{16}$ & $3.56 \times 10^{16}$ &  & $5.04 \times 10^{15}$ & $7.11 \times 10^{15}$ & $1.13 \times 10^{16}$ &  & $1.27 \times 10^{15}$ & $2.00 \times 10^{15}$ & $3.18 \times 10^{15}$ \\
\ce{HCN} & $1.79 \times 10^{16}$ & $2.25 \times 10^{16}$ & $3.18 \times 10^{16}$ &  & $4.00 \times 10^{15}$ & $5.65 \times 10^{15}$ & $7.98 \times 10^{15}$ &  & $1.13 \times 10^{15}$ & $1.59 \times 10^{15}$ & $2.83 \times 10^{15}$ \\
\ce{NH3} & $1.59 \times 10^{16}$ & $1.79 \times 10^{16}$ & $2.25 \times 10^{16}$ &  & $4.00 \times 10^{15}$ & $5.04 \times 10^{15}$ & $6.34 \times 10^{15}$ &  & $1.13 \times 10^{15}$ & $1.42 \times 10^{15}$ & $2.00 \times 10^{15}$ \\
\ce{H2S} & $1.59 \times 10^{16}$ & $2.00 \times 10^{16}$ & $2.52 \times 10^{16}$ &  & $3.18 \times 10^{15}$ & $4.49 \times 10^{15}$ & $5.65 \times 10^{15}$ &  & $7.98 \times 10^{14}$ & $1.13 \times 10^{15}$ & $1.42 \times 10^{15}$ \\
\ce{SiO} & $1.79 \times 10^{16}$ & $2.52 \times 10^{16}$ & $4.00 \times 10^{16}$ &  & $4.00 \times 10^{15}$ & $5.65 \times 10^{15}$ & $1.01 \times 10^{16}$ &  & $1.13 \times 10^{15}$ & $1.59 \times 10^{15}$ & $3.18 \times 10^{15}$ \\
\ce{SiS} & $2.83 \times 10^{16}$ & $3.56 \times 10^{16}$ & $4.49 \times 10^{16}$ &  & $7.98 \times 10^{15}$ & $1.01 \times 10^{16}$ & $1.42 \times 10^{16}$ &  & $2.52 \times 10^{15}$ & $3.18 \times 10^{15}$ & $5.04 \times 10^{15}$ \\
\ce{CS} & $2.52 \times 10^{16}$ & $4.49 \times 10^{16}$ & $7.98 \times 10^{16}$ &  & $6.34 \times 10^{15}$ & $1.13 \times 10^{16}$ & $2.52 \times 10^{16}$ &  & $2.00 \times 10^{15}$ & $4.00 \times 10^{15}$ & $1.01 \times 10^{16}$ \\
\ce{HCl} & $1.79 \times 10^{16}$ & $2.25 \times 10^{16}$ & $3.56 \times 10^{16}$ &  & $4.00 \times 10^{15}$ & $5.65 \times 10^{15}$ & $1.01 \times 10^{16}$ &  & $1.01 \times 10^{15}$ & $1.59 \times 10^{15}$ & $3.18 \times 10^{15}$ \\
\ce{HF} & $3.18 \times 10^{16}$ & $4.00 \times 10^{16}$ & $5.04 \times 10^{16}$ &  & $8.96 \times 10^{15}$ & $1.13 \times 10^{16}$ & $1.59 \times 10^{16}$ &  & $4.00 \times 10^{15}$ & $5.04 \times 10^{15}$ & $7.11 \times 10^{15}$ \\
\ce{CH4} & $2.00 \times 10^{16}$ & $2.52 \times 10^{16}$ & $3.18 \times 10^{16}$ &  & $4.49 \times 10^{15}$ & $5.65 \times 10^{15}$ & $8.96 \times 10^{15}$ &  & $1.13 \times 10^{15}$ & $1.79 \times 10^{15}$ & $2.83 \times 10^{15}$ \\
\ce{C2H2} & $1.42 \times 10^{16}$ & $1.79 \times 10^{16}$ & $2.25 \times 10^{16}$ &  & $3.18 \times 10^{15}$ & $4.00 \times 10^{15}$ & $5.65 \times 10^{15}$ &  & $8.96 \times 10^{14}$ & $1.27 \times 10^{15}$ & $2.00 \times 10^{15}$ \\
\ce{C2H4} & $1.42 \times 10^{16}$ & $1.79 \times 10^{16}$ & $2.52 \times 10^{16}$ &  & $3.18 \times 10^{15}$ & $4.00 \times 10^{15}$ & $5.04 \times 10^{15}$ &  & $7.98 \times 10^{14}$ & $1.01 \times 10^{15}$ & $1.42 \times 10^{15}$ \\
\ce{SiC2} & $3.18 \times 10^{16}$ & $4.00 \times 10^{16}$ & $5.65 \times 10^{16}$ &  & $7.98 \times 10^{15}$ & $1.13 \times 10^{16}$ & $2.00 \times 10^{16}$ &  & $2.52 \times 10^{15}$ & $4.00 \times 10^{15}$ & $7.11 \times 10^{15}$ \\
\ce{HCP} & $2.00 \times 10^{16}$ & $2.83 \times 10^{16}$ & $5.04 \times 10^{16}$ &  & $5.04 \times 10^{15}$ & $7.98 \times 10^{15}$ & $1.59 \times 10^{16}$ &  & $1.42 \times 10^{15}$ & $2.52 \times 10^{15}$ & $5.65 \times 10^{15}$ \\
    \hline 
    \noalign{\smallskip}
     \multicolumn{12}{c}{Oxygen-rich}  \\
    \noalign{\smallskip}
    \hline
    \noalign{\smallskip}
	 &     \multicolumn{3}{c}{$\mathrm{10^{-5}\ M_\odot\ yr^{-1}}$} & & \multicolumn{3}{c}{$\mathrm{10^{-6}\ M_\odot\ yr^{-1}}$}  & & \multicolumn{3}{c}{$\mathrm{10^{-7}\ M_\odot\ yr^{-1}}$}  \\  
    \cline{2-4} \cline{6-8} \cline{10-12}
    \noalign{\smallskip}
    Species & Min & Mean & Max & & Min & Mean & Max & & Min & Mean & Max  \\
    \cline{2-4} \cline{6-8} \cline{10-12}
    \noalign{\smallskip}
\ce{CO} & $1.27 \times 10^{17}$ & $1.59 \times 10^{17}$ & $3.56 \times 10^{17}$ &  & $4.00 \times 10^{16}$ & $5.65 \times 10^{16}$ & $1.27 \times 10^{17}$ &  & $1.59 \times 10^{16}$ & $2.25 \times 10^{16}$ & $5.65 \times 10^{16}$ \\
\ce{N2} & $4.00 \times 10^{16}$ & $5.65 \times 10^{16}$ & $1.27 \times 10^{17}$ &  & $1.01 \times 10^{16}$ & $1.79 \times 10^{16}$ & $4.49 \times 10^{16}$ &  & $3.18 \times 10^{15}$ & $5.65 \times 10^{15}$ & $1.79 \times 10^{16}$ \\
\ce{H2O} & $2.00 \times 10^{16}$ & $2.83 \times 10^{16}$ & $4.49 \times 10^{16}$ &  & $5.04 \times 10^{15}$ & $7.11 \times 10^{15}$ & $1.27 \times 10^{16}$ &  & $1.42 \times 10^{15}$ & $2.25 \times 10^{15}$ & $4.00 \times 10^{15}$ \\
\ce{HCN} & $1.79 \times 10^{16}$ & $2.52 \times 10^{16}$ & $4.00 \times 10^{16}$ &  & $4.00 \times 10^{15}$ & $5.65 \times 10^{15}$ & $1.13 \times 10^{16}$ &  & $1.13 \times 10^{15}$ & $1.59 \times 10^{15}$ & $3.57 \times 10^{15}$ \\
\ce{NH3} & $1.79 \times 10^{16}$ & $2.00 \times 10^{16}$ & $2.52 \times 10^{16}$ &  & $4.00 \times 10^{15}$ & $5.04 \times 10^{15}$ & $6.34 \times 10^{15}$ &  & $1.13 \times 10^{15}$ & $1.42 \times 10^{15}$ & $2.00 \times 10^{15}$ \\
\ce{H2S} & $1.42 \times 10^{16}$ & $1.79 \times 10^{16}$ & $2.25 \times 10^{16}$ &  & $3.18 \times 10^{15}$ & $4.00 \times 10^{15}$ & $5.04 \times 10^{15}$ &  & $7.98 \times 10^{14}$ & $1.01 \times 10^{15}$ & $1.42 \times 10^{15}$ \\
\ce{SiO} & $2.25 \times 10^{16}$ & $3.56 \times 10^{16}$ & $5.65 \times 10^{16}$ &  & $5.04 \times 10^{15}$ & $1.01 \times 10^{16}$ & $1.79 \times 10^{16}$ &  & $1.59 \times 10^{15}$ & $3.18 \times 10^{15}$ & $5.65 \times 10^{15}$ \\
\ce{SiS} & $2.25 \times 10^{16}$ & $2.83 \times 10^{16}$ & $4.49 \times 10^{16}$ &  & $7.11 \times 10^{15}$ & $1.13 \times 10^{16}$ & $2.25 \times 10^{16}$ &  & $2.00 \times 10^{15}$ & $3.57 \times 10^{15}$ & $8.96 \times 10^{15}$ \\
\ce{CS} & $7.98 \times 10^{15}$ & $1.27 \times 10^{16}$ & $1.27 \times 10^{17}$ &  & $1.42 \times 10^{15}$ & $2.52 \times 10^{15}$ & $4.49 \times 10^{16}$ &  & $3.57 \times 10^{14}$ & $6.34 \times 10^{14}$ & $6.34 \times 10^{15}$ \\
\ce{HCl} & $1.79 \times 10^{16}$ & $2.25 \times 10^{16}$ & $4.00 \times 10^{16}$ &  & $4.00 \times 10^{15}$ & $5.65 \times 10^{15}$ & $1.13 \times 10^{16}$ &  & $1.01 \times 10^{15}$ & $1.59 \times 10^{15}$ & $3.57 \times 10^{15}$ \\
\ce{HF} & $3.18 \times 10^{16}$ & $4.49 \times 10^{16}$ & $5.65 \times 10^{16}$ &  & $1.01 \times 10^{16}$ & $1.27 \times 10^{16}$ & $2.00 \times 10^{16}$ &  & $5.04 \times 10^{15}$ & $7.11 \times 10^{15}$ & $8.96 \times 10^{15}$ \\
\ce{CO2} & $2.25 \times 10^{16}$ & $3.18 \times 10^{16}$ & $6.34 \times 10^{16}$ &  & $5.04 \times 10^{15}$ & $7.98 \times 10^{15}$ & $1.79 \times 10^{16}$ &  & $1.59 \times 10^{15}$ & $3.18 \times 10^{15}$ & $7.11 \times 10^{15}$ \\
\ce{SO} & $1.59 \times 10^{16}$ & $2.83 \times 10^{16}$ & $4.00 \times 10^{16}$ &  & $3.57 \times 10^{15}$ & $7.11 \times 10^{15}$ & $1.13 \times 10^{16}$ &  & $1.01 \times 10^{15}$ & $2.25 \times 10^{15}$ & $4.00 \times 10^{15}$ \\
\ce{SO2} & $1.59 \times 10^{16}$ & $2.25 \times 10^{16}$ & $3.56 \times 10^{16}$ &  & $3.57 \times 10^{15}$ & $5.65 \times 10^{15}$ & $1.01 \times 10^{16}$ &  & $8.96 \times 10^{14}$ & $1.59 \times 10^{15}$ & $3.18 \times 10^{15}$ \\
\ce{PO} & $2.52 \times 10^{16}$ & $4.00 \times 10^{16}$ & $6.34 \times 10^{16}$ &  & $7.11 \times 10^{15}$ & $1.27 \times 10^{16}$ & $2.25 \times 10^{16}$ &  & $2.25 \times 10^{15}$ & $4.00 \times 10^{15}$ & $7.98 \times 10^{15}$ \\
\ce{PN} & $2.00 \times 10^{17}$ & $5.04 \times 10^{17}$ & $1.01 \times 10^{18}$ &  & $8.96 \times 10^{16}$ & $2.52 \times 10^{17}$ & $1.01 \times 10^{18}$ &  & $4.00 \times 10^{16}$ & $1.13 \times 10^{17}$ & $5.65 \times 10^{17}$ \\
    \hline 
    \end{tabular}%
    % }
    \label{table:uncert-parents}    
\end{table*}

\newpage
%%%%%%%%%%%%%%%%%%%%%%%%%%%%%%%%%%%%%%%%%%%%%%%%%%%%%%%%%%%%%%%%%%%%%%%%%%%%%%%%%%%%%%%%%%%%%%%%%%%%%%%%%%%%%%
\section{Dispersion all daughters}				\label{app:daughters}
%%%%%%%%%%%%%%%%%%%%%%%%%%%%%%%%%%%%%%%%%%%%%%%%%%%%%%%%%%%%%%%%%%%%%%%%%%%%%%%%%%%%%%%%%%%%%%%%%%%%%%%%%%%%%%

Table \ref{table:RCCs} lists the Spearman rank correlation coefficients between the mean peak fractional abundance and its error, the mean column density and its error, and the dispersion of the column density and the dispersion of the peak fractional abundance for all daughters and abundant daughters for all mass-loss rates and for C-rich and O-rich outflows. 
These are shown in Figs \ref{fig:disp-relevant}, \ref{fig:app-crich-relevant}, \ref{fig:app-orich-relevant}, \ref{fig:app-crich-all}, and \ref{fig:app-orich-all}.

A moderate negative correlation exists between peak fractional abundance and column density and their errors for all daughter species in both O-rich and C-rich outflows (RCCs between $-0.39$ and $-0.53$).
The correlation is weak for the abundant C-rich daughters (RCC between $-0.29$ and $-0.39$).
For the abundant O-rich daughters, the correlation is statistically non-significant, which could be due to the smaller number of abundant daughters ($\sim 25$).
We find a strong positive correlation between the dispersion of the peak fractional abundance and the dispersion of the column density (RCC $\geq 0.60$), except for all C-rich daughters, which has a moderate correlation (RCC $\sim 0.44$).
For each type of chemistry, the correlation is stronger for the abundant daughters than for all daughters.

\begin{table}
	\centering
	\caption{Spearman rank correlation coefficients (RCCs) between the mean peak fractional abundance and its error (\emph{top}), the mean column density and its error (\emph{middle}), and the dispersion of the column density and the dispersion of the peak fractional abundance (\emph{bottom}) for all daughters and abundants daughters for all mass-loss rates and for C-rich and O-rich outflows. p-values larger than 0.05 are boldfaced. 
	}
	\begin{tabular}{c c c c c cc c c c c} 
		\hline
		\noalign{\smallskip}
\multicolumn{6}{c}{Mean peak fractional abundance - error peak fractional abundance }\\
	\hline
%     & \multicolumn{4}{c}{Carbon-rich} && \multicolumn{4}{c}{Oxygen-rich}  \\  
%    \cline{2-3} \cline{5-6} 
    \noalign{\smallskip}
    $\dot{M}$ & RCC & p-value && RCC & p-value \\  
	\hline
	\noalign{\smallskip}
	 &  \multicolumn{5}{c}{Carbon-rich}  \\     
	\cline{2-6} 
    \noalign{\smallskip}
	 &  \multicolumn{2}{c}{All} && \multicolumn{2}{c}{Relevant}  \\   
	 \cline{2-3} \cline{5-6} 
    \noalign{\smallskip}
$10^{-5}\ \mathrm{M}_\odot\ \mathrm{yr}^{-1}$	& $-0.53$	&	$1.56\times 10^{-39}$	&&	$-0.29$	&	$1.09\times 10^{-2}$	\\
$10^{-6}\ \mathrm{M}_\odot\ \mathrm{yr}^{-1}$	& $-0.53$	&	$3.56\times 10^{-38}$	&&	$-0.39$	&	$8.37\times 10^{-4}$	\\
$10^{-7}\ \mathrm{M}_\odot\ \mathrm{yr}^{-1}$	& $-0.51$	&	$1.54\times 10^{-34}$	&&	$-0.32$	&	$1.17\times 10^{-2}$	\\
    \noalign{\smallskip}
	\hline
    \noalign{\smallskip}
	 &  \multicolumn{5}{c}{Oxygen-rich}  \\     
	\cline{2-6} 
    \noalign{\smallskip}
	 &  \multicolumn{2}{c}{All} && \multicolumn{2}{c}{Relevant}  \\   
	 \cline{2-3} \cline{5-6} 
    \noalign{\smallskip}
$10^{-5}\ \mathrm{M}_\odot\ \mathrm{yr}^{-1}$	& $-0.46$	&	$1.02\times 10^{-11}$	&&	$-0.08$	&	\textbf{\boldmath{$6.82\times 10^{-1}$}}	\\
$10^{-6}\ \mathrm{M}_\odot\ \mathrm{yr}^{-1}$	& $-0.44$	&	$1.56\times 10^{-10}$	&&	$-0.07$	&	\textbf{\boldmath{$7.45\times 10^{-1}$}}	\\
$10^{-7}\ \mathrm{M}_\odot\ \mathrm{yr}^{-1}$	& $-0.39$	&	$5.26\times 10^{-08}$	&&	$-0.17$	&	$4.46\times 10^{-1}$	\\
    \noalign{\smallskip}
		\hline
    \noalign{\smallskip}
\multicolumn{6}{c}{Mean column density - error column density }\\
	\hline
    \noalign{\smallskip}
    $\dot{M}$ & RCC & p-value && RCC & p-value \\  
	\hline
	\noalign{\smallskip}
	 &  \multicolumn{5}{c}{Carbon-rich}  \\     
	\cline{2-6} 
    \noalign{\smallskip}
	 &  \multicolumn{2}{c}{All} && \multicolumn{2}{c}{Relevant}  \\   
	 \cline{2-3} \cline{5-6} 
    \noalign{\smallskip}
$10^{-5}\ \mathrm{M}_\odot\ \mathrm{yr}^{-1}$	& $-0.54$	&	$2.48\times 10^{-41}$	&&	$-0.47$	&	$1.72\times 10^{-5}$	\\
$10^{-6}\ \mathrm{M}_\odot\ \mathrm{yr}^{-1}$	& $-0.55$	&	$2.46\times 10^{-42}$	&&	$-0.43$	&	$1.81\times 10^{-4}$	\\
$10^{-7}\ \mathrm{M}_\odot\ \mathrm{yr}^{-1}$	& $-0.53$	&	$2.22\times 10^{-38}$	&&	$-0.37$	&	$2.27\times 10^{-3}$	\\
    \noalign{\smallskip}
	\hline
	\noalign{\smallskip}
	 &  \multicolumn{5}{c}{Oxygen-rich}  \\     
	\cline{2-6} 
    \noalign{\smallskip}
	 &  \multicolumn{2}{c}{All} && \multicolumn{2}{c}{Relevant}  \\   
	 \cline{2-3} \cline{5-6} 
    \noalign{\smallskip}
$10^{-5}\ \mathrm{M}_\odot\ \mathrm{yr}^{-1}$	& $-0.63$	&	$3.49\times 10^{-23}$	&&	$-0.47$	&	$1.52\times 10^{-2}$	\\
$10^{-6}\ \mathrm{M}_\odot\ \mathrm{yr}^{-1}$	& $-0.58$	&	$5.17\times 10^{-19}$	&&	$-0.40$	&	$4.76\times 10^{-2}$	\\
$10^{-7}\ \mathrm{M}_\odot\ \mathrm{yr}^{-1}$	& $-0.53$	&	$2.53\times 10^{-14}$	&&	$-0.27$	&	\textbf{\boldmath{$2.01\times 10^{-1}$}}	\\
    \noalign{\smallskip}
		\hline
    \noalign{\smallskip}
\multicolumn{6}{c}{Dispersion of column density - dispersion of peak fractional abundance }\\
	\hline
%     & \multicolumn{4}{c}{Carbon-rich} && \multicolumn{4}{c}{Oxygen-rich}  \\  
%    \cline{2-3} \cline{5-6} 
    \noalign{\smallskip}
    $\dot{M}$ & RCC & p-value && RCC & p-value \\  
	\hline
	\noalign{\smallskip}
	 &  \multicolumn{5}{c}{Carbon-rich}  \\     
	\cline{2-6} 
    \noalign{\smallskip}
	 &  \multicolumn{2}{c}{All} && \multicolumn{2}{c}{Relevant}  \\   
	 \cline{2-3} \cline{5-6} 
    \noalign{\smallskip}
$10^{-5}\ \mathrm{M}_\odot\ \mathrm{yr}^{-1}$	& $0.44$	&	$4.60\times 10^{-27}$	&&	$0.70$	&	$1.37\times 10^{-12}$	\\
$10^{-6}\ \mathrm{M}_\odot\ \mathrm{yr}^{-1}$	& $0.44$	&	$1.07\times 10^{-25}$	&&	$0.83$	&	$1.84\times 10^{-19}$	\\
$10^{-7}\ \mathrm{M}_\odot\ \mathrm{yr}^{-1}$	& $0.44$	&	$6.96\times 10^{-26}$	&&	$0.83$	&	$4.55\times 10^{-17}$	\\
    \noalign{\smallskip}
	\hline
	\noalign{\smallskip}
	 &  \multicolumn{5}{c}{Oxygen-rich}  \\     
	\cline{2-6} 
    \noalign{\smallskip}
	 &  \multicolumn{2}{c}{All} && \multicolumn{2}{c}{Relevant}  \\   
	 \cline{2-3} \cline{5-6} 
    \noalign{\smallskip}
$10^{-5}\ \mathrm{M}_\odot\ \mathrm{yr}^{-1}$	& $0.63$	&	$3.34\times 10^{-23}$	&&	$0.79$	&	$1.58\times 10^{-6}$	\\
$10^{-6}\ \mathrm{M}_\odot\ \mathrm{yr}^{-1}$	& $0.67$	&	$6.67\times 10^{-27}$	&&	$0.80$	&	$1.47\times 10^{-6}$	\\
$10^{-7}\ \mathrm{M}_\odot\ \mathrm{yr}^{-1}$	& $0.64$	&	$2.11\times 10^{-22}$	&&	$0.81$	&	$2.78\times 10^{-6}$	\\
    \noalign{\smallskip}
		\hline
	\end{tabular}
    \label{table:RCCs}    
\end{table}

\begin{figure*}
\centering

\begin{subfigure}[b]{0.85\textwidth}
  \centering
	\includegraphics[width=\textwidth]{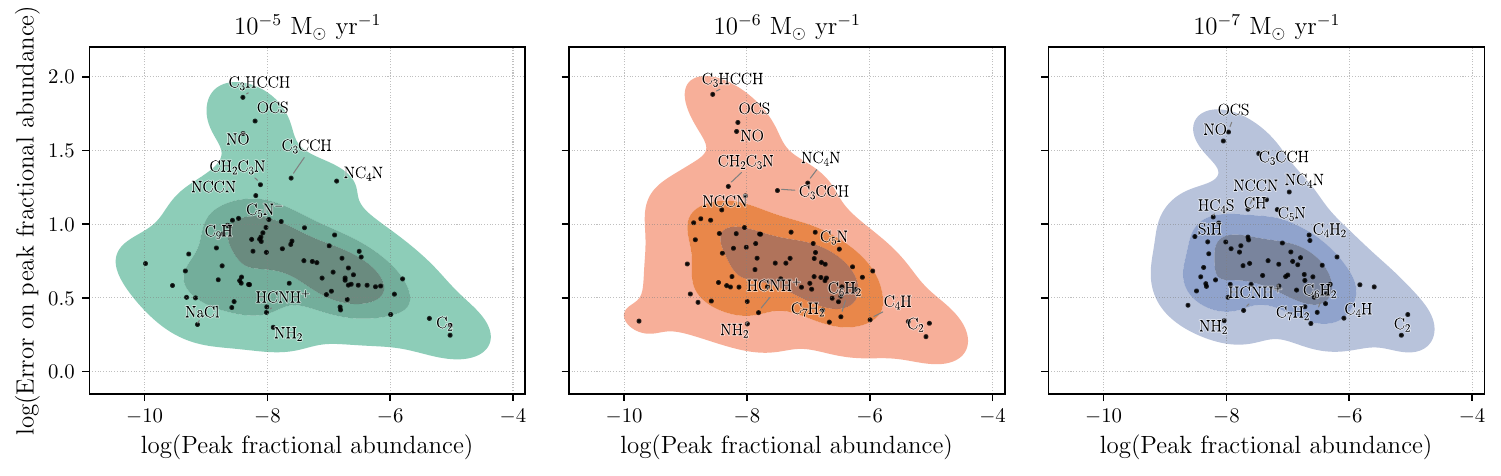}
%  \caption{1a}
  \label{fig:sfig1}
\end{subfigure}%

\vspace{-1em}

\begin{subfigure}[b]{0.85\textwidth}
  \centering
	\includegraphics[width=\textwidth]{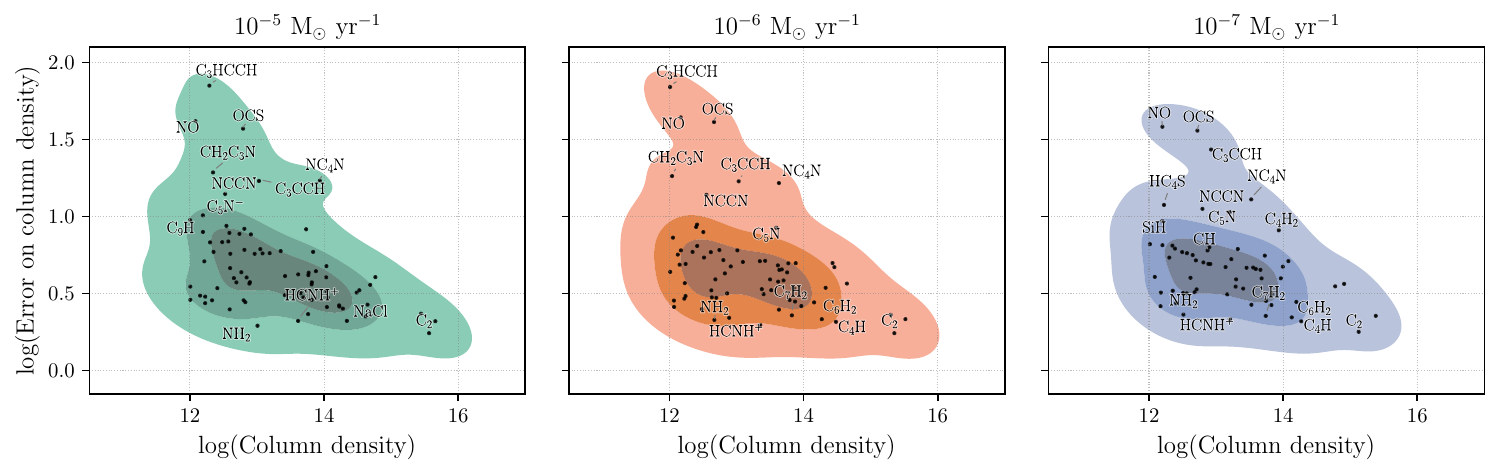}
%  \caption{1a}
  \label{fig:sfig1}
\end{subfigure}%
\caption{Scatter plots for all abundant species in C-rich outflows across all mass-loss rates.
\emph{Top:} mean peak fractional abundance versus the error on the peak abundance. 
\emph{Middle:} mean column density versus its error. 
\emph{Bottom:} dispersion of the column density versus the dispersion of the peak fractional abundance.
Shaded contours show the kernel density estimation of the point distribution, with three intensity levels indicating regions of increasing point concentration.
Least and most precise daughters( dispersion of both peak abundance and column density larger/smaller than their mean plus standard deviation) are labeled.
}
\label{fig:app-crich-relevant}
\end{figure*}

\begin{figure*}
\centering

\begin{subfigure}[b]{0.85\textwidth}
  \centering
	\includegraphics[width=\textwidth]{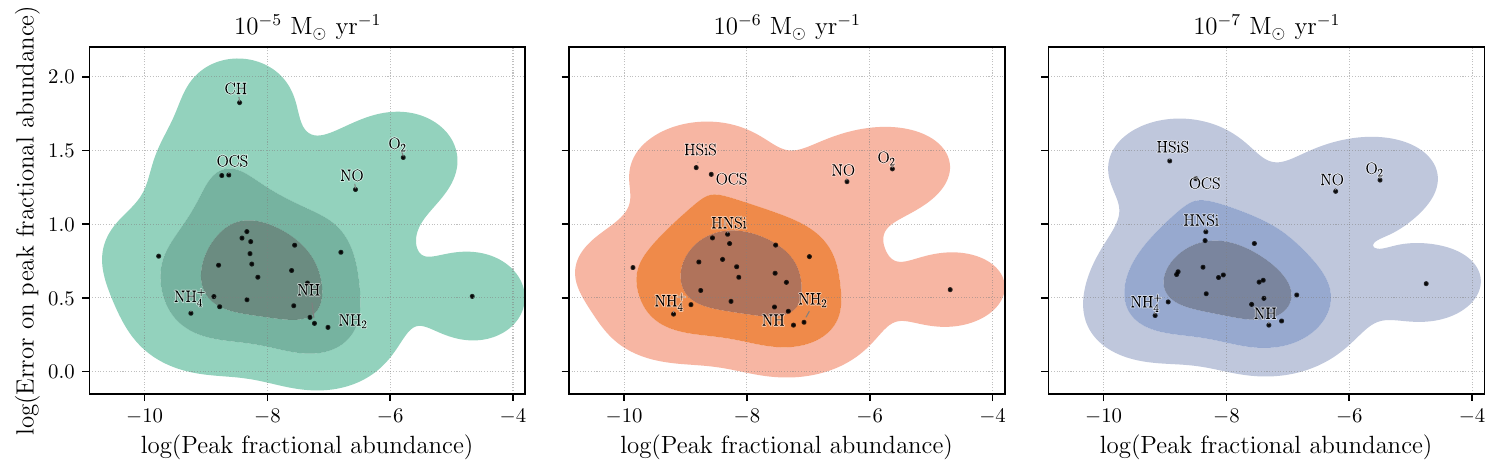}
%  \caption{1a}
  \label{fig:sfig1}
\end{subfigure}%

\vspace{-1em}

\begin{subfigure}[b]{0.85\textwidth}
  \centering
	\includegraphics[width=\textwidth]{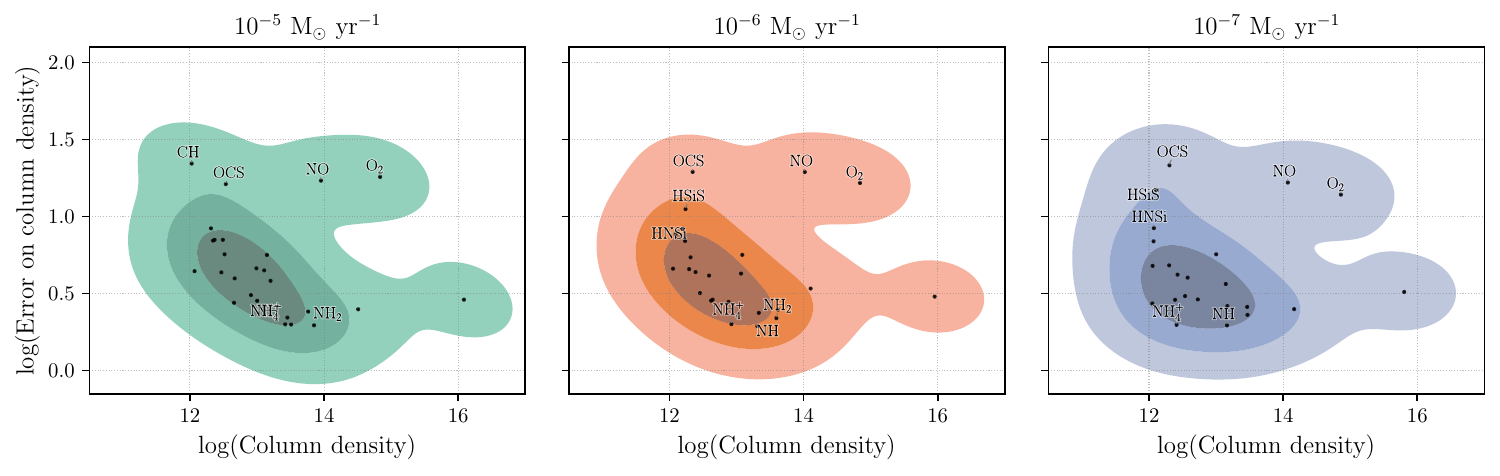}
%  \caption{1a}
  \label{fig:sfig1}
\end{subfigure}%
\caption{Same as Fig. \ref{fig:app-crich-relevant}, but for O-rich outflows across all mass-loss rates.
}
\label{fig:app-orich-relevant}
\end{figure*}

\begin{figure*}
\centering

\begin{subfigure}[t]{1.0\textwidth}
  \centering
	\includegraphics[width=\textwidth]{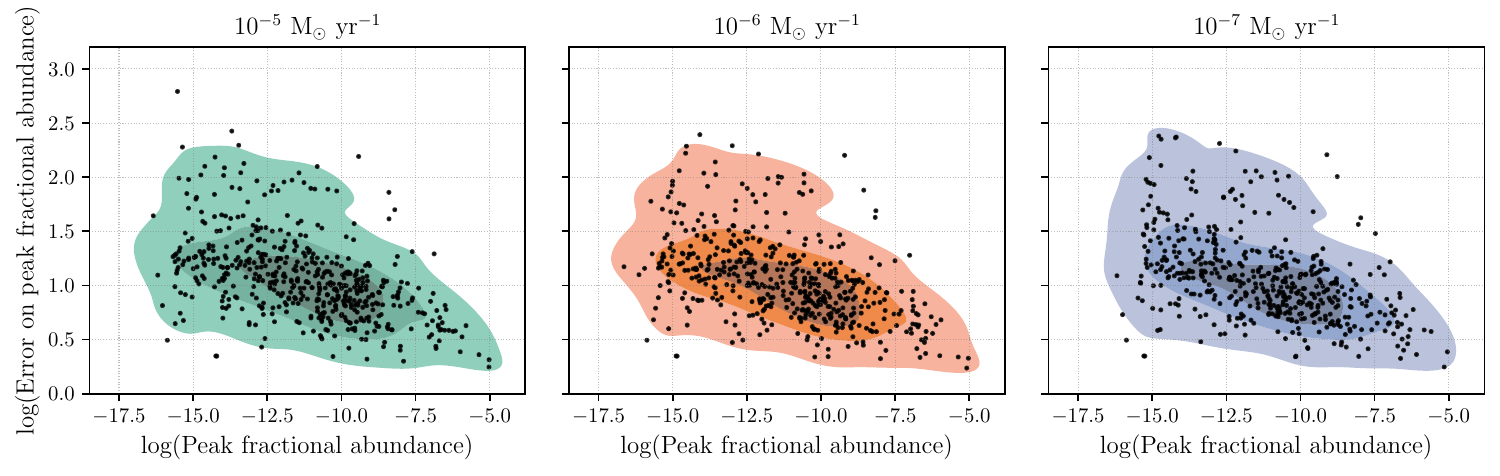}
%  \caption{1a}
  \label{fig:sfig1}
\end{subfigure}%

\vspace{-1em}

\begin{subfigure}[t]{1.0\textwidth}
  \centering
	\includegraphics[width=\textwidth]{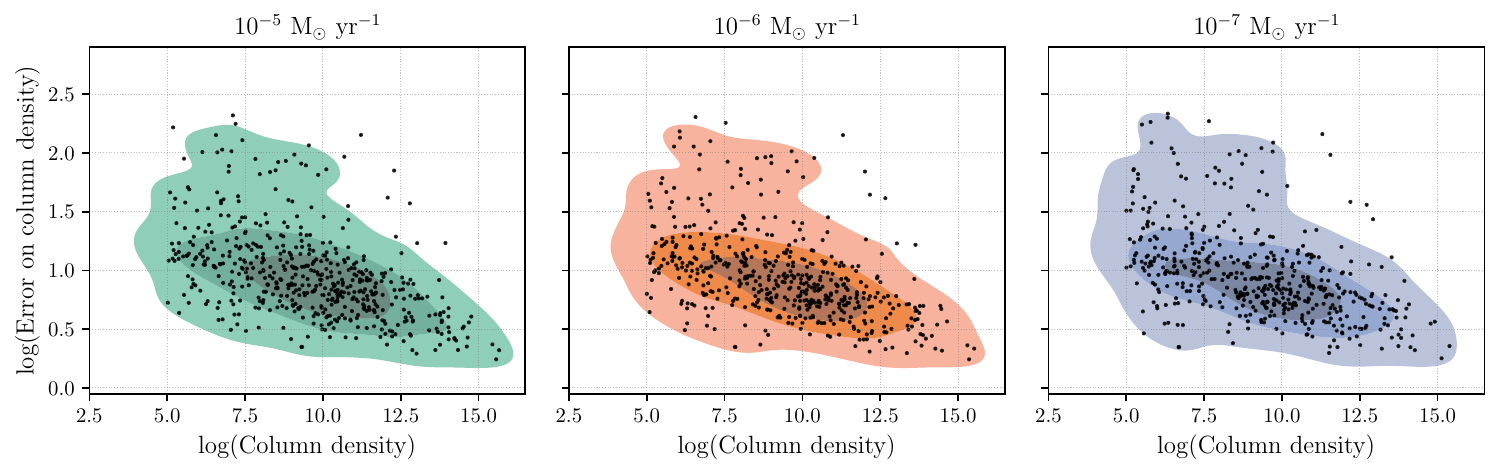}
%  \caption{1a}
  \label{fig:sfig1}
\end{subfigure}%

\vspace{-1em}

\begin{subfigure}[t]{1.0\textwidth}
  \centering
	\includegraphics[width=\textwidth]{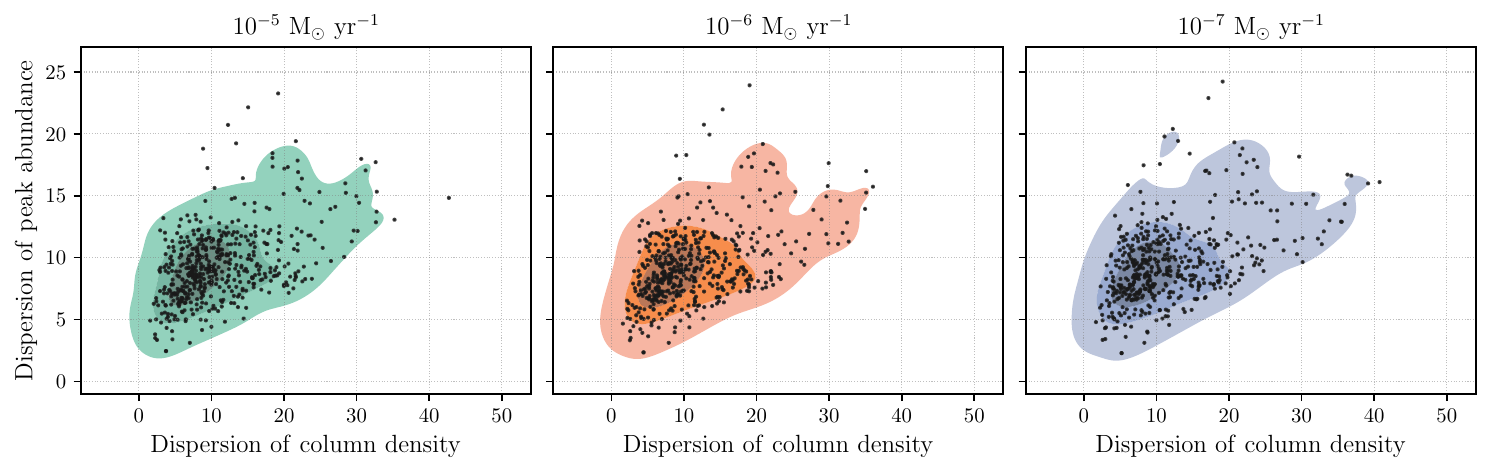}
%  \caption{1a}
  \label{fig:sfig1}
\end{subfigure}%

\caption{Scatter plots for all daughter species in a C-rich outflow across all mass-loss rates.
\emph{Top:} mean peak fractional abundance versus the error on the peak abundance. 
\emph{Middle:} mean column density versus its error. 
\emph{Bottom:} dispersion of the column density versus the dispersion of the peak fractional abundance.
The daughter species have a column density $\geq 10^{5}$ cm$^{-2}$.  
}
\label{fig:app-crich-all}
\end{figure*}

\begin{figure*}
\centering

\begin{subfigure}[t]{1.0\textwidth}
  \centering
	\includegraphics[width=\textwidth]{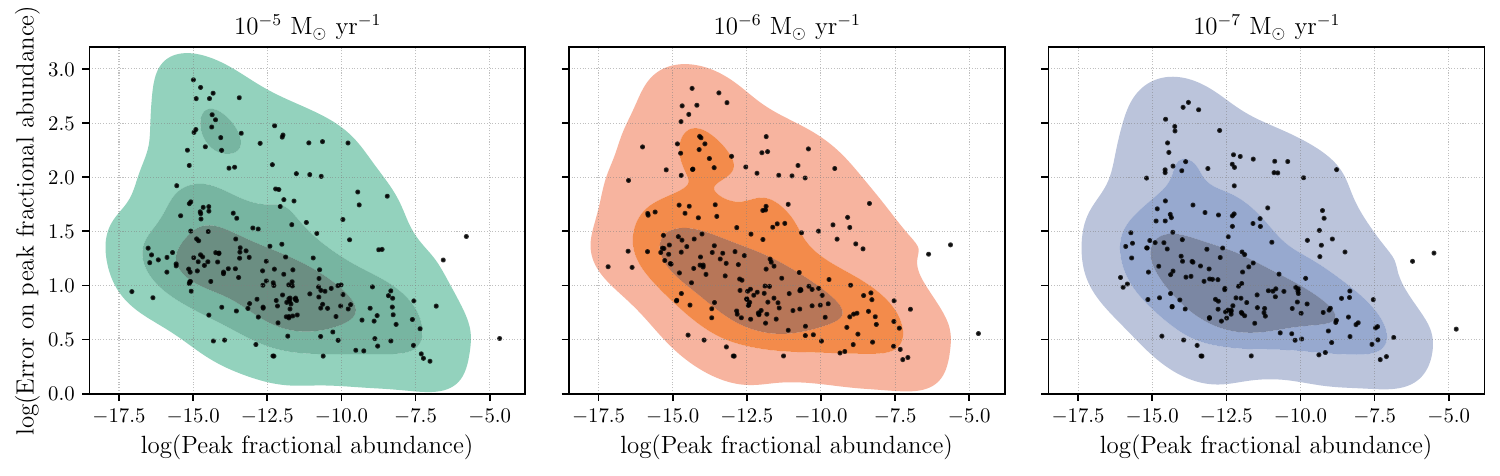}
%  \caption{1a}
  \label{fig:sfig1}
\end{subfigure}%

\vspace{-1em}

\begin{subfigure}[t]{1.0\textwidth}
  \centering
	\includegraphics[width=\textwidth]{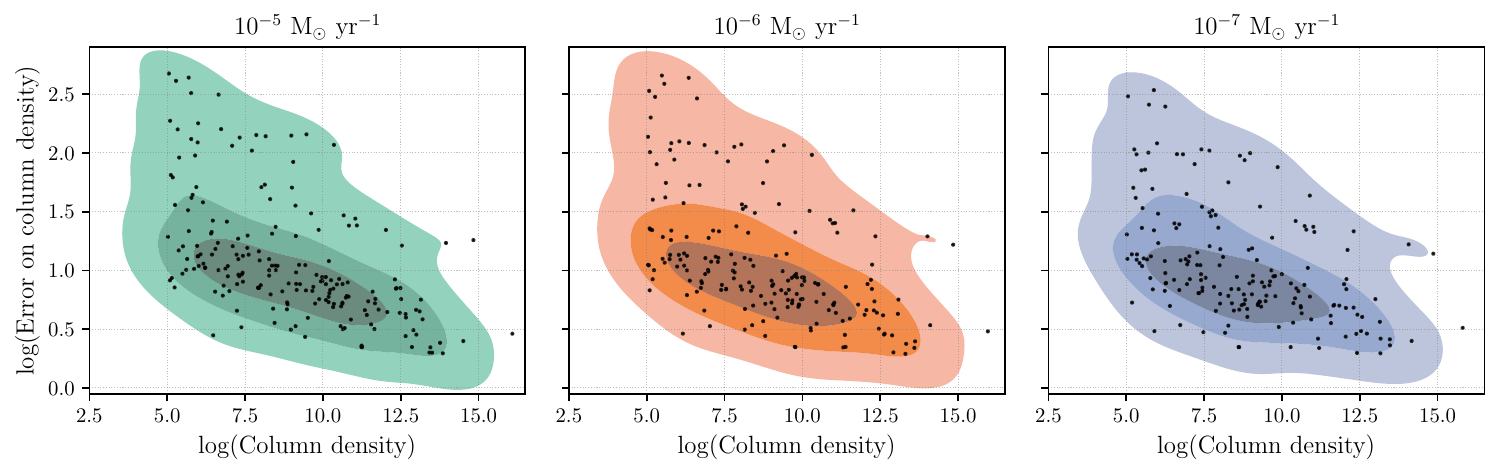}
%  \caption{1a}
  \label{fig:sfig1}
\end{subfigure}%

\vspace{-1em}

\begin{subfigure}[t]{1.0\textwidth}
  \centering
	\includegraphics[width=\textwidth]{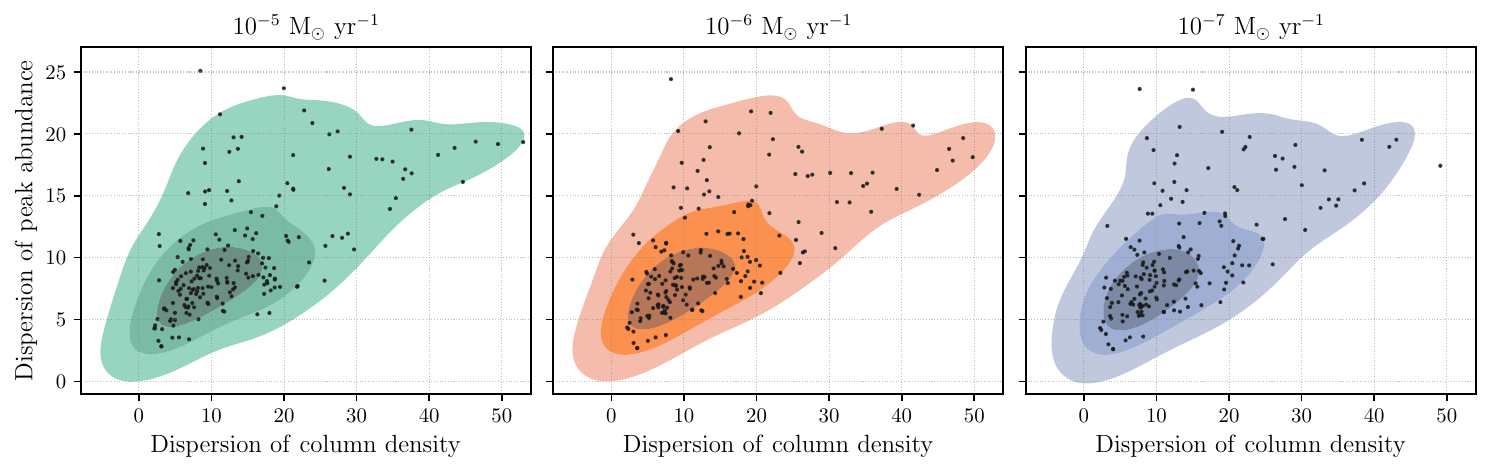}
%  \caption{1a}
  \label{fig:sfig1}
\end{subfigure}%

\caption{Scatter plots for all daughter species in a O-rich outflow across all mass-loss rates.
\emph{Top:} mean peak fractional abundance versus the error on the peak abundance. 
\emph{Middle:} mean column density versus its error. 
\emph{Bottom:} dispersion of the column density versus the dispersion of the peak fractional abundance.
The daughter species have a column density $\geq 10^{5}$ cm$^{-2}$.  
}
\label{fig:app-orich-all}
\end{figure*}

%%%%%%%%%%%%%%%%%%%%%%%%%%%%%%%%%%%%%%%%%%%%%%%%%%%%%%%%%%%%%%%%%%%%%%%%%%%%%%%%%%%%%%%%%%%%%%%%%%%%%%%%%%%%%%
\section{CO radiative transfer models}				\label{app:RT}
%%%%%%%%%%%%%%%%%%%%%%%%%%%%%%%%%%%%%%%%%%%%%%%%%%%%%%%%%%%%%%%%%%%%%%%%%%%%%%%%%%%%%%%%%%%%%%%%%%%%%%%%%%%%%%

The \citet{Mamon1988} prescription parametrises the CO abundance profile using
\begin{equation}		\label{eq:mamon}
	x(r) = x_0 \exp\left( -\ln 2\ \left(\frac{r}{r_{1/2}}\right)^\alpha\right),
\end{equation}
with $x$ the CO fractional abundance, $x_0$ its initial abundance, $r_{1/2}$ the radius in cm where its abundance is half its initial value, and $\alpha$ a measure of the profile's steepness.
The values of $r_{1/2}$ and $\alpha$ used in the radiative transfer modelling, for the mass-loss rates and expansion velocities given in Table \ref{table:model-params} and initial abundances given in Table \ref{table:model-parents}, were retrieved using the equations described in \citet{Schoier2001} and are listed in Table \ref{table:mamon}.

Radiative transfer models were calculated using the smallest and largest predicted CO envelope sizes (Table \ref{table:uncert-parents}) along with the standard \citet{Mamon1988} sizes.
This is illustrated in Fig.~\ref{fig:RT-Orich-low}, which shows the CO abundance profiles and calculated lines for the O-rich outflow with $\dot{M} = 10^{-7}\ \mathrm{M}_\odot\ \mathrm{yr}^{-1}$.
Table \ref{table:RT} lists the CO $J=1 \to 0$ and $J = 2 \to 1$ integrated intensities and the mass-loss rates predicted from these values, along with the percentage difference of these values, determined as (predicted value from maximum extent - predicted value from minimum extent)/predicted value from \citet{Mamon1988}.

\begin{table}
	\centering
	\caption{Values of $r_{1/2}$ [cm] and $\alpha$ of the \citet{Mamon1988} CO envelope size (Eq. \ref{eq:mamon}) used in the radiative transfer modelling.
	}
	\begin{tabular}{c c c} 
	\hline
	\noalign{\smallskip}
		$\dot{M}$ & $r_{1/2}$ & $\alpha$ \\
    \noalign{\smallskip}
	\hline
      \noalign{\smallskip}

     \multicolumn{3}{c}{Carbon-rich} \\  
	\hline
	 \noalign{\smallskip}
    $10^{-5}\ \mathrm{M}_\odot\ \mathrm{yr}^{-1}$	& $2.80 \times 10^{17}$ & $2.79$ \\
    $10^{-6}\ \mathrm{M}_\odot\ \mathrm{yr}^{-1}$	& $8.13 \times 10^{16}$ & $2.35$ \\
    $10^{-7}\ \mathrm{M}_\odot\ \mathrm{yr}^{-1}$	& $2.46 \times 10^{16}$ & $2.04$ \\
    \noalign{\smallskip}
	\hline
    \noalign{\smallskip}
	\multicolumn{3}{c}{Oxygen-rich} \\  
	\hline
	 \noalign{\smallskip}
%	$\dot{M}$ & $r_{1/2}$ & $\alpha$ \\
%	\hline
    $10^{-5}\ \mathrm{M}_\odot\ \mathrm{yr}^{-1}$	& $1.48 \times 10^{17}$ & $2.79$ \\
    $10^{-6}\ \mathrm{M}_\odot\ \mathrm{yr}^{-1}$	& $4.44 \times 10^{16}$ & $2.35$ \\
    $10^{-7}\ \mathrm{M}_\odot\ \mathrm{yr}^{-1}$	& $t1.54 \times 10^{16}$ & $2.03$ \\
	\hline
	\end{tabular}
    \label{table:mamon}    
\end{table}

\begin{figure*}
\centering
\includegraphics[width=\textwidth]{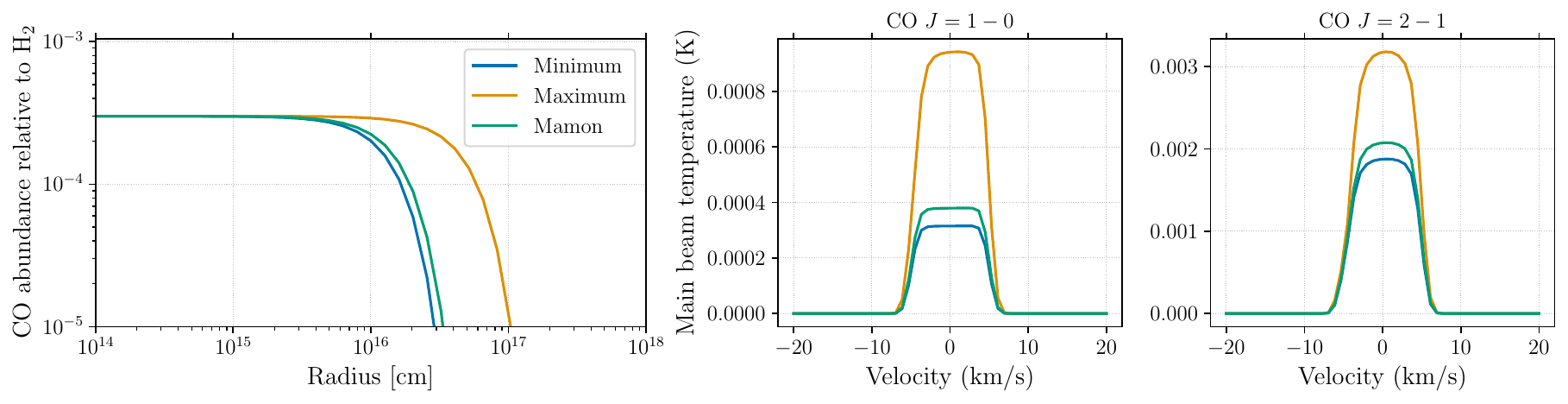}
\caption{
\emph{Left}: CO abundance profiles used in the radiative transfer modelling of the O-rich outflow with $\dot{M} = 10^{-7}\ \mathrm{M}_\odot\ \mathrm{yr}^{-1}$; the minimum and maximum predicted CO envelope sizes (Table \ref{table:uncert-parents}) and the standard \citet{Mamon1988} size.
\emph{Right panels}: corresponding CO $J = 1 \to 0$ and $J=2-1$ line profiles calculated by the radiative transfer routine.
}
\label{fig:RT-Orich-low}
\end{figure*}

\begin{table*}
	\centering
	\caption{Radiative transfer results for the C-rich outflows (top) and O-rich outflows(bottom), for all mass-loss rates (\emph{1st column}) using the minimum and maximum extent (Table \ref{table:uncert-parents}) and the \citet{Mamon1988} envelope size (Table \ref{table:mamon}, \emph{2nd column}).
\emph{3rd \& 6th columns}: integrated intensity of the CO $J = 2\to 1$ and $J = 3 \to 2$ lines, respectively.
\emph{4th \& 7th columns}: percentage difference of the integrated intensities.
\emph{5th \& 8th columns}: predicted mass-loss rates based on these intensities, following \citet{Ramstedt2008}.
\emph{6th \& 10th columns}: percentage difference of the predicted mass-loss rates.
Percentage differences are calculated as (predicted value from maximum extent - predicted value from minimum extent)/predicted value from \citet{Mamon1988}.
	}
	\begin{tabular}{C L C C C C C C C C} 
	\hline
	\noalign{\smallskip}
	\mathrm{Mass-loss\ rate} & \mathrm{Envelope\ size} & \mathrm{CO\ }J=1 \to 0  & \mathrm{Ratio} & \mathrm{Predicted\ } \dot{M} &\mathrm{Ratio\ }\dot{M}& \mathrm{CO\ } J=2 \to 1  & \mathrm{Ratio} & \mathrm{Predicted\ } \dot{M} &   \mathrm{Ratio\ }\dot{M}\\ 
		&  & \mathrm{[K \ km/s]}  & \mathrm{intensities} & \ &  & \mathrm{[K \ km/s]}  & \mathrm{intensities} & & \\ 
	\hline
	\noalign{\smallskip}
     \multicolumn{10}{c}{Carbon-rich} \\  
	\noalign{\smallskip}
	\hline
	\noalign{\smallskip}
\multirow{3}{*}{$10^{-5}\ \mathrm{M}_\odot\ \mathrm{yr}^{-1}$}  & \mathrm{Minimum} &	6.69\times 10^{-1}	&	\multirow{3}{*}{0.00\%}	&	4.38\times 10^{-6}	&	\multirow{3}{*}{-0.03\%}	&	1.57				&	\multirow{3}{*}{0.00\%}	&	4.40\times 10^{-6}	&	\multirow{3}{*}{0.00\%}	\\
& \mathrm{Maximum} &	6.69\times 10^{-1}	&			&	4.38\times 10^{-6}	&			&	1.57				&			&	4.40\times 10^{-6}	&		\\
& \mathrm{Mamon} &	7.06\times 10^{-1}	&			&	4.54\times 10^{-6}	&			&	1.57				&			&	4.41\times 10^{-6}	&		\\
\noalign{\smallskip}
\multirow{3}{*}{$10^{-6}\ \mathrm{M}_\odot\ \mathrm{yr}^{-1}$} & \mathrm{Minimum} &	9.26\times 10^{-2}	&	\multirow{3}{*}{37.50\%}	&	8.99\times 10^{-7}	&	\multirow{3}{*}{25.92\%}	&	3.01\times 10^{-1}	&	\multirow{3}{*}{15.60\%}	&	9.44\times 10^{-7}	&	\multirow{3}{*}{12.87\%}	\\
& \mathrm{Maximum} &	1.38\times 10^{-1}	&			&	1.18\times 10^{-6}	&			&	3.53\times 10^{-1}	&			&	1.08\times 10^{-6}	&		\\
& \mathrm{Mamon} &	1.20\times 10^{-1}	&			&	1.07\times 10^{-6}	&			&	3.37\times 10^{-1}	&			&	1.04\times 10^{-6}	&		\\
\noalign{\smallskip}
\multirow{3}{*}{$10^{-7}\ \mathrm{M}_\odot\ \mathrm{yr}^{-1}$} & \mathrm{Minimum} &	1.10\times 10^{-2}	&	\multirow{3}{*}{68.80\%}	&	1.40\times 10^{-7}	&	\multirow{3}{*}{44.85\%}	&	4.33\times 10^{-2}	&	\multirow{3}{*}{28.10\%}	&	1.40\times 10^{-7}	&	\multirow{3}{*}{22.86\%}	\\
& \mathrm{Maximum} &	2.02\times 10^{-2}	&			&	2.12\times 10^{-7}	&			&	5.66\times 10^{-2}	&			&	1.75\times 10^{-7}	&		\\
& \mathrm{Mamon} &	1.34\times 10^{-2}	&			&	1.60\times 10^{-7}	&			&	4.75\times 10^{-2}	&			&	1.51\times 10^{-7}	&		\\
	\hline
	\noalign{\smallskip}
     \multicolumn{10}{c}{Oxygen-rich} \\  
	\noalign{\smallskip}
	\hline
	\noalign{\smallskip}
\multirow{3}{*}{$10^{-5}\ \mathrm{M}_\odot\ \mathrm{yr}^{-1}$}&	\mathrm{Mininum}	&	3.91\times 10^{-1}	&	\multirow{3}{*}{16.70\%}	&	6.67\times 10^{-6}	&	\multirow{3}{*}{11.50\%}	&	1.08	&	\multirow{3}{*}{7.60\%}	&	5.79\times 10^{-6}	&	\multirow{3}{*}{6.30\%}	\\
&	\mathrm{Maximum}	&	4.65\times 10^{-1}	&		&	7.50\times 10^{-6}	&		&	1.17	&		&	6.17\times 10^{-6}	&		\\
&	\mathrm{Mamon}	&	4.43\times 10^{-1}	&		&	7.25\times 10^{-6}	&		&	1.14	&		&	6.07\times 10^{-6}	&		\\
\noalign{\smallskip}
\multirow{3}{*}{$10^{-6}\ \mathrm{M}_\odot\ \mathrm{yr}^{-1}$}&	\mathrm{Mininum}	&	1.08\times 10^{-1}	&	\multirow{3}{*}{122.00\%}	&	2.18\times 10^{-6}	&	\multirow{3}{*}{57.62\%}	&	3.38\times 10^{-1}	&	\multirow{3}{*}{38.00\%}	&	1.85\times 10^{-6}	&	\multirow{3}{*}{26.75\%}	\\
&	\mathrm{Maximum}	&	1.59\times 10^{-1}	&		&	2.85\times 10^{-6}	&		&	3.98\times 10^{-1}	&		&	2.12\times 10^{-6}	&		\\
&	\mathrm{Mamon}	&	4.23\times 10^{-2}	&		&	1.16\times 10^{-6}	&		&	1.58\times 10^{-1}	&		&	9.93\times 10^{-7}	&		\\
\noalign{\smallskip}
\multirow{3}{*}{$10^{-7}\ \mathrm{M}_\odot\ \mathrm{yr}^{-1}$}&	\mathrm{Mininum}	&	3.14\times 10^{-3}	&	\multirow{3}{*}{154.60\%}	&	1.31\times 10^{-7}	&	\multirow{3}{*}{91.96\%}	&	1.73\times 10^{-2}	&	\multirow{3}{*}{57.50\%}	&	1.18\times 10^{-7}	&	\multirow{3}{*}{45.71\%}	\\
&	\mathrm{Maximum}	&	8.97\times 10^{-3}	&		&	2.68\times 10^{-7}	&		&	2.82\times 10^{-2}	&		&	1.76\times 10^{-7}	&		\\
&	\mathrm{Mamon}	&	3.77\times 10^{-3}	&		&	1.48\times 10^{-7}	&		&	1.90\times 10^{-2}	&		&	1.27\times 10^{-7}	&		\\	\hline
	\end{tabular}
    \label{table:RT}    
\end{table*}

%%%%%%%%%%%%%%%%%%%%%%%%%%%%%%%%%%%%%%%%%%%%%%%%%%

% Don't change these lines
\bsp	% typesetting comment
\label{lastpage}
\end{document}